
\documentclass[twocolumn,times,tighten]{aastex63}
\usepackage{color}
\usepackage{enumitem}
\usepackage{hyperref}
\usepackage{verbatim}
\usepackage{amsmath,amstext,mathrsfs}
\usepackage[all]{hypcap} 
\usepackage{afterpage}    
\usepackage{marginnote}
\usepackage{makecell}
\usepackage{subfigure}
\usepackage{xfrac}

\definecolor{gg}{RGB}{0,89,76}

\DeclareFontFamily{OMS}{oasy}{\skewchar\font48 }
\DeclareFontShape{OMS}{oasy}{m}{n}{%
         <-5.5> oasy5     <5.5-6.5> oasy6
      <6.5-7.5> oasy7     <7.5-8.5> oasy8
      <8.5-9.5> oasy9     <9.5->  oasy10
      }{}
\DeclareFontShape{OMS}{oasy}{b}{n}{%
       <-6> oabsy5
      <6-8> oabsy7
      <8->  oabsy10
      }{}
\DeclareSymbolFont{oasy}{OMS}{oasy}{m}{n}
\SetSymbolFont{oasy}{bold}{OMS}{oasy}{b}{n}

\DeclareMathSymbol{\smallleftarrow}     {\mathrel}{oasy}{"20}
\DeclareMathSymbol{\smallrightarrow}    {\mathrel}{oasy}{"21}
\DeclareMathSymbol{\smallleftrightarrow}{\mathrel}{oasy}{"24}

\defcitealias{LP00}{LP00}
\defcitealias{LP12}{LP12}

\newcommand*{\kh}{} % This is KH's color
\newcommand*{\re}{}

\newcommand*{\torefereeone}{}
\newcommand*{\torefereetwo}{}

\graphicspath{{./}{}}

\shorttitle{B-Field Strength with DMA}
\shortauthors{Lazarian, Yuen \& Pogosyan}
\submitjournal{ApJ}

\begin{document}

\title{Magnetic field strength from turbulence theory (I): Using differential measure approach (DMA)}

\author[0000-0002-7336-6674]{A. Lazarian}
\affiliation{Department of Astronomy, University of Wisconsin-Madison, USA}
\affiliation{Centro de Investigación en Astronomía, Universidad Bernardo O’Higgins, Santiago, General Gana 1760, 8370993,
Chile}
\email{alazarian@facstaff.wisc.edu}

\author[0000-0003-1683-9153]{Ka Ho Yuen}
\affiliation{Department of Astronomy, University of Wisconsin-Madison, USA}
\email{kyuen@astro.wisc.edu}

\author[0000-0002-7998-6823]{Dmitri Pogosyan}
\affiliation{Department of Physics, University of Alberta, Edmonton, Canada}
\affiliation{Korea Institute for Advanced Studies, Seoul, Republic of Korea}
\email{pogosyan@ualberta.ca}

\begin{abstract}
The mean plane-of-sky magnetic field strength is traditionally obtained from the combination of polarization and spectroscopic data using the Davis-Chandrasekhar-Fermi (DCF) technique. However, we identify the major problem of the DCF to be its disregard of the anisotropic character of MHD turbulence. On the basis of the modern MHD turbulence theory we introduce a new way of obtaining magnetic field strength from observations. Unlike the DCF, the new technique uses not the dispersion of the polarization angle and line of sight velocities, but increments of these quantities given by the structure functions. To address the variety of the astrophysical conditions for which our technique can be applied, we consider the turbulence in both media with magnetic pressure larger than the gas pressure corresponding e.g. to molecular and the gas pressure larger than the magnetic pressure corresponding to the warm neutral medium. We provide general expressions for arbitrary admixture of Alfv\'en, slow and fast modes in these media and consider in detail the particular cases relevant to diffuse media and molecular clouds. We successfully test our results using synthetic observations obtained from MHD turbulence simulations. We demonstrate that our Differential Measure Approach (DMA), unlike the DCF, can be used to measure the distribution of magnetic field strengths, can provide magnetic field measurements with limited data and is much more stable in the presence of large scale variations induces of non-turbulent nature. In parallel, our study uncover the deficiencies of the earlier DCF research. 
\end{abstract}


\section{INTRODUCTION}

The role of magnetic fields in astrophysics is difficult to overestimate. Magnetic force is the second most important force in the present day Universe after gravity. The magnetic field plays an important role at different stages of star formation (e.g. \citealt{1956MNRAS.116..503M,2006ApJ...647..374G,2006ApJ...646.1043M,2007IAUS..243...31J}). In view of astrophysical flows with large Reynolds numbers the magnetic fields are turbulent (see  \citealt{2004ARA&A..42..211E,MO07}, McKee \& Stone 2021). The evidence of turbulent magnetic field is coming from observations of density structure of the interstellar medium (e.g. \citealt{1995ApJ...443..209A,CL09}) and velocity fluctuation studies (\citealt{1981MNRAS.194..809L,2004ApJ...615L..45H,2010ApJ...710..853C}, Chepurnov et al. 2016, \citealt{VDA,cattail,spectrum}) and synchrotron polarization studies (\citealt{LY18b,2020NatAs.tmp..174Z}). In particular the detailed properties of the turbulent velocity can be obtained using using the theory of mapping of velocity fluctuations in real space into the fluctuations of intensity in the Position-Position-Velocity (PPV) space in Lazarian \& Pogosyan (2000) and subsequent publications (Lazarian \& Pogosyan 2004, 2006, 2008, Kandel et al. 2017, 2018). The theory is used in \cite{VDA} to develop Velocity Decomposition Algorithm (VDA) approach that further improves the mapping of velocity statistics. The grand velocity cascade spanning from the large galactic scales ($10^4$ to $10^{-2}$ pc reported in \cite{spectrum} is the example of what sort of detailed information is {\torefereetwo becoming} available combining the advances of theory with the modern spectroscopic measurements.  

It is also important that improvement in resolution and sensitivity of instruments (see Cortes et al. 2021, Henley et al. 2021) allows to map polarization of hundreds of molecular clouds with unprecedented accuracy, obtaining from dozens to hundreds of polarization vectors per cloud. The distribution of these vectors reflects POS distribution of magnetic field. The latter is determined by the interplay of the randomizing effects of turbulence and aligning effect of magnetic tension.  The quantitative description of the two effects should relate the magnetic field strength with the intensity of turbulent motions. The latter can be obtained from the measurements of turbulent velocity fluctuations. The availability of detailed information of the distribution of both magnetic field fluctuations and turbulent velocity fluctuations calls for the development of quantitative techniques that can provide the detailed distribution of magnetic field strength. The latter being essential for the quantitative understanding of the role of magnetic field in molecular clouds, including the processes of star formation (see McKee \& Ostriker 2007).   

The well-known attempt to get the information using the logic above, unfortunately, provides not a distribution of B-strength on the plane of sky, but a single estimate for a map. In addition, this estimate is known to be of insufficient accuracy.  The technique that employs the observed fluctuations of polarization together with the fluctuations of the Doppler-shifted velocities of gas was proposed by  \cite{1951PhRv...81..890D} and, independently, \cite{CF53}.  Their technique termed Davis-Chandrasekhar-Fermi technique (henceforth DCF), assumes that magnetic field fluctuations and those of the velocity are directly related through the averaged Alfv\'en relation \citep{1942Natur.150..405A}:
\begin{equation}
\delta B=\delta v  \sqrt{ 4\pi \langle \rho\rangle },
\label{eq:alf}
\end{equation}
where $\delta B$ is assumed to be perpendicular to the mean magnetic field and $\langle \rho\rangle$ is the mean density, which can be obtained through independent observations. It is natural to assume that velocity fluctuations with $\delta v<V_A$, where $V_A$ is the Alfv\'en velocity, induce the deviations of the underlying field direction by an angle $\delta \phi \approx \delta B/B_{mean}$. If we associate $\delta v$ and $\delta \phi$ with, respectively, the dispersion of velocities $\sigma_v$ and angles $\sigma_\phi$ available from observations, Eq.(\ref{eq:alf}) can be used to express the mean magnetic field strength through these observables: 
\begin{equation}
B_{mean}= f_{DCF} \sqrt{4\pi \langle \rho \rangle} \frac{\sigma_v}{\sigma_\phi}
\label{eq:BB}
\end{equation}
where $f_{DCF}$ is an adjustable factor that traditionally assumed to be a constant. Another derivation that does not directly refer to Alfv\'en modes, but assumes that magnetic turbulence and kinetic energies are at {\torefereetwo equipartition} is provided in Li et al. (2021). In fact, this assumption also corresponds to Eq.~(\ref{eq:alf}). For nearly incompressible runs this assumption corresponds to numerical studies of sub-Alfv\'enic turbulence \citep{2004PhRvE..70c6408H} at the injection scale where the DCF measurements are done. However, even at this situation, one should ask a question what fraction of energy is being in the fluctuations of magnetic field parallel to mean magnetic field and which part corresponds to the fluctuations perpendicular to the mean magnetic field. In the DCF, only the dispersion of the latter is being measured. According to Cho \& Lazarian (2003), the energies in the aforementioned components are nearly equal for incompressible turbulence, which gives the factor $f_{DCF}\approx 1/2$ in Eq. (\ref{eq:BB}). When the compressibility gets important, more energy is being transferred into gas compression.

As a result, $f_{DCF}$ is, in fact, a function that depends on the actual properties of turbulence. This oversimplification results in DCF having a reputation of an inaccurate technique that can provide only order of magnitude estimates (see \citealt{2021MNRAS.tmp.3119L}). Naturally, this decreases the scientific output of polarization studies.  Nevertheless, attempts to improve the DCF accuracy have been numerous  (e.g. \citealt{2001ApJ...561..800H,2004Ap&SS.292..225C,2004ApJ...616L.111H,2006Sci...313..812G,Fal08}), but the foundations of the technique stayed the same. More recent notable suggestions include \cite{2009ApJ...696..567H,CY16} and Skalidis \& Tassis (2021), and we shall discuss further in our paper why we do not believe that these attempts can solve the challenges that the DCF faces.  

The inability of the DCF to capitalize on modern high resolution polarization and spectroscopic maps stems both from its use of dispersion of angle $\phi$, which implies averaging over the large areas of sky. The inaccuracy of the DCF arises mostly from the inadequate model of turbulence that is adopted in the derivation of the technique. Indeed, the DCF assumes that the perturbations of magnetic field direction arise from the collection of Alfv\'en waves moving along the mean magnetic field. This assumption, however, disregards the actual properties of \textit{MHD turbulence} (see Beresnyak \& Lazarian 2019 and ref. therein). Anisotropy of MHD turbulence is its fundamental property and the disregard of this property is a serious deficiency of the DCF approach.\footnote{In addition to theoretical deficiencies, the DCF suffers also from the fact that it relies on dispersions measured at large scales. In fact, the magnetic field evaluation using Eq. (\ref{eq:BB}) are affected by motions and magnetic field distortions of non-turbulent origin, For instance, large scale galactic shear can significantly modifies both $\delta v$ and $\delta \phi$ (See e.g., Cho 2019) and this decreases the accuracy of the magnetic field determination.} 

The issue of angle $\gamma$ between the magnetic field and the line of sight is usually not discussed within the DCF technique and the numerical testing are done for $\gamma=\pi/2$. However, it is easy to see that the DCF expressions should change as $\gamma$ changes. In particular, repeating the arguments that brought Eq. (\ref{eq:BB}), but taking into account that the magnetic field is inclined in respect to the line of sight, we can obtain a modified expression for the DCF in the form:
\begin{equation}
B_{\bot}=B\sin \gamma= f_{DCF} \sqrt{4\pi \langle \rho \rangle} \frac{\sigma_v}{\sigma_\phi},
\label{eq:BBsin}
\end{equation}
where $B_\bot$ is the POS component of magnetic field. This is still an oversimplification that does not account for the actual properties of MHD turbulence.

To deal with the aforementioned deficiencies of the DCF, in this paper we propose an alternative technique of measuring magnetic field strength. Our approach employs a realistic description of velocity and magnetic field turbulent fluctuations based on the modern theory of MHD turbulence \citep{LP12,KLP17a} and suggests measurements of velocity and angle variations on small scales using the structure functions.

Apart from addressing the issue of magnetic field for the evolution of molecular clouds and star formation, the present study that opens avenues for providing the {\it distribution} of magnetic field strength is important for understanding the properties of galactic foregrounds. Indeed, the recent progress of cosmological studies raise the importance of the detailed information of polarized galactic foreground to an unprecedented level. This is, first of all, is motivated by the search of the polarization arising from the enigmatic B-modes related to the gravitational waves in the early Universe (Kaminkowsky \& Kovetz 2016). An additional motivation comes from the attempts to study emission of polarized atomic hydrogen lines (e.g. Zhang et.al 2020, 2021) potentially at higher redshifts. The latter emission is not polarized at the source, but the effects of polarized foreground, nevertheless, seriously complicates the procedure of the foreground removal. Although the latter effect is related to the instrumental polarization, it is extremely challenging to get the signal when the unknown polarized foreground is a million times stronger than the signal.

In \S \ref{ref:ISM} we discuss the properties of turbulent interstellar medium to which we seek the application of our technique. In \S \ref{sec:theory} we introduce the observables that we deal with,  as well as a new way of  obtaining of magnetic field strength employing structure functions measured at the scales less than the turbulence injection scale. We demonstrate the advantages of using our differential measures compared to the global measures employed in the DCF. 
In \S \ref{sec:analytical_numerical} we introduce our approach of using the theory of MHD turbulence to accurately calculate the structure functions of velocity and positional angle $\phi$. There we also explain how the anisotoropy of turbulence induces "anomalous" scaling of structure functions of $\phi$ with media magnetization. The projection of basic MHD modes are discussed in \S \ref{sec:modes}. The exact expressions for obtaining $B$-strength for pure Alfv\'enic and weakly compressible MHD turbulence are provided in \S \ref{sec:pure_alfven}. Obtaining magnetic field strength in the media similar to molecular clouds is quantified in \S \ref{subsec:lowbeta}. In \S \ref{sec:numerical} we provide the input of our compressible MHD simulations to guide us in finding which of the theoretically discussed cases are applicable to actual interstellar settings. We provide a set of simplified equations and practical recommendations for the analysis of the observational data in \S \ref{sec:practical}. In \S \ref{sec:compare} we compare our technique to other existing techniques.  We discuss the advantages and prospects for the our new  technique in \S \ref{sec:discussion} and in \S \ref{sec:conclusion} we summarize our work. Technical details are collected in Appendices. In particular, our list of notations can be found in Table \ref{tab:notations}, while our extensive set of numerical simulations used to test our analytical derivations is presented in Appendix \ref{sec:method}.

\section{MODELING INTERSTELLAR TURBULENT MEDIA}
\label{ref:ISM}

Similar to the original DCF, we adopt a model of isothermal turbulent media. The assumption of isothermality is applicable to the molecular clouds as well as to the idealized phases of the ISM (see Draine 2006)\footnote{For warm and cold phases the equation of state (EoS) is monoatomic- adiabatic (See, e.g. Kritsuk et.al 2017, \citealt{VDA,HO21}. For turbulence with adiabatic EoS one could express the mean sonic Mach number $c_s^2 \propto \partial P \partial \rho$. Simulation community have studied adiabatic turbulence for decades, e.g. Nolan et.al 2015.}. This does not apply to the Unstable HI phase that constitutes a significant portion of the interstellar hydrogen (\citealt{HO21,cattail}). 

The turbulence is assumed to be injected at a large scale $L_{inj}$ and cascades to the small dissipation scale $l_d$ determined by viscosity, ion-neutral collisions or resistivity, depending on the media we study. The scales between $L_{inj}$ and $l_d$ correspond to the turbulence inertial range and the properties of turbulence at this range do not depend on the physics of the dissipation at $l_d$. However, the properties of turbulence are affected by the properties of the driving that is given by the sonic Mach number $M_s=V_L/V_s$ and Alfv\'en Mach number $M_A=V_L/V_A$, where $V_L$ is the turbulent injection velocity at the scale $L$, while $V_s$ and $V_A$ the sound and Alfv\'en velocities respectively. The response of the media to turbulence depends on the ratio of the gaseous to magnetic pressure $\beta\sim V_s^2/V_A^2$. The warm diffuse medium has $\beta>1$, while molecular clouds typically have $\beta<1$.

\begin{deluxetable}{c c c c }
\tablecaption{\label{tab:ISMtable}The typical $M_s,M_A,\beta$ values for interstellar media and molecular clouds \label{tab:beta}, $n_H$, $\delta v$ for molecular clouds from Draine (2011), WNM/CNM from Kalberla et.al (2018), \citealt{HO21,VDA,cattail}, $\beta$ is from \citealt{HO21}.}
\tablehead{ & $n_H (cm^{-3})$ & $\delta v (km/s) $  & $\beta$  }
\startdata \hline
WNM                  & $0.1-1 $ & $10-17$ & $\sim 100$\\
UNM                  & $1-10$   & $6-10$ & $\sim 1$\\
CNM                  & $10-50$  & $3-5$ & $\sim0.1$\\
GMC Complex ($H_2$)  & $50-300$   & $4-17$ & $\sim 0.01$  \\
\enddata
\end{deluxetable}

For magnetic field strength studies in this paper we are interested in $M_A\leq 1$, as for $M_A\gg 1$ the turbulence is super-Alfv\'enic and marginally affected by magnetic field at scales larger than $L_{inj} M_A^{-3}$ (see Appendix \ref{app:mhdturb}). In this paper we focus on studying magnetic field strength in sub-Alfv\'enic and trans-Alfv\'enic turbulence. This automatically entails that turbulence in high $\beta$ media is subsonic with $M_s<1$, meaning that it can be approximated by nearly incompressible turbulence. The compressible turbulence in our studies takes place in low $\beta$, i.e. $\beta<1$, media. 

The compressible turbulence in isothermal MHD turbulence has been extensively studied both theoretically and numerically (see Lithwick \& Goldreich 2001, Cho \& Lazarian 2002, 2003, Kowal \& Lazarian 2010, \citealt{2020PhRvX..10c1021M}, also a monograph by Beresnyak \& Lazarian 2019). The backbone of the theory of MHD turbulence is the model of Alfv\'enic turbulence by Goldreich \& Sridhar (1995, henceforth GS95), that was formulated for $M_A=1$ and then generalized for smaller $M_A$ in Lazarian \& Vishniac 1999, henceforth LV99). In the latter study it was shown that for scales larger than $l_{tr}=L_{inj} M_A^2$ the Alfv\'enic turbulence is in "weak" regime, while for scales smaller than $l_{tr}$ the turbulence is in the "strong" regime. The "weak" and "strong" characterize the rate of non-linear interactions of turbulent motions and are not related to the amplitude of the turbulent perturbations. The properties of Alfv\'enic turbulence in weak and strong regimes are very different as we discuss in Appendix \ref{app:mhdturb}

The Alfv\'enic turbulence is very anisotropic and it imprints its properties on slow modes both in high-$\beta$ and low-$\beta$ media (GS95, Cho \& Vishniac 2000, Lithwick \& Goldreich 2001, Cho \& Lazarian 2002), while the fast modes have their own cascade that is isotropic both in high-$\beta$ medium (GS95) and in low-$\beta$ medium (Cho \& Lazarian 2002). 

The anisotropic scaling of Alfv\'enic turbulence that we provide in Appendix \ref{app:mhdturb}  is defined in the frame of reference that is termed "local frame". This is the frame related to the local direction of magnetic field surrounding the Alfv\'enic perturbations at hand. The understanding of local frame is most natural in the eddy model of turbulence proposed in LV99.\footnote{The concept of the local magnetic field direction was not a part of the original GS95 picture. It was brought in and established through the later research (LV99, Cho \& Vishniac 2000).}  Indeed, due to fast turbulent reconnection, in strong Alfv\'enic turbulence, there exist magnetic field eddies that mix up magnetic field perpendicular to the direction of magnetic field in the vicinity of the eddies. The local direction of magnetic field varies through the 3D volume and therefore the statistics of magnetic field is different for the observer.

These properties MHD turbulence in the frame of the observer were described in  \citeauthor{LP12}(\citeyear{LP12}, henceforth LP12). The latter and the subsequent studies \citep{KLP16,KLP17a} provide the theoretical framework for describing our efforts in obtaining magnetic field strength from observations. 

The energy injection for different modes depends on the properties of the turbulent driving at $L_{inj}$. The driving is expected to significantly affect the dispersions of turbulent velocities $\delta v$ and $\delta \phi$, but general theoretical considerations (see Beresnayk \& Lazarian 2019) suggest that for sufficiently extended cascade, as it is the case of the cascade of the interstellar medium, the equipartition of energy between different modes tends to establish. The numerical simulations, even the largest ones (see Federath et.al 2021), have more limited inertial range and therefore the properties of the properties of turbulence driving may have imprint over the entire range of scales. Note, that the most robust in the cascade are Alfv\'en modes, while slow and fast modes are subject to collisionless damping (see Yan \& Lazarian 2002, 2004, Brunitti \& Lazarian 2007). Therefore, in the presence of the collisionless damping the fraction of Alfv\'enic waves and their importance is expected to increase with the decrease of the scale. In our study we consider an arbitrary admixture of Alfv\'en, fast and slow modes. 

Numerical issues present a serious problem that have not been properly considered in the framework of DCF studies. As MHD turbulence presents different regimes, the interpretation of numerical simulations and their relation to astrophysical reality may not be trivial. Indeed, even for the simplest case of incompressible MHD turbulence, the scaling of velocity and magnetic field fluctuations are different for weak, strong and super-Alfv\'enic turbulence. The relation between numerical dissipation scale the scales of the transfer from one regime to another must be carefully accounted for. Indeed, while in typical astrophysical conditions the dissipation scale of turbulence is much smaller than the $L_{inj}M_A^2$ for the transfer from weak to strong MHD turbulence (see Appendix ), for numerical studies for sufficiently small $M_A<1$ this is not true. Thus relating the numerical results with astrophysical measurements may be misleading. 

In addition, our study includes an extensive use of compressible MHD turbulence simulations that are listed in Appendix~\ref{app:mhdturb}.  These simulations are employed to test our theoretical predictions and the expressions that we obtain for the magnetic field strength.

Similar to DCF, we assume that the observed polarization represent the variations of projected magnetic field. The polarization can be of different origin, not only from dust, as we discuss in 
Appendix~\ref{sec:waystomeasure}.

\section{INTRODUCTION OF NEW TECHNIQUE: DIFFERENTIAL MEASURE ANALYSIS (DMA)}
\label{sec:theory}

\subsection{The Input Information for Obtaining Magnetic Field Strength}
\label{sec:input}

Our technique employs two types of observables, the polarization to represent the magnetic field direction, and the velocity centroids for measuring turbulent velocities. \linebreak

\noindent{\bf Polarization as a tracer of the magnetic field direction:} 
The magnetic field acting on dust grains, molecules and atoms induces polarization of the emitted radiation (see Appendix~\ref{sec:waystomeasure}).  In polarization observations the Stokes parameters for, e.g., thermal dust emission are given by
\begin{equation}
\label{eq.stokes}
    \begin{aligned}
    Q&\propto\int dz n \cos(2\theta)\sin^2\gamma\\
    U&\propto\int dz n \sin(2\theta)\sin^2\gamma \\
    \phi &= \frac{1}{2} \tan_2^{-1}(U/Q)
    \end{aligned}
\end{equation}
where for the case of the dust polarization $n$ is the number density {\torefereeone of dust grains}, while $\theta$ and $\gamma$ are the POS positional angle of the magnetic field and the inclination angle of magnetic field with respect to the line of sight, respectively. Similar expressions are available for the polarization arising from atoms and molecules (see Appendix \ref{sec:waystomeasure}). 

The statistics of polarization fluctuations in turbulent media is described in LP12. This is the study results of which our paper heavily relies upon. 

\noindent{\bf Velocity centroids :}
The velocity information on astrophysical turbulent volume is available from spectroscopic observations in the form of Position-Position-Velocity (PPV) cubes where intensity {\torefereeone $Em (\mathbf{X},v)$} is given as a function of the plane of sky (POS) coordinate ${\bf X}$ and the Doppler-shifted line-of-sight (LOS) velocity $v$. The first velocity moment of the intensity distribution is 
\begin{equation}
C(\mathbf{X}) \propto \int_a^b  dv v Em (\mathbf{X},v)/ \int_a^b dv Em (\mathbf{X},v) ,
\label{centroid}
\end{equation}
where we depending on the choice of the integration limits $a,b$ one can get different measures. For instance, integrating over the entire spectral line width one gets a measure known as a velocity centroid.\footnote{If the integration limits are chosen over a part of the line, we are dealing with the {\it reduced centroids} \citep{LY18a}. The reduced centroids are valuable for probing turbulence in the presence of galactic rotational curve.} One of the advantages of using velocity centroid is that such measures are not affected by thermal broadening. 

Changing the integration from the velocity to real space, it is possible to show (see Lazarian \& Esquivel 2003) that centroids are can be written as
\begin{equation}
   C(\mathbf{X}) \propto \int_{\cal L}  dz v (\mathbf{X},z)  \rho (\mathbf{X},z)/ \int_{\cal L} dz \rho (\mathbf{X},z) 
\end{equation}
where ${\cal L}$ is the optical depth of the turbulent object under study. 
In what follows, we disregard that the fluctuations of $\rho$, assuming that their contribution can be significantly reduced in observation using the recipe in \cite{VDA}. 
 The velocity centroids have been used extensively for studies of turbulence statistics (see Scalo \& Elmegreen 2004). A detailed numerical study of their properties can be found in Esquivel \& Lazarian (2005). The quantitative analytical study of the statistics of velocity fluctuations was initiated in \cite{LP00} and continued in the subsequent papers. In our study we will rely on the results obtained for the analytical description of statistics of velocity centroids obtained in \citep{KLP17a}, henceforth KLP17.\footnote{An interesting possibility of measuring turbulent velocities using the positions and velocities of stars was suggested in Ha et al (2021). Potentially, this allows to measure not only parallel to line of sight velocity, but the POS turbulent velocities. We do not discuss this in this paper, however.}

Within the DCF study the use of centroids to measure the dispersion of velocities was suggested in \cite{CY16}. Though using similar information as DCF, this paper presents a new technique that focuses on short scale turbulent information and does not require assumptions about behaviour of the magnetic field and velocities at scales approaching injection scale.

\subsection{Sampling Small Scale Turbulence}

The technique that we introduce in this paper relies heavily on the statistical properties of MHD turbulence in the observer's frame. Before getting to the detailed calculations, we discuss the advantage of using measurements of magnetic and velocity fluctuations at small scales, rather than employing the dispersion of these quantities as it is done in the DCF. For the sake of simplicity, we do this first in the DCF framework, i.e. without accounting for the actual anisotropic properties of MHD turbulence. 

Let us initially consider a toy problem assuming that the mean magnetic field $B_{mean,\bot}$ is perpendicular to the line of sight, i.e. $\gamma=\pi/2$. Consider the variations of the observed magnetic field direction within a volume with size ${\cal L}$ measured along the line of sight and the turbulence injection scale $L_{inj}$. The fluctuations of the magnetic field angle are 
\begin{equation}
\tan\delta\phi \approx \delta \phi \approx \frac{\int \delta B_\bot dz}{ B_{mean,\bot} {\cal L}}
\label{eq:tantheta}
\end{equation}
where the integration is done along the line of sight and where, without losing generality, we assumed that $\delta B_\bot$ is a perturbation in the magnetic field component in the plane of the sky. Later, in this paper we will abandon the assumptions both about $\gamma$ and toy model turbulence.

Instead of following the DCF path and considering the dispersions of angles and velocities at the turbulence injection scale, we focus on the turbulence statistics at small scales that can be retrieved by using differential measures. In particular, if we are interested in the spatial variations of the observed magnetic field directions at the scale $l$, e.g. angle dispersions of magnetic field, those can be obtained using the second-order structure functions ($D$) of the polarization angle $\phi$\footnote{For the sake of simplicity we do not distinguish here the magnetic field angle $\theta$ and  the polarization angle $\phi$. Those quantities are, in general, differ, as discussed  e.g. in \cite{LY18a}.}:
\begin{equation}
D_{2D}\{\phi\}({\bf R} )= \langle [\delta\phi ( {\bf X}+{\bf R}) -\delta \phi( {\bf X})]^2\rangle_{{\bf X} }
\label{d_theta}
\end{equation}
where ${\bf X}$ is a two dimensional vector on the Plane of Sky (POS), $\langle...\rangle_{{\bf X }}$ denotes the averaging over the position ${\bf X }$. For practical applications, this means averaging for different ${\bf X}$ over the area $\gg l^2$.  Substituting Eq. (\ref{eq:tantheta}) in  Eq. (\ref{d_theta}) one gets
\begin{equation}
 D_{2D}\{\phi\}({\bf R} )\approx \frac{1}
 {B_{mean,\bot}^2 {\cal L}^2} D_{2D}\{B\} ({\bf R }), 
\label{eq:D_D}
\end{equation}
where the $D_{2D}\{B\} ({\bf R })$ is structure function of the POS projected magnetic field. The latter is related to the 3D structure of the magnetic field $D_{3D}\{b\}({\bf r})$ as 
\begin{equation}
D_{2D}\{B\} ({\bf R })=\iint  D_{3D}\{b\}({\bf r}) dz_1 dz_2
\label{st_turb}
\end{equation}
 where as it is usual in turbulence studies, is assumed to be magnetic field in the turbulent volume is homogeneous, i.e. dependent only on the 3D lag between the points ${\bf r}$ (Monin \& Yaglom 1976):
\begin{equation}
D_{3D}\{b\}({\bf r})=\langle [b_\bot ({\bf x}+{\bf r})- b_\bot ({\bf x})]^2\rangle.
\label{struc_b}
\end{equation}
with ${\bf x}$ being the 3D position vector and $z_1$ and $z_2$ denoting the pair of lines of sight along which the integration of the structure function of 3D fluctuating magnetic field $b$ is performed. 

For the sake of simplicity, let us discuss structure functions averaged over the positional angle, which will make these functions only dependent on the line of sight distance $l$ separating the points.  Observing that fluctuations of turbulent field are accumulated along the line of sight ${\cal L}$ in a random walk fashion one gets the structure function of the polarization angle $\phi$ 
\begin{equation}
D_{2D}\{B\} (l) \approx D_{3D}\{b\} (l) l {\cal L}
\label{eq:sf2d3d}
\end{equation}
where $l$ is the separation between the lines of sight\footnote{To understand this, from Eq.\eqref{st_turb} the Jacobian $dz_1dz_2$ gives roughly $dl d{\mathcal{L}}$. See LP12 for the rigorous mathematical formalism. }. In statistical sense, the turbulence has the axial symmetry around the direction of the mean magnetic field (see discussion in \citealt{LP12}). However, for the sake of simplicity within this simplified treatment we do not consider the complications arising from the anisotropy in Eq. (\ref{eq:sf2d3d}). 

In fact, the problem at hand has 3 length scales - separation on the sky $l$, integration/cloud depth $\mathcal{L}$ and the injection scale $L_{inj}$, which is also the line-of-sight correlation length. In our discussion, it is assumed that $l<L_{inj}$, thus the random walk takes place on with the step $l$ and $L_{inj}$ does not enter the problem. If $\mathcal{L} \gg l$, the random walk results in $D_{2D}\{B\}(l) \propto D_{3D}\{b\}(l) l \mathcal{L}$, which makes the slope of the 2D structure function steeper by unity compared to the 3D structure function of magnetic field. Expressing $D_{2D}\{B\}(l)$ through the structure function for the polarization angle (see Eq.~\ref{eq:D_D}), one obtains  
\begin{equation}
D_{2D}\{\phi\}(l) \approx \frac{D_{3D}\{b\}}{ B_{mean,\bot}^2} \frac{l}{\cal L},
\label{struc_main}
\end{equation}
where $D_{2D}\{\phi\}(l)$ is available from observations.

Let us now look at velocity fluctuations. Here we only consider the simplest case of pure Alfv\'en turbulence. For Alfv\'enic turbulence the fluctuations of velocity and magnetic field are symmetric {\it in 3D}. Therefore $\delta v_{turb}=\delta B_{\bot}/\sqrt{4\pi \rho}$ , where $\rho$ is the medium density. Notice that for pure Alfv\'en mode the medium density is constant since Alfv\'en wave does not contribute to any density fluctuations (see Biskamp 2003). This means that the structure function of velocity
\begin{equation*}
D_{3D}\{v\}(l)=\langle [v_{turb} (\mathbf{x}+\mathbf{r})-
v_{turb}(\mathbf{x})]^2\rangle_\mathbf{x}.
\end{equation*}
is related to the structure function of the magnetic field in Eq.~(\ref{struc_b}) as
\begin{equation}
D_{3D}\{B\}=4\pi \langle \rho \rangle D_{3D}\{v\}.
\label{struc_rel}
\end{equation}

With observational spectral line data, one can measure the (constant density)\footnote{One could use the approach in \cite{VDA} to decrease the effects of density fluctuations within the centroid in spectroscopic maps.} structure function of velocity centroids:
\begin{equation}
D_{2D}\{C\} ({\bf R}) =\langle [C ({\bf X}+{\bf R})-C({\bf X})]^2\rangle_{\bf X},
\end{equation}
which presents the proxy of the structure function of the velocities, averaged along the line of sight. In this  procedure, the addition of velocity fluctuations is similar to summing up of magnetic perturbations $\delta b_{turb}$ that we dealt with earlier. As the result, the summation process of the velocity fluctuations is a random walk process, i.e.
\begin{equation}
D_{2D}\{C\}({l})\approx \frac{1}{\cal L}\int_{\cal L} D_{3D}\{v\}\approx D_{3D}\{v\}\frac{l}{\cal L}
\label{struc_centr}
\end{equation}
Combining Eqs.(\ref{struc_main}), (\ref{struc_rel}) and  (\ref{struc_centr}), one gets the expression for the mean magnetic field:
\begin{equation}
 B_{mean, \bot}\approx f \sqrt{4\pi \langle \rho \rangle}\frac{D_{2D}^{1/2}\{C\}(l)}{D_{2D}^{1/2}\{\phi\}(l)}
\label{mean_B}
\end{equation}
where, similarly to DCF, in our toy problem the factor $f$ is constant. The actual calculations of this factor, based on the properties of anisotropic MHD turbulence, is done further in the paper and demonstrate it dependence on the Alfv\'en Mach number and the structure of turbulent modes.

Reader should recognize that, while our Eq.~(\ref{mean_B}) might look similar to the DCF relation in the RMS of the form, the sampling requirement is very different for these two methods. The radical difference between the DCF approach and our present approach is that we measure the structure functions \textit{ at scales $l$ significantly smaller than $L_{inj}$} which allows to use data sets covering regions smaller than $L_{inj}$. Whereas DCF method requires sampling over $l \sim L_{inj}$.
More importantly, with the use of structure functions, we can connect the magnetic field estimation to the detailed statistical theory of MHD turbulence.

We would like to stress that Eq. (\ref{mean_B}) is applicable to situations when the turbulence injection scale $L_{inj}$ is either larger or smaller than the line-of-sight depth ${\cal L}$, as long as $l \ll {\cal L}$. The only requirement is that ${\cal L}$ should be the same for the calculations of $D_{2D}^{1/2}\{C\}(l)$ and $D_{2D}^{1/2}\{\phi\}(l)$. This requirement is automatically fulfilled if the same spectral line data is used for obtaining both the polarization and Doppler broadening. This is the case for velocity gradients (see Yuen \& Lazarian 2017, Lazarian \& Yuen 2018) or Ground State Alignment (GSA) (see Yan \& Lazarian 2007). The case of dust-induced polarization requires more care to be sure that the polarization is collected from the same column of gas that contributes to the line emission. For instance, if the used line is 13CO, it is necessary to make sure that the column density of gas associated with CO emission is much larger than the column density of the HI along the same line of sight .

Finally, we can consider a limiting case of  $l\gg \mathcal{L}$, when we get $D_{2D}(l) = D_{3D}(l) \mathcal{L}^2$ for both the structure functions of magnetic field direction and centroids. This is a special case of studies when only a shallow surface area of the turbulent volume is being observed. This case can be realized in the presence of strong dust absorption as discussed in \cite{KLP18}. However, it is easy to see that Eq. (\ref{mean_B}) is also applicable to this limiting case.

\subsection{Role of instrumental beam in DCF and DMA}

Both DCF and differential DMA approaches compare the statistical measures of two distinct sky maps - one for polarization and the other related to LOS velocity. It is important for the technique that the polarization and the spectroscopic data sample the same regions. The differences can emerge due to the difference in the volumes sampled along the line of sight as well as due to the differences arising from differences of telescope beams. Within a modification of the DCF suggested in \cite{2009ApJ...696..567H} , the careful study of the effects of the telescope beam is provided in \cite{2009ApJ...706.1504H}. \cite{KLP16} analyzed the isotropizing effect of a beam  on the measured anisotropy of the velocity structure functions. The current paper deals mostly with the monopole part of the structure functions and therefore the beam-related suppression of higher multipoles reported in \cite{KLP16} is not so relevant. 

The issues related to the beam size affect the practical data handling.
In general, the two data sets used in DMA or DCF are obtained in different experiments, with different instruments, and as such the sky maps that serve as an input are convolved with different instrumental beams. This difference must be accounted for when numerical
comparison between two measures is used to deduce the value of the magnetic field. This can be achieved by a detailed modelling of the beams through the formalism, though a more practical remedy is to deconvolve the beams from the datasets, and then smooth the datasets again with an equal synthetic beam
 to suppress the noise and artifacts induced by the instrumental
beam deconvolution.  The benefit of this procedure is the ability to control 
the properties of the final map smoothing, in particular, to suppress possible anisotropy of the instrumental beams that interferes with tracking the anisotropic turbulence effects, and
to have a sufficiently effective noise reduction at small scales. Utilizing simple isotropic Gaussian
smoothing is one example of such a well-behaved synthetic beam.  

The downside is,
of course, the reduced resolution of the final maps, since the synthetic smoothing window must be larger than either of
instrumental beams. For differential measures, this limits the scale at which DMA can be applied and
the appropriate optimal balance needs to be found when analyzing the data.
Overall, controlling the smoothing of 
sky maps is more critical to DMA, as it is sensitive to small scales, than to DCF, as it operates
primarily on large scales.  In what follows, we assume that the input data has been pre-processed accordingly and smoothed to the same resolution both for polarization and centroid maps. 

\subsection{Advantages of Eq. (\ref{mean_B}) Compared to the DCF}

Eq.~(\ref{mean_B}) presents the first step in formulating the technique that we term the {\bf Differential Measure Analysis (DMA)}. Below we describe the advantages of our approach.

\subsubsection{Obtaining B-strength with Smaller Data Sets and B-field Distribution}

\begin{figure}
\includegraphics[width=0.45\textwidth]{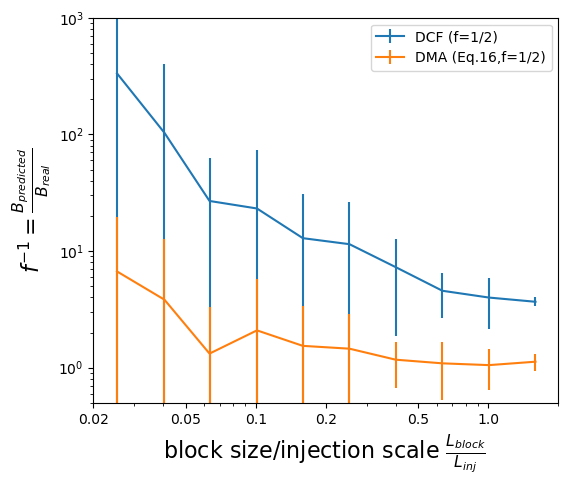}
\caption{The variations of $f$ as a function of sampling size for the DCF as compared to the variations of $f$ in Eq.~(\ref{mean_B}) for the simulation "huge-0" (See Table \ref{tab:sim}). The block sizes are given as a fraction of the turbulence injection scale $L_{inj}$. The DMA method that we employed is from Eq.\ref{eq:DMA_eq} at $R=L_{inj}/198$. Here, both DCF and DMA are applied assuming $f=\onehalf$. }
\label{fig:f_trad}
\end{figure}

While the DCF employs global dispersion of velocities and magnetic field directions defined on the turbulence injection scale, the DMA uses the local measurements by structure functions that can be applied at much smaller scales. The first notable advantage of this approach is that the value of magnetic field can be obtained over smaller areas of the sky. For instance,  in Fig.~\ref{fig:f_trad} we show the coefficient $f$ in DCF  (Eq.~\ref{mean_B}) as a function of the sample size normalized by the injection scale $L_{inj}$, using "run-2 ($M_A=0.6$)"  (see Table~\ref{tab:sim}) numerical simulation. It is clear that  the DCF significantly overestimates the strength of magnetic field in the resolutions dependent manner, which is not straightforward to calibrate away. Only when the sampling size of the data is  larger than the injection scale of turbulence $L_{inj}$ does the correction factor become constant. In the interstellar medium this scale is around 100 pc (see Chepurnov \& Lazarian 2010) and exceeds the sizes of molecular clouds.\footnote{The recent study by \cite{spectrum} confirmed that the turbulence cascade  continues from the large diffuse media scales to the sizes of molecular cloud cores.} Therefore, DCF is bound to overestimate magnetic field if limited area over a molecular cloud is used for observations. Naturally, using DCF it is impossible to provide the detailed distribution of the magnetic field strength over the image of a molecular cloud.

In contrast, Fig.~\ref{fig:f_trad} testifies that the factor $f$ in Eq.~(\ref{mean_B}) does not change much. This allows using relatively small patches of data to obtain the magnetic field strength, in particular get the distribution of the projected magnetic field strength over the molecular cloud map. Thus, one can productively use the improved resolution of the polarization and spectroscopic maps that are getting available.    

Apart from the regular change of $f$, we observe a significant statistical dispersion of its values within the classical DCF approach. We explain this effect in Appendix~\ref{sec:variations}.  The dispersion of values when the Eq. (\ref{mean_B}) is employed is significantly smaller, which reflects the fact that the structure functions are more focused on small scale statistics and less subject to large scale variations. 

The overestimation of the magnetic field strength in patches less than $L_{inj}\times L_{inj}$ is not the only reason why DCF is inaccurate. As we discuss in later sections, the DCF has no inclusion on the nature of MHD turbulence, in particular, the anisotropic character embedded in some of the fundamental modes in MHD turbulence. Accounting for this differences is very difficult within the DCF, as the technique deals with large scale dispersions. At the same time, the properties of MHD turbulence change along the cascade, e.g. due to the transition from weak to strong turbulence that takes place at $L_{inj} M_A^2$ (Lazarian \& Vishniac 1999). The DMA focuses only on the small scale turbulent fluctuations by choosing the lag of the structure functions less than this scale, i.e. $R<L_{inj}M_A^2$. This makes the DMA results that we derive further in this paper robust.\footnote{At the first glance, if turbulence is uniform and homogeneous, Eq.~(\ref{mean_B}) is reminiscent of Eq.~(\ref{eq:alf}) as for $R\rightarrow\infty$ the structure functions are proportional to the squared dispersion. However, due to the aforementioned transition from weak to strong turbulence the expected scaling of $f$ will change with $R$ and it will be different for $R<L_{inj}M_A^2$ and much larger $R>L_{inj}$ for which the transition of the structure functions to the dispersion is justifiable. }

\subsubsection{Realistic Inhomogeneity of Data on Large Scales}
\label{sec:multipoint}

On large scales the structure of observed magnetic and velocity field is strongly affected by regular shear, gravity, outflows and other galactic processes. Therefore the dispersion $\delta v$ and $\delta \phi$ that enter Eq.~(\ref{mean_B}) can be significantly distorted, decreasing the accuracy of the DCF technique. In comparison, the DMA employs structure functions, that are not sensitive to the poorly controlled shifts of the mean background values. In fact, within our approach it is easy to remove any regular  magnetic field and velocity distortions by using multi-point structure functions as we demonstrate further.

If necessary, the removal of regular contributions can be practically obtained with higher order structure functions which for the astrophysical context are discussed in  \cite{LP08}, \cite{CL09} (see also \citealt{2007PhRvL..98o4501F,LP08,Cho19}). 

For our approach there is no explicit dependence on how many points should we use in computing the structure functions for the estimation of magnetic field strength, as the magnetic field strength enters the expression via the Alfv\'enic relation between the perturbations of magnetic field and velocity. Therefore with the multi-point structure functions we can use the equations in the main text (\S \ref{sec:modes}) to determine the magnetic field strength.

The three \& four point second order structure functions are defined as
\begin{equation}
\begin{aligned}
    D_{3pt}\{A\}({\bf r}) &\propto\langle [A({\bf x}+{\bf r})-2A({\bf x})+A({\bf x}-{\bf r})]^2\rangle \\
    D_{4pt}\{A\}({\bf r}) & \propto \langle [A({\bf x} + 2{\bf r}) -3A ({\bf x} +{\bf r}) \\
    & ~~~~~~~ +3A({\bf x})-A({\bf x}-{\bf r})]^2\rangle ~.
    \label{eq:multiptSF2}
\end{aligned}
\end{equation}
\cite{Cho19} has discussed how the multi-point structure functions can possibly remove the shear velocity field and recover the turbulent part.
In a similar manner to 2-point structure function that cancels a constant contribution, 3-point and higher order structure functions cancel any linear and respectively higher order regular contributions to the field. The cost of using them is an increased noise contribution when applied to noisy data.
 
\begin{figure}[th]
\centering
\includegraphics[width=0.48\textwidth]{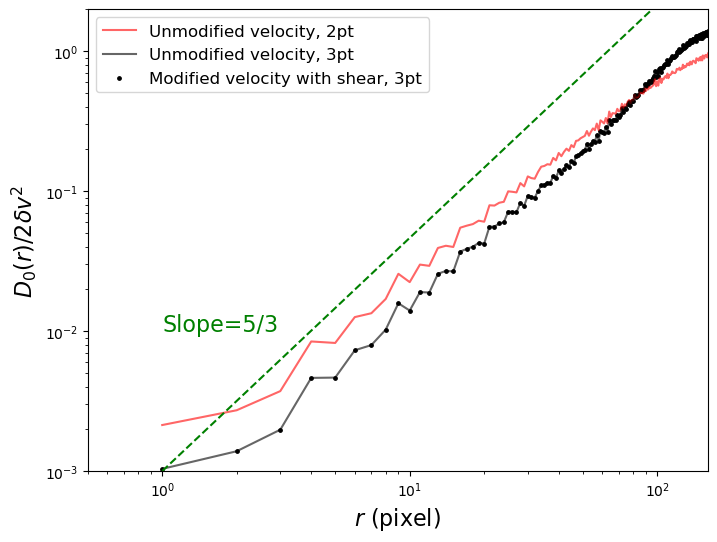}
\caption{\label{fig:sf_multipoint} A figure showing the behavior of {\torefereetwo the unmodified 2 (red) and 3 point (black)} projected velocity structure function with and without the presence of shear in the incompressible cube. 
}
\end{figure}

Here we test whether this is the case. Fig.~\ref{fig:sf_multipoint} shows how the 2-point and 3-point {\it angular averaged} structure functions of the projected velocities of incompressible cube is affected by a large-scale {\it coherent shear}.  Here we only perform a simple test assuming that the velocity experienced a constant randomly oriented shear:
\begin{equation}
    v_{z,modified} = v_{z,original} + {\bf A}\cdot \hat{z}
\end{equation}
where $A$ is some random vector representing the shear strength. We compare both the structure functions before and after shear modifications for the case of 2- and 3-point. We can see from Fig.~\ref{fig:sf_multipoint} that the modifications of velocity has no effect on the amplitudes on the 3-point structure functions. However, for the 2-point structure function the introduction of the shear changes the structure function dramatically, both the slope and the shape of it. Therefore, the multi-point structure functions are indeed not altered by the presence of constant velocity shear, confirming the suggestions from \cite{Cho19}.

The dispersions of the observables, $\delta v$ and $\delta \phi$,  that are employed by the DCF are one point statistics.  Such measures are less demanding in terms of the number of measurements compared to the two point statistics of structure functions. The multi-point structure functions require even richer statistics. Thus the DCF could be applicable to the poorly sampled data, but at the expense of having low reliability results.

\subsection{Common Problem of Eq. (\ref{mean_B}) and DCF: Failure to Account for Anisotropy and mode composition of MHD Turbulence}

The obvious limitation of Eq. (\ref{mean_B}) is that, similar to the DCF approach, it ignores both the anisotropic properties of MHD turbulence. This anisotropy changes with the media magnetization $M_A$, the angle $\gamma$ between the mean magnetic field and LOS, as well as the ratio of energy in Alfv\'en, slow and fast modes of turbulent motions. 

One effect of the anisotropy is that the contribution of the basic MHD modes to the variations of polarization angles and the variations of velocities changes with $\gamma$.

\begin{figure}[th]
\centering
\includegraphics[width=0.48\textwidth]{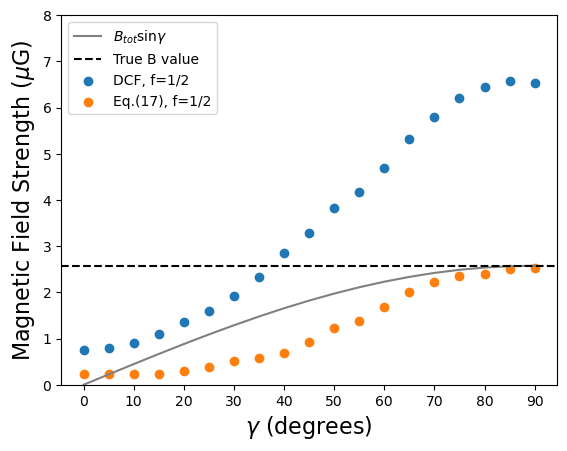}
\caption{\label{fig:blos} The variation of the Plane of Sky (POS) magnetic field strength $B_\perp$ via evaluated with the DCF (blue dots) and (Eq.~\eqref{mean_B})  (orange dots) as a function of $\gamma$ obtained with the supersonic trans-Alfv\'enic numerical simulation ("huge-3" in Table.\ref{tab:sim}). The actual value of POS magnetic field is given by a grey line. }
\end{figure}

Fig. \ref{fig:blos} demonstrates that  Eq. (\ref{mean_B}) provides a better approximation to the actual $B_\bot$ compared to the DCF. However, we see that the fit is not perfect. This is one of the consequence of the anisotropic character of MHD turbulence that is {\torefereetwo missing in the derivation of } Eq. (\ref{mean_B}).

We are not aware of any attempts to quantitatively account for the anisotropy of the MHD turbulence. When accounting for the anisotropy effect, one should keep in mind that the Alfv\'enic turbulence changes from weak to strong regime at a particular scale $L_{inj} M_A^2$, with a significant changes of its properties. Within the DCF approach one deals with the properties of turbulence at large scales, i.e. mostly with the weak turbulence. On the contrary, using the DMA we are focused on the motions at the small scales, i.e. dealing with Alfv\'enic turbulence at the strong regime.   

In addition, as we discuss in Appendix \ref{app:mhdturb},
the realistic MHD turbulence is compressible and contains Alfv\'en, slow and fast modes, each having its own anisotropy. 
In \S \ref{sec:modes} we outline our approach to dealing with the contributions of the modes and further in the paper we obtain the functional dependence of $f$ that enters Eq.~(\ref{mean_B}).

\section{DMA: INTRODUCTION TO THE FORMALISM}
\label{sec:analytical_numerical}

The high resolution of polarization and spectroscopic data sets ushers a new era where more sophisticated analysis that provides high quality output is possible.  The goal of DMA is to increase the accuracy of $B$-strength predictions by accounting for the actual properties of MHD turbulence. 
Introducing the differential measures in Eq. (\ref{mean_B}) is the first step in our derivation of the DMA formalism. 
Our next step is to provide a realistic description of turbulence at sufficiently small scales that are sampled by structure functions that describe magnetic field fluctuations as well as the fluctuations of turbulent velocities. 

In \citeauthor{LP12} (\citeyear{LP12}, henceforth LP12) we described the statistics of magnetic fluctuations arising from MHD turbulence. In the subsequent study by \citeauthor{KLP17a} (\citeyear{KLP17a}, henceforth KLP17) the fluctuations of velocities have been described following the approach in LP12. These papers provide the basis for our detailed calculations. 
 
\subsection{Structure Function of Polarization Angles}

For angle of polarization signal that traces the magnetic field direction (synchrotron, dust polarization, synthetic polarization from gradient maps)  we can generally write 
\begin{equation}
\cos(2 \phi) = Q/(p I) ~, \quad 
\sin(2 \phi) = U/(p I) 
\label{eq:stokes1}
\end{equation}
where $I,Q,U$ are the Stokes parameters (See \S \ref{sec:input}) and $pI$ is the polarized intensity
\begin{equation}
pI = \sqrt{ Q^2 + U^2 }
\end{equation}

For these quantities we can construct the structure function for polarization angle $\phi$: 
\begin{align}
D^\phi(\mathbf{R}) &\equiv \frac{1}{4}
\left\langle \left( \frac{Q_1}{p I_1}-\frac{Q_2}{p I_2}\right)^2\right\rangle
+
\left\langle \left( \frac{U_1}{p I_1}-\frac{U_2}{p I_2}\right)^2\right\rangle 
\nonumber \\
&= \frac{1}{2} \left\langle 1 - \cos(2(\phi_1-\phi_2) \right\rangle 
\label{eq:correct_dtheta}
\end{align}
where indexes $1$ and $2$ refer to two LOS separated by the 2D vector $\mathrm{\mathbf{R}}=\mathrm{\mathbf{X}_1}-\mathrm{\mathbf{X}_2}$ on the sky.
Note, that our measure differs from the measure introduced for the polarization angle in \citet{2009ApJ...706.1504H} by the multiplier $2$ in the cosine argument, and a factor $1/2$ in front.
The difference stems from the nature of the polarization direction that has a period of $\pi$ rather than $2\pi$. Naturally, in the small angle approximation, i.e. for $|\phi_1-\phi_2|\ll 1$, this difference does not play a role. In this small angle case, our expression, as well as that of \citet{2009ApJ...706.1504H}, transfers to the "typical version" of the structure function of angle 
\begin{equation}
D^\phi(\mathbf{R}) \approx \left\langle  (\phi_1 - \phi_2)^2 \right\rangle
\label{eq:phi_simple}
\end{equation}
that coincides with $D_{2D}\{\phi\}$ given by Eq.~(\ref{d_theta}) and that
was also previously explored in \cite{Fal08}.
However, $D^\phi$ in Eq.~(\ref{eq:correct_dtheta})  is a more general expression that is better defined from observations. $D^\phi$ is also applicable to the case when angle fluctuations are large, e.g., when the Alfv\'enic Mach number is large.

In the case of dust polarization, which arises due to dust grain alignment by the magnetic field, we can model sky signal of dust emission as 
\begin{equation}
\begin{aligned}
Q &\propto  \int\! dz \; n_{dust} \; \frac{ (B_x^2 - B_y^2)}{ (B_x^2 + B_y^2 + B_z^2)} \\
U & \propto  \int\! dz \; n_{dust} \; \frac{ 2 B_x B_y}{ (B_x^2 + B_y^2 + B_z^2 )} 
\label{eq:stokes_dust}
\end{aligned}
\end{equation}
where $B_x,B_y$ are the magnetic field components in x,y directions and z is along the line of sight. Eq.~(\ref{eq:stokes_dust}) is a combination of the local polar angle $\cos 2\theta
= (B_x^2-B_y^2)/(B_x^2+B_y^2)$, $\sin 2\theta = 2 B_x B_y /(B_x^2+B_y^2)$, and the angle between the magnetic field and LOS, $\sin^2\gamma = (B_x^2+B_y^2)/(B_x^2+B_y^2+B_z^2)$.

Eq.~(\ref{eq:stokes_dust}) relates the properties of statistics of polarization directions that is available from observations and the underlying statistics of magnetic field, in particular, allowing to study the statistics of magnetic turbulence (see \citealt{LP16}). 

Let us consider the coordinate system where the mean field lies in in $(x,z)$ plane, 
$\overline{B}_x = \overline{B} \sin\gamma, \overline{B}_y=0, \overline{B}_z=\overline{B} \cos\gamma$.
The average in Eq.~(\ref{eq:correct_dtheta}) can be computed rigorously in a series expansion in $\delta B/\bar{B}_x$. The leading contributions to Stokes parameters from Eq.~(\ref{eq:stokes_dust}) are
\begin{equation}
\label{eq:no_wandering_approx}
\begin{aligned}
Q &\sim \sin^2\gamma \int \! dz\; n_{dust} \\
U &\sim 2 \sin^2\gamma  \int \! dz \; n_{dust} \frac{\delta B_y}{\overline{B}}
\end{aligned}
\end{equation}
which tells us that $p_I \sim Q + \mathcal{O} \left( \frac{\delta B^2}{B^2}\right)$ and that the leading term in the structure function Eq.~(\ref{eq:correct_dtheta}) is provided by the $U$ part
\begin{equation}
D^\phi(\mathbf{R}) \approx \frac{1}{4}
\left\langle \left( \frac{U_1}{Q_1}-\frac{U_2}{Q_2}\right)^2\right\rangle 
\label{eq:dtheta_approxU}
\end{equation}

If we assume that fluctuations of the dust density are of the same order as the fluctuations of the magnetic field $\frac{\delta n_{dust}}{n_{dust}} \sim \mathcal{O}\left(\frac{\delta B}{B} \right)$, then we find
\begin{equation}
\frac{U}{Q} \sim \frac{2}{\mathcal{L}} \int \! dz \frac{\delta B_y}{\overline{B}}
+ \mathcal{O} \left( \frac{\delta B^2}{B^2} \right)
\end{equation}
and
\begin{equation}
D^\phi(\mathbf{R}) \approx \frac{1}{\mathcal{L} \overline{B}^2}  \widetilde{D}_{yy}(\mathbf{R})
\label{eq:Dphi_def}
\end{equation}
where
\begin{equation}
 \widetilde{D}_{yy}(\mathbf{R}) \equiv 
\int \! dz \left( D_{yy}(\mathbf{R},z) - D_{yy}(0,z) \right) 
\end{equation} 
is the regularized projection of the 3D structure function $D_{yy}(\mathbf{r}) = \left\langle \left( B_y(\mathbf{r}_1) - B_y(\mathbf{r}_2) \right)^2 \right\rangle$ for the magnetic field y-component that is orthogonal to both LOS and the direction of the mean field. Remarkably, in this leading order perturbations of the dust density (assumed to be of the same order as perturbations in the magnetic field) do not affect the result.

Our numerical tests show that  when the field is perpendicular to the line of sight, Eq.~(\ref{eq:Dphi_def}) is accurate to under 1\% for
$M_A=0.15$ (b21 in Table \ref{tab:sim}). At higher $M_A=0.66$ (b15 in Table \ref{tab:sim}) accuracy
varies from 1\% when structure functions are measured small lag to $\sim 10\%$ at $R \approx L_{inj}$, which points to better accuracy when measuring angle
differences at short separation instead of the angle variance as in DCF.
For $M_A=1.1$ (b42 in Table \ref{tab:sim}), which is near the limit of approximations in Eq.~(\ref{eq:Dphi_def}), the accuracy drops to $\sim 40\%$. Note, that for magnetic strength determination, the uncertainties are roughly half of the quoted ones, since $D^\phi$ enters in a square root. 

Our leading order approximation, that starts with Eq.~(\ref{eq:no_wandering_approx}),
features a constant angle of the magnetic field $\sin\gamma$ to LOS, whereas its local value
$\sin^2\gamma = (B_x^2+B_y^2)/(B_x^2+B_y^2+B_z^2)$ varies along the line-of-sight as reflected
in Eq.~(\ref{eq:stokes_dust}). Further we will account for this next-order effect within
the model of ``magnetic field wandering'' along the LOS.

\subsection{Effect of Turbulence Anisotropy on the Angle Fluctuations}

According to Eq.~(\ref{eq:Dphi_def}), the most important component that determines the polarization angle fluctuation is the y-component of magnetic field. The y-component magnetic field fluctuations contain both spatial power distribution and also anisotropic information of magnetic field fluctuations in the sky.  Following description of magnetic field fluctuations in LP12, Appendix~\ref{sec:statistics_LP} contains general derivation of $\widetilde{D}_{yy}(\mathbf{R})$. However, for our context here we focus on the case when the y-axis fluctuations are described by Alfv\'en mode power spectra only $A=F=E$, which, in particular, is the case for strong turbulence in high-$\beta$ regime. 
We shall also average our structure functions over all directions of  $R$, i.e we will measure the monopole of the structure functions $D_0(R) = \int d\phi_R D({\bf R}=(R,\phi_R))$. Then, from Eq.~(\ref{eq:app_tildeDyy}) :
\begin{equation}
\begin{aligned}
\widetilde{D}_{yy}(R) = & \frac{1}{2\pi^2} \! \int \!\! KdK \left( 1 - J_0(K R)\right)  \\
& \times \int d \phi_K E(K,\sin\gamma \cos\phi_K) \cos^2\phi_K
\label{eq:Dyy_approx}
\end{aligned}
\end{equation}
where $\gamma$ is the line of sight angle $\cos\gamma=\hat{B}\cdot \hat{z}$. Let us compare Eq.\eqref{eq:Dyy_approx} to the total power in POS components given, as seen from Eqs.~(\ref{eq:app_tildeDyy},
\ref{eq:app_tildeDxx}), by
\begin{eqnarray}
&& \widetilde{D}_{B_\perp}(R)  \equiv \widetilde{D}_{xx}(R) + \widetilde{D}_{yy}(R) \\
&& = \frac{1}{2\pi^2} \! \int \!\! KdK \left( 1 - J_0(K R)\right) 
\int d \phi_K E(K,\sin\gamma \cos\phi_K) 
\nonumber
\end{eqnarray}
which differs by the absence of $\cos^2\phi_k$ factor.

{\bf The presence of the $\cos^2\phi_k$ is the most important reason why the $f$ parameter should contain a natural dependence on $M_A$.} If the spectrum of turbulence was isotropic, we would have $\widetilde{D}_{yy}(R)=\frac{1}{2} 
\widetilde{D}_{B_\perp}(R) $. However, for a typical anisotropic MHD power
distribution, the power is concentrated in the wave modes orthogonal to the mean magnetic field,
with suppression of power for modes aligned with it. Thus, $E(K,\sin\gamma \cos\phi_K)$ is peaked at $\cos\phi_K = 0$.
In particular, for solenoidal  magnetic field that its perturbations are orthogonal to the wave vector in 3D which results into the projection of $\cos^2\phi_K$ factor in Eq.~(\ref{eq:Dyy_approx}) for the $y$-component. This projection will be zero when the modes power $E(K,\sin\gamma \cos\phi_K)$ attains its maximum. As a result, the contribution of structure function fluctuations $\widetilde{D}_{yy}(R)$ due to solendoial projection is suppressed, so as the polarization angle perturbations, in comparison to expectation from the general level of perturbations in $\delta B_\perp$.  Mathematically, we can express the polarization angle fluctuations as a function of $\widetilde{D}_{yy}(R)/\widetilde{D}_{B_\perp}(R)$:
\begin{equation}
D^\phi(R) \approx \frac{1}{\mathcal{L} \overline{B}_\perp^2} \left( \frac{\widetilde{D}_{yy}(R)}{\widetilde{D}_{B_\perp}(R)} \right) \widetilde{D}_{B_\perp}(R)
\end{equation}
From our argument above, we find that the factor $ \frac{\widetilde{D}_{yy}(R)}{ \widetilde{D}_{B_\perp}(R)} \ll \frac{1}{2}$, and is smaller, when the MHD power spectrum is more anisotropic. Moreover, its magnitude is turbulent model dependent, but as we shall see further $\frac{\widetilde{D}_{yy}(R)}{ \widetilde{D}_{B_\perp}(R)} \propto M_A^2$ for many cases of sub-Alfv\'enic turbulence.

\subsection{Multipole Expansion of Polarization Structure Function}

Let us now return to a general exposition of the angle structure function. We follow the description of magnetic fluctuations in LP12 and Appendix~\ref{sec:statistics_LP}. In Appendix~\ref{sec:Appendix_Dyy}, we obtain Eq.~(\ref{eq:app_tildeDRgen}) for
$ \widetilde{D}_{yy}(\mathbf{R})$ for sub-Alfv\'enic turbulence in the strong regime at sufficient short scales $R < L_{trans} = L_{inj} M_A^2$ (See Eq.\eqref{trans}) and where the LOS depth $\mathcal{L} \gg R$ as well. Using it in Eq.~(\ref{eq:Dphi_def}) gives the following expression for the coefficients of the multipole expansion $D^\phi(R,\phi_R) = \sum_n D^{\phi}_n(R) e^{in\phi_R}$ of the polarization angle structure function
\begin{equation}
\label{eq:phimultipole}
D_n^\phi(R) =   \frac{\left\langle \delta B^2\right\rangle}{\overline{B}_\perp^2} \frac{\mathcal{I}_n(R)}{\mathcal{L}}
\sum_{p=-\infty}^\infty \!\! \widehat{E}_p^{2D} (\gamma)\; G_{n-p}^{(I,A,F)}(\gamma)
\end{equation}
where $\mathcal{I}_n(R)$ are scaling functions defined in Eq.~(\ref{eq:scalingfcn}). Coefficients $\widehat{E}_p^{2D}$ are POS angular harmonic decomposition of the projected power spectrum of the magnetic field which depend on the angle $\gamma$ of 3D orientation of the mean field relative to LOS and the Alfv\'en Mach number $M_A$.  This additional dependence on $M_A$ reflects the change of the anisotropy of MHD turbulence with the change of media magnetization and, as we just argued for in the previous section, can significantly change the DMA approach compared to the naive treatment given by Eq.~(\ref{mean_B}).
The geometrical functions $G^{(I,A,F)}(\gamma)$ depend on the mode structure of the turbulence, with several particular cases discussed in \S~\ref{sec:modes}. From KLP16, KLP17 the higher order multipoles encode the information of $\gamma$ since $\gamma$ and $M_A$ enters to the multipole contributions in different power. However for our purpose in this paper, we will only consider the case when $n=0$ in the RMS  of B-field estimations.

\subsection{Multipole Expansion of Velocity Centroid Structure Functions}

The next step in DMA is to evaluate the structure function of velocity centroids in the nominator of Eq.~(\ref{mean_B}). The statistics of centroids was discussed in \cite{KLP17a}. There, for the sake of theoretical convenience the definition of centroids was modified compared with the standard one given by Eq.~(\ref{centroid}). In particular, the density was set to be constant in the computation of velocity centroids. This approach is now possible to realize in observational data analysis, as the recently developed Velocity Decomposition Algorithm (VDA, \citealt{VDA}) allows one to significantly mitigate the effects of the density fluctuations.  Thus, the theoretical model in \cite{KLP17a} coupled with VDA becomes applicable to the observational data that is frequently strongly affected by density inhomogeneities. Therefore, we proceed with the same constant-density assumption in velocity centroids assuming that we can always remove the density effects from observation using the VDA.

Velocity centroids structure function has a very similar behaviour to the polarization angle one. The multiple moments of the structure function of centroids KLP17, normalized by the mean column intensity of the gas along the line of sight $\overline{Em} = \epsilon \mathcal{L} \bar\rho$, where $\epsilon$ is the emissivity coefficient, can be written in the form similar to Eq.~(\ref{eq:phimultipole}) :
\begin{equation}
\begin{aligned}
\widetilde{\mathcal{D}}_n^v & \equiv
\frac{\mathcal{D}_n}{(\epsilon \bar{\rho} \mathcal{L})^2} = \\
& =\langle\delta v^2\rangle 
\frac{\mathcal{I}_n(R)}{\mathcal{L}}
\sum_{p=-\infty}^\infty \!\! \widehat{E}_p^{2D}(\gamma) \mathcal{W}^{(I,A,F)}_{n-p}(\gamma)
\label{eq:csf} 
\end{aligned}
\end{equation}
where we have expressed the amplitude of velocity fluctuations $\hat{\mathcal{A}}_p$ of KLP17 via the variance $\langle \delta v^2 \rangle$. In Appendix~\ref{sec:app_turbv} we show that the scaling function $\mathcal{I}_n(R)$ here is the same as in Eq.~(\ref{eq:phimultipole}) and give the geometrical functions $\mathcal{W}_s $  for the velocity centroid structure functions. 

Velocity centroids structure functions are less affected by the anisotropy of the power spectrum than
the angle structure functions. None of geometrical weights $\mathcal{W}$,
as Eq.~(\ref{eq:app_tildeDzzIAv}) attests, have a suppression of the projected modes orthogonal
to the magnetic field via  $\cos^2\phi_K$ factor that $G^{(I,A,F)}$ functions contained. Thus, 
no additional small parameter $\propto M_A$ arises in the evaluation, e.g., of $\widetilde{\mathcal{D}}_0^v$.

\subsection{The Ratio of the Multipole Coefficients of Magnetic Field Angle and Velocity Centroid Structure Functions}

Eq~(\ref{eq:phimultipole}) and Eq.~(\ref{eq:csf}) are purposefully written in the form to highlight the common and distinct parts of the angle and centroid structure functions.   From these two equations the ratio of the centroids and angle structure function multipole coefficients is clearly
\begin{equation}
\frac{\widetilde{\mathcal{D}}_n^v}{D_n^\phi} =
\overline{B}_\perp^2 \; 
\frac{\left\langle \delta v^2 \right\rangle}{\left\langle \delta B^2 \right\rangle }
\frac{\sum_{p}\widehat{E}_p^{2D}\mathcal{W}_{n-p}^{(I,A,F)}}
{\sum_{p} \widehat{E}_p^{2D} G_{n-p}^{(I,A,F)}}
\label{eq:Dc_over_Dtheta}
\end{equation}

If we consider Alfv\'en and slow modes for $\beta\ll 1$, or consider Alfv\'en and fast modes for $\beta\gg 1$ , the velocities and magnetic field values are connected through Alfv\'enic relation given by Eq.~(\ref{eq:alf}) the variances are
\begin{equation}
\frac{\left\langle \delta v^2 \right\rangle }{\left\langle \delta B^2 \right\rangle} \approx
\frac{1}{4 \pi \langle \rho \rangle}.
\label{eq:alf2}
\end{equation}

Then we can obtain for all even $n$:
\begin{eqnarray}
    \label{eq:Dc_over_Dtheta_rho_alt}
    \overline{B}_\perp^2 &\approx& 4 \pi \langle \rho \rangle f_n^2 \frac{\widetilde{\mathcal{D}}_n^v}{D_n^\phi}  \\ 
    f_n^2(M_A,\gamma) &\equiv& \frac{\sum_{p} \widehat{E}_p^{2D} G_{n-p}^{(I,A)}}{\sum_{p}\widehat{E}_p^{2D}\mathcal{W}_{n-p}^{(I,A)}}
    \label{eq:fn_def}.
\end{eqnarray}

{Within our assumption of the similarity of velocity and magnetic field scaling, the factor $f_n$ in Eq.(\ref{eq:fn_def}) does not depend on the POS lag $R$.} This is true for the velocity and magnetic field fluctuations for Alfv\'enic turbulence, turbulence of slow modes and the admixture of Alfv\'en and slow modes. We deferred the study of fast modes that have different scaling (\citealt{CL02,CL03}, Kowal \& Lazarian 2009) to \S \ref{subsec:lowbeta}.
In general, for an arbitrary mixture of 3 modes with different spectra, $\mathcal{A},\mathcal{S},\mathcal{F}$, one can write:
\begin{equation}
\begin{aligned}
    & f_n^2(M_A,\gamma) = \\ 
    & = \frac{\sum_{p} \left[ 
    \mathcal{\widehat{A}}_p^{2D} G_{n-p}^\mathcal{A} + \frac{\langle \delta B_\mathcal{S}^2 \rangle \mathcal{I}_0^\mathcal{S}}{\langle \delta B_\mathcal{A}^2 \rangle \mathcal{I}_0^\mathcal{A}} \widehat{\mathcal{S}}_p^{2D} G_{n-p}^\mathcal{S} + \frac{\langle \delta B_\mathcal{F}^2 \rangle \mathcal{I}_0^\mathcal{F}}{\langle \delta B_\mathcal{A}^2 \rangle \mathcal{I}_0^\mathcal{A}} \widehat{\mathcal{F}}_p^{2D} G_{n-p}^\mathcal{F} \right]}
    {\sum_{p}\left[ \mathcal{\widehat{A}}_p^{2D} \mathcal{W}_{n-p}^\mathcal{A} + \frac{\langle v_\mathcal{S}^2 \rangle \mathcal{I}_0^\mathcal{S}}{\langle v_\mathcal{A}^2 \rangle \mathcal{I}_0^\mathcal{A}} \widehat{\mathcal{S}}_p^{2D} \mathcal{W}_{n-p}^\mathcal{S} + \frac{\langle v_\mathcal{F}^2 \rangle \mathcal{I}_0^\mathcal{F}}{\langle v_\mathcal{A}^2 \rangle \mathcal{I}_0^\mathcal{A}} \widehat{\mathcal{F}}_p^{2D} \mathcal{W}_{n-p}^\mathcal{F} \right]}
\end{aligned}
\label{eq:totalf}
\end{equation}
where we have factorized the Alfv\'en mode power spectrum and have used the relation Eq.~(\ref{eq:alf2}) for the ratio of the magnitudes of $\delta B$ and $v$ fluctuations in the Alfv\'en mode.

We note that the evaluation of the magnetic field strength depends via $f_n$ on $M_A$ as well as the orientation of the magnetic field with respect to the LOS. These dependencies arises due to difference of anisotropy of line-of-sight velocities and perpendicular magnetic field fluctuations in MHD turbulence. { We explore the properties of $f_n$ further in \S~\ref{sec:modes}.}

It is important to stress once more that {\bf in our studies we focus on the small scale asymptotic behavior of MHD turbulence}. The description of MHD turbulence is much less certain in the vicinity of the injection scale and can be significantly affected by the turbulent energy injection mechanisms. We also assume that we sample turbulence at scales less than the transition from the weak to strong turbulence $L_{trans}=L_{inj} M_A^2$. This condition is especially crucial to check with for simulations with very low $M_A$. Therefore our calculations are not applicable to larger scales, e.g. beyond $L_{inj}$, and the transition from the DMA expressions to the DCF ones is not straightforward.

\section{Projection of MHD modes}
\label{sec:modes}

Before considering particular admixtures of Alfv\'en, slow and fast modes of MHD turbulence
\footnote{In a number of sources the combination of slow and fast modes is called {\it compressible} to distinguish them from an {\it incompressible} Alfv\'en mode. This is misleading as both fast and slow modes have compressible and incompressible components. The actual distinction can be done on the basis of the Helmholtz decomposition (see Kowal et al. 2010). Thus, we always put "compressible" in "..." while talking about fast + slow modes.}, in this section we first discuss the observed properties common to all feasible magnetic fluctuations. More specifically, we will discuss first the basis of the frame vectors  which decide the directions of the modes fluctuations. We will discuss how both the velocity and magnetic field angle fluctuations are accumulated along the line of sight. In particular, we discuss an important but easily overlooked suppression of the projected magnetic angle fluctuations when the depth of integration is significantly larger than the correlation scale of the magnetic field fluctuations. 

\subsection{Orthogonal Modes of Magnetic and Velocity Vector Fields in the Presence of a Globally Preferred Direction}
\label{sec:global_frame}

In the system with the preferred direction $\hat\lambda_0$ set, in our case, by the mean magnetic field direction, the fluctuations of any solenoidal vector field such as $\delta\mathbf{b}$ can be decomposed, for each Fourier mode $\mathbf{k}$, into two orthogonal components; of $A$-type (Alfv\'en) along $\widehat\zeta^A  = (\widehat{\mathbf{k}} \times \hat\lambda_0)/|\widehat{\mathbf{k}} \times \hat\lambda_0|$, and of $F$-type along $\widehat{\zeta}^F = \widehat{\mathbf{k}} \times \left( \widehat{\mathbf{k}} \times \hat\lambda_0\right)/|\widehat{\mathbf{k}} \times \hat\lambda_0|$. In the context of MHD turbulence, Alfv\'enic magnetic field perturbations are of $A$-type; while both slow and fast modes induce the same $F$-type component in the magnetic field perturbations, though they differ in the power distribution among different wave vectors $\mathbf{k}$. 

\begin{figure}[th]
\centering
\includegraphics[width=0.4\textwidth]{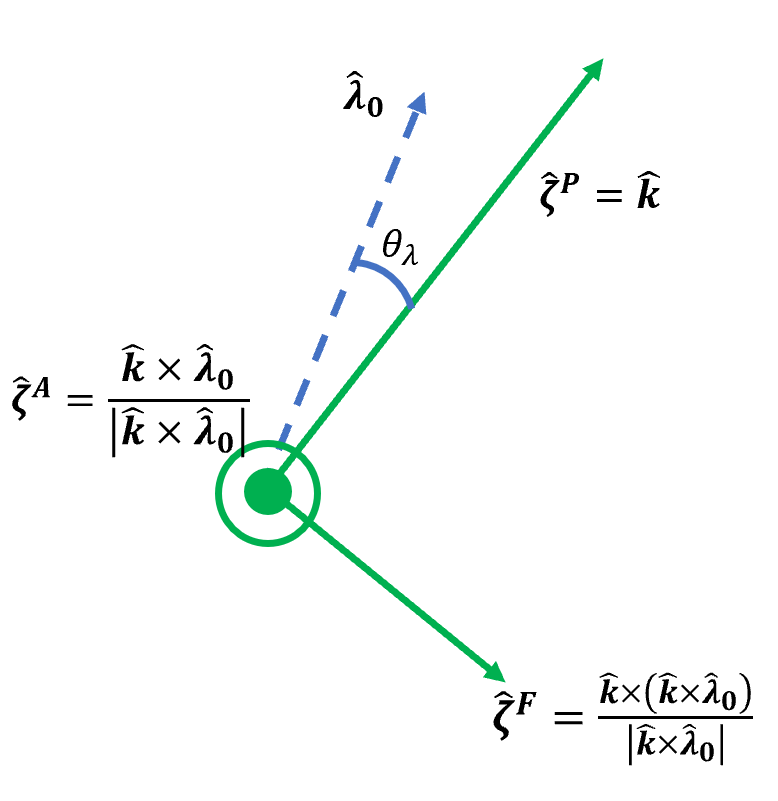}
\caption{\label{fig:frame} An illustration of the frame vector system {\it in the global frame of reference}. The vectors $\hat{\zeta}^{A,F,P}$ correspond to the A,F,P-modes. Readers should be careful that A-mode do correspond to the Alfv\'en contributions, but F-mode {\it is not necessarily} the fast mode. In particular, for the displacement vector the fast mode actually lies in between $\hat{\zeta}^F$ and $\hat{\zeta}^P$, while for magnetic field fluctuations both slow and fast modes are along $\hat{\zeta}^F$. See \cite{CL03,LP12} and Appendix~\ref{sec:statistics_LP}.
 }
\end{figure}

A general vector field, such as the vector of turbulent velocity, also has the third component, orthogonal to both $A$ and $F$ ones, that is along $\widehat{\zeta}_P = \widehat{\mathbf{k}}$, which we'll call $P$ (potential) one (See Fig.\ref{fig:frame}).  This is where slow and fast turbulent modes structurally differ by generating velocities orthogonal to each other at each Fourier mode.  

As we discussed in Section~\ref{ref:ISM} for the description of MHD turbulence it is convenient to distinguish the case of magnetically dominated media with magnetic pressure $P_{mag}$ larger than the gas pressure $P_{gas}$, i.e. the case of $\beta=P_{gas}/P_{mag}<1$, and the gas pressure dominated media, i.e. with $\beta>1$. The properties of basic modes of MHD turbulence are different in low and high $\beta$ cases (see \citealt{CL03}). 

In the limiting case of high $\beta$ plasma, where $\beta$ is the ratio of the gas and magnetic pressures, slow mode velocities are of $F$-type, while fast mode ones are purely potential $P$ ones.  In the opposite limit of low-$\beta$ plasma, this pair is rotated, each being a mixture of $F$ and $P$ motions, with slow mode velocities aligned along the magnetic field, while fast modes perpendicular to the magnetic field.  {\torefereetwo Alfv\'en} mode velocities remain to be of $A$-type.

In the \textit{global} reference frame given by the mean magnetic field direction $\widehat{\lambda}_0$, one can always formally decompose a vector field into $A$, $F$, $P$ components. In the physical sense the {\torefereetwo Alfv\'en}, fast and slow modes in the turbulent cascade are defined with respect to the \textit{local} direction of the magnetic field $\hat{\lambda}$ (see Lazarian \& Vishniac 1999, Cho \& Vishniac 2000), while large scale fluctuations in the magnetic field direction are viewed as fluctuation of the preferred direction $\hat{\lambda}$. \footnote{Potentially, some instabilities, e.g. streaming instability of cosmic rays or gyroresonance instability (see Kulsrud 2005) can induce Alfv\'en waves which are exactly parallel to local direction of magnetic field with which the cosmic rays are interacting. In this paper we do not consider this rather special case of generating Alfv\'enic perturbations.} Such locally defined modes are then found to be mixed between $AFP$ types when viewed in the global reference frame. This effect is discussed in more details in \cite{leakage}.

\subsection{Projection Effects due to LOS Signal Accumulation}
\label{sec:projections}

Applying the DMA to observations one deals with the signal integrated along the line of sight.  Let us summarize how this projection affects the contribution of individual basic MHD modes on the quantities of interest - fluctuations of the direction of the polarization angle that in the leading order is given by $\delta B_y$, and the LOS component of velocity in velocity centroids $v_z$. As the DMA samples fluctuations at small scales $R<L_{inj}$, the effects of the averaging along the line of sight ${\cal L}$ can be more important for the technique than for the DCF technique that deals with the dispersion of fluctuations at scales larger than $L_{inj}$. Nevertheless, when the techniques are tested with the numerical simulations obtained with periodic boundary conditions, the integration is formally over infinite scale for both the DCF and the DMA.  

By choosing lag $R$ in the DMA analysis we identify the eddies of the scale $L_{corr}=R$ that we sample. At a minimum level, integrating over the LOS depth $\mathcal{L}$ beyond the correlation length $L_{corr}$ we expect a decay of the relative magnetic field fluctuations $\delta B/\bar{B}$ due to random walk as $(L_{corr}/\mathcal{L})^{1/2}$. However, our signals depend only on specific components $\delta B_y$ and $v_z$. Moreover, the structure of global modes in Fourier space leads to an additional suppression that depends on the mode, angle of the mean magnetic field to LOS and the depth of integration $\mathcal{L}$.

Namely, one can see that in the limiting case when the integration length $\mathcal{L}$ is larger than the POS scale $R$, the LOS projection results in averaging out of all Fourier modes with $k_z > \mathcal{L}^{-1}$. This asymptotically leaves only the Fourier components with wavevectors lying in POS,  $\widehat{\mathbf{k}} = ( \widehat{\mathbf{K}},0)$. Consider now a setting with the mean magnetic field perpendicular to the line of sight $\hat\lambda=(\hat\Lambda,0)$ and {\torefereetwo Alfv\'en} mode fluctuations. Since A-mode excites only the field component simultaneously perpendicular to the wave vector and the mean field, $\zeta^A$ is along LOS, i.e. after the projection of the A-mode in this configuration no $\delta B_y$ is remaining. Thus, in the leading order angle fluctuations of the projected magnetic field in POS are absent. As another example, F-mode lies in a plane spanned by the wave vector and the mean field, so since the projection restricts the wavevector to POS,  the F-mode will not produce any $\delta B_y$ when the mean field is along LOS.  

This illustrates the geometrical suppression of the contributions from $A$ and $F$ modes when the mean magnetic field is either perpendicular or parallel to the LOS. Table~\ref{tab:proj} summarizes the asymptotic limit of the projection effects.  We also note that the P mode does not contribute to DMA in this regime, since magnetic field is solenoidal, and the velocity  contribution is averaged out. In Table~\ref{tab:proj},  the numerical factors are given by the fraction of the mode total projected power,  that is carried by the component that affects the observable signal.

\begin{deluxetable}{c c c c c}
\tablecaption{Contributions from A, F and P parts to fluctuations of velocity and magnetic field in the limiting case of infinite LOS integration depth and perpendicular or parallel orientation of the mean magnetic field. \label{tab:proj}.  The numerical factors are given by the fraction of the modes' total power that are seen after projection. }
\tablehead{Mode & $\delta B_y, \gamma=\pi/2 $ & $\delta B_y, \gamma=0 $ & $V_z, \gamma=\pi/2 $ & $V_z, \gamma=0 $ }
\startdata \hline
A                  & 0  & 1/2 & 1 & 0 \\
F                  & 1/2  & 0 & 0 & 1 \\
P                  & \multicolumn{2}{c}{not~present} & 0 & 0 \\
\enddata
\end{deluxetable}

Let us investigate how the projection changes the variances of the angles and velocities as in DCF, when the signal accumulates over the depth $\mathcal{L}$ as it varies relative to the injection scale $L_{inj}$. We perform our analysis numerically, on synthetic representation of the turbulent magnetic field and velocity motion.  

\begin{figure*}
\label{fig:Ldependence}
\centering
\includegraphics[width=0.48\textwidth]{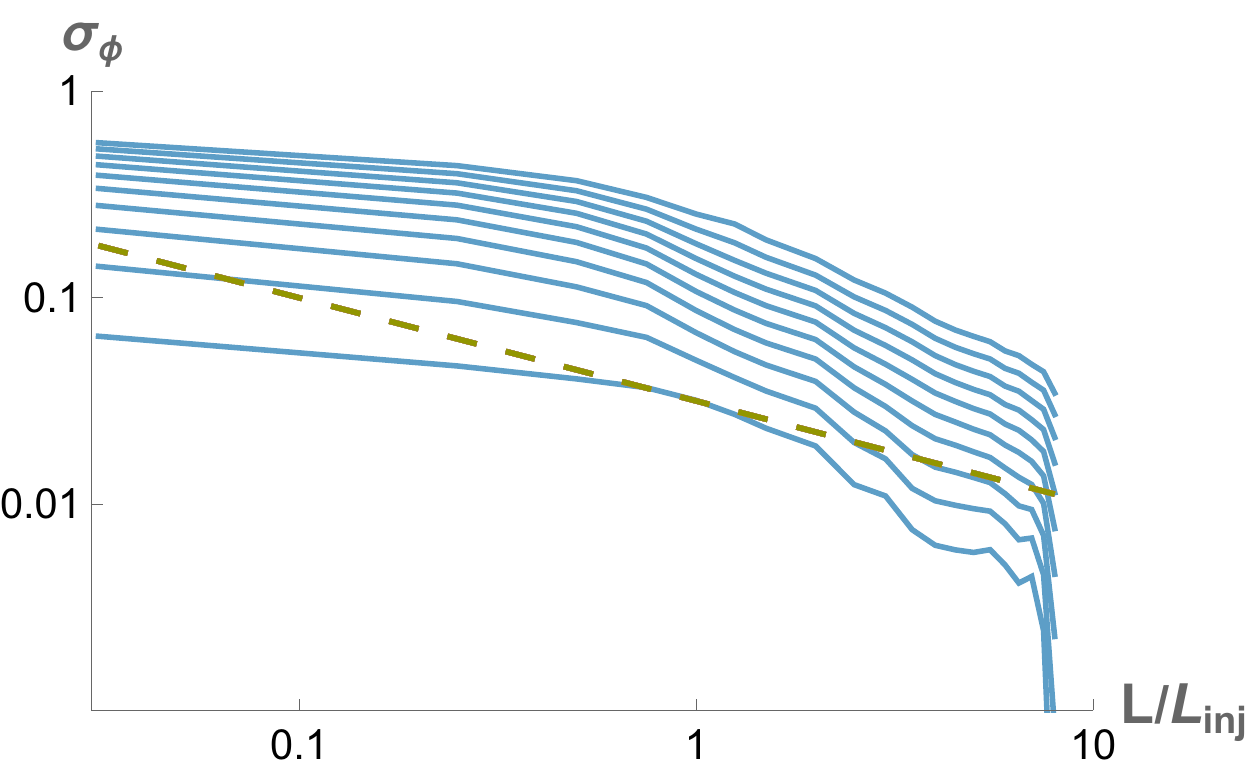}
\includegraphics[width=0.48\textwidth]{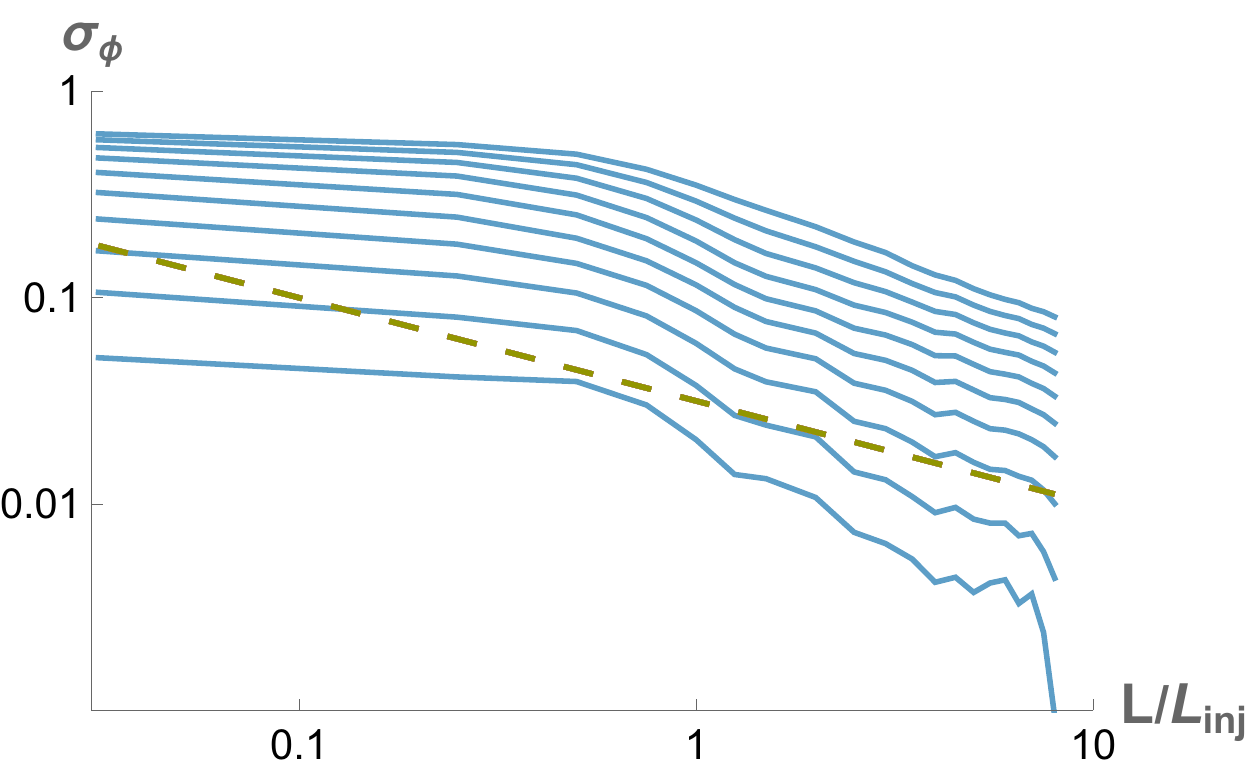}\\
\includegraphics[width=0.48\textwidth]{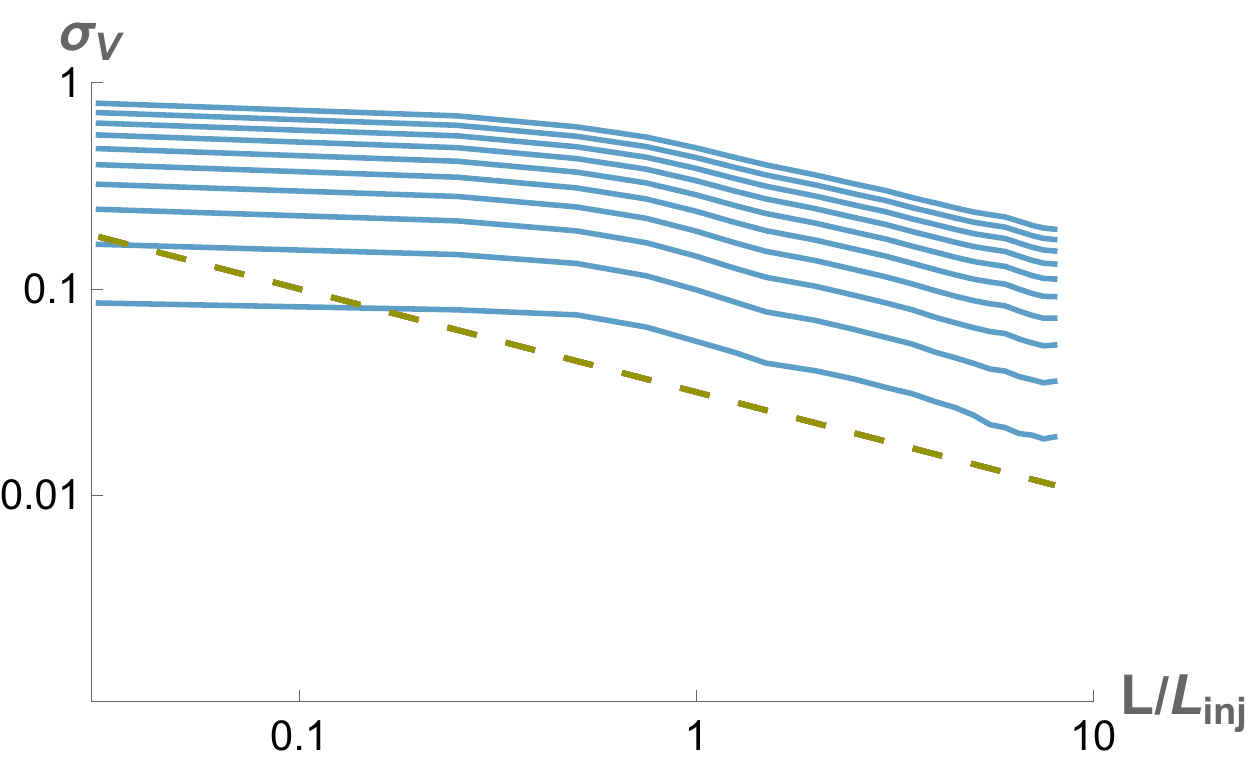}
\includegraphics[width=0.48\textwidth]{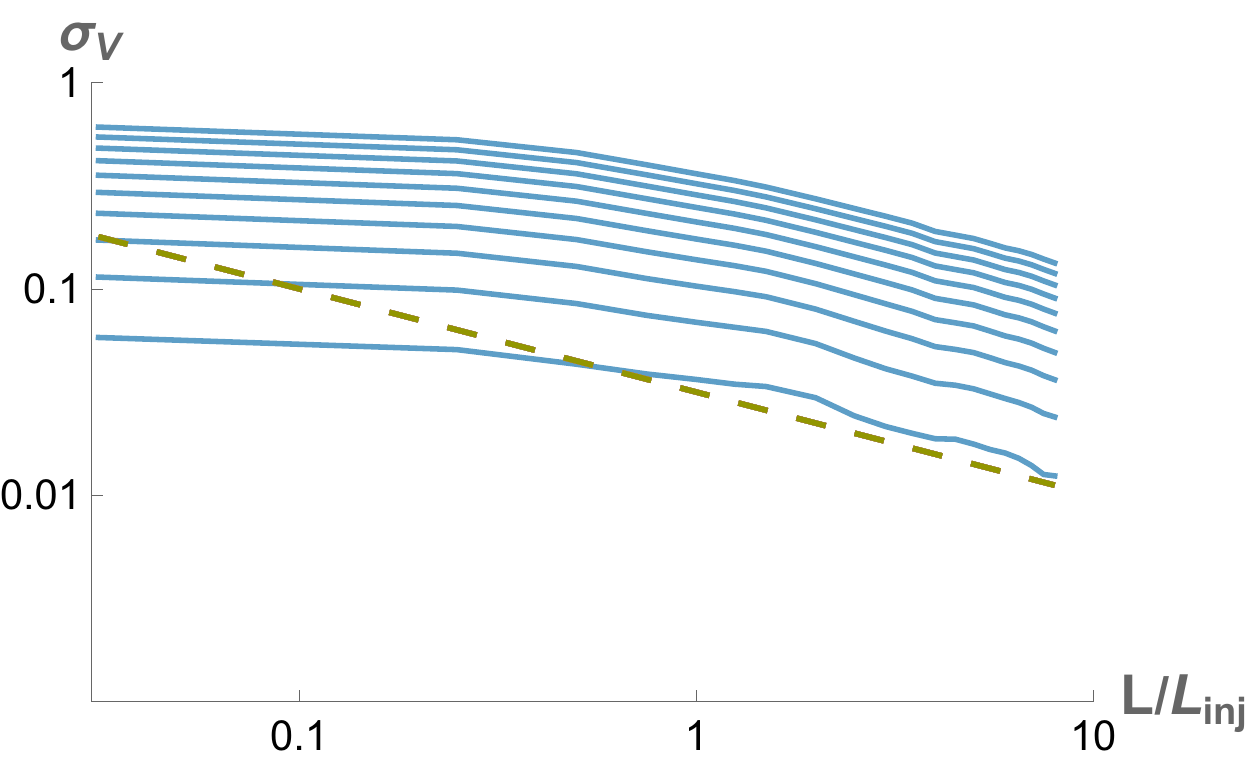}
\caption{Top row: \textit{RMS} fluctuations (radians) of the POS angle of magnetic field as traced by the dust polarization. Bottom row: RMS fluctuation of velocity centroids (arbitrary units) divided by $\mathcal{L}$. Left column: Alfv\'en mode.  Right column: equal mix of {\torefereetwo Alfv\'en} and slow modes, characteristic of the strong incompressible regime. Individual curves correspond to different $M_A$ ranging from $M_A=0.1$ (bottom curves) to $M_A=1$ (top curves). Dashed lines mark $\propto (L_{inj}/\mathcal{L})^{1/2}$ uncorrelated random walk scaling. }
\end{figure*}

In Fig.~\ref{fig:Ldependence}, we show the dependence of the standard deviation for the POS magnetic field angle and velocity centroid  with the LOS integration depth in units of the injection scale. For $\mathcal{L} > L_{inj}$  centroid \textit{rms} fluctuations (divided by $\mathcal{L}$) decrease as $(L_{inj}/\mathcal{L})^{1/2}$ in accordance with the random walk expectation, that is shown in both upper and lower panels with dashed lines.
However angle fluctuations decrease faster both for strong turbulence and even more so for purely Alfv\'en turbulence, due to projection effects. The last point on the graphs corresponds to full projection over the periodic simulation box which is $10L_{inj}$, which is formally equivalent to the infinite integration range. In this limit, the fluctuations of magnetic field angle due to Alfv\'en mode formally vanish.

The effect of the suppression of Alfv\'enic mode contribution has been missed, as far as we know, by the researchers studying the DCF. In fact, the suppression $\sim N^{-1}$ has a simple explanation. The fluctuations of Alfv\'en modes as viewed at $\gamma =\pi/2$ can be modelled by a simple sine wave $\sin(kL_{max})$. This approximation is only valid for Alfv\'en modes in this geometry since if $\gamma \neq \pi/2$ the Alfv\'en mode has additional contributions that we discussed in the Appendix. In this geometry it is easy to see that the cumulative RMS fluctuation accumulated along the line of sight is $\int_{k_{los}=0} dk \sin (kL_{max}) \sim \cos(kL_{max})/L_{max} \sim 1/N$ instead of $N^{-1/2}$.

\section{Pure Alfv\'en Case and Incompressible Turbulence} 
\label{sec:pure_alfven}

Alfv\'en and incompressible MHD turbulence are valuable idealization to be considered. In both cases the Alfv\'enic relation between magnetic field and velocity fluctuations is valid. The case of Alfv\'en mode is useful to discuss for both low and high $\beta$ cases as the Alfv\'en modes are known to be dominant in inducing variations in 3D magnetic field directions, as was shown in LV99.  Our discussion will focus on geometric factors $G$ for angle fluctuations and $\mathcal{W}$ for velocity centroids, that enter determination of the magnetic field strength given by  Eq.~(\ref{eq:Dc_over_Dtheta_rho_alt}). Here we will formally show that in 2D projection Alfv\'en mode contribution to angle fluctuations may be suppressed when the mean magnetic field is {\it nearly perpendicular to the line of sight}. However the Alfv\'en mode remains after projection if the magnetic field has significant LOS component.

\subsection{The Geometric Factors $G_n^A$ and $\mathcal{W}_n^A$}
In the global reference frame we will be speaking about A-mode as Alfv\'en mode.  The contribution of A-mode to the structure function of the projected magnetic field angles with $\mathcal{L} \gg R$ is represented by the geometrical weight in the structure function of $\delta B_y$ component, reproduced here from  Eq.~(\ref{eq:app_tildeDijIA})
\begin{equation}
\begin{aligned}
G^A_{n} & = \frac{1}{2\pi} \int d \phi_K e^{i n \phi_K}  \frac{\cos^2\gamma \; \cos^2\phi_K }{1-\sin^2\gamma\cos^2\phi_K} \\
& = \frac{\cos\gamma}{\sin^2\gamma} \left( -\cos\gamma \; \delta_{n0} + 
\left(\frac{1-\cos\gamma}{1+\cos\gamma}\right)^{|n|/2} \right) 
\end{aligned}
\label{eq:GAn}
\end{equation}

Contribution of the Alfv\'en modes to the (unnormalized) \footnote{The justification of using unnormalized, constant density velocity centroid is that we can use \cite{VDA} to remove the density fluctuations in observational data. In this case, the denominator of the normalized centroid is a constant factor.}
velocity centroids is governed, in turn, by (KLP17)
\begin{equation}
\mathcal{W}_n^A = \delta_{n0} - \cos\gamma \left(\frac{1-\cos\gamma}{1+\cos\gamma}\right)^{|n|/2}
\label{eq:WAn}
\end{equation}

We note that $G_n^A$ vanishes at $\gamma=\pi/2$, thus global A-type mode is not reflected in the fluctuations of the projected magnetic field angles when the mean magnetic field is perpendicular to LOS. At this orientation A-mode is projected out, or highly suppressed if we account for finite LOS depth $\mathcal{L}$. 

Similarly, $\mathcal{W}^A_n$ vanishes in the opposite limit when $\widehat{\lambda}$ is along the line of sight, $\gamma=0$. However, the case of $\gamma =0$ is less of an issue since $\langle B_{\perp}\rangle=0$ in this case and it is problematic to determine the magnetic field strength in this configuration anyway.

\begin{figure}
\label{fig:GWAlfven}
\includegraphics[width=0.43\textwidth]{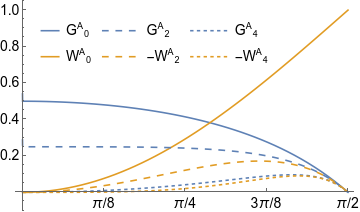}
\caption{Geometrical factors $G^A_n$ (blue) and $W^A_n$ (orange) for $n=0$ (solid), $n=2$ (dashed) and $n=4$ (dotted) as functions of the mean magnetic field orientation $\gamma$. }
\end{figure}

When using both  Eq~(\ref{eq:GAn}) and Eq.~(\ref{eq:WAn}), $n$ is to be taken even. For most angles both $G^A_n$ and $\mathcal{W}^A_n$ decrease substantially for $|n| > 2$ (see Fig.~\ref{fig:GWAlfven} as well as  LP12, KLP16, KLP17a). In particular, for $\pi/8 < \gamma < 3\pi/8$ the leading RMS fluctuations in both geometrical functions are the monopole ($n=0$) and the quadrupole ($n=2$):
\begin{equation}
\begin{array}{ll}
G^A_0 = \frac{\cos\gamma ( 1 - \cos\gamma)}{\sin^2\gamma}, ~ &
\mathcal{W}^A_0 = 1 - \cos\gamma  \\
G^A_2 = \frac{\cos\gamma ( 1 - \cos\gamma)^2}{\sin^4\gamma}, ~ &
\mathcal{W}^A_2 = - \frac{\cos\gamma (1 - \cos\gamma)^2}{\sin^2\gamma}
\end{array}
\end{equation}
As $\gamma \to \pi/2$ all orders of $G_n^A$ become present but are equally suppressed, while $W_n^A$ retains only the monopole
\begin{equation}
 G^A_n \sim \pi/2-\gamma, ~ 
\mathcal{W}^A_n \sim \delta_{n0} - ( \pi/2 - \gamma) 
\end{equation}

\subsection{Power Spectrum in Strong Turbulent Regime}

For sub-Alfv\'enic $M_A < 1$ strong turbulent regime, the power spectrum can be concisely described by Eq.~(\ref{eq:hatEmu_averaged}) which gives the following ratio of the projected 2D multipoles in terms of modified Bessel functions $I_{p}$ of the variable $M_{A\perp}= M_A/\sin\gamma$
\begin{equation}
\label{eq:44}
\frac{\widehat{E}^{2D}_p}{\widehat{E}^{2D}_0}  = 
(-1)^{p/2}\; \frac{I_{p/2}\left(\sfrac{1}{2} M_{A\perp}^{-2}\right)}{I_0\left(\sfrac{1}{2} M_{A\perp}^{-2}\right)} 
\end{equation}
with the latter expression being a reasonable fit even for $M_{A\perp}$ exceeding unity.

At small $M_{A\perp}$ we may utilize the expansion 
\begin{equation}
\frac{\widehat{E}^{2D}_p}{\widehat{E}^{2D}_0}  \sim (-1)^{p/2} \left(1 - (p/2)^2 M_{A\perp}^2 \right)
\label{eq:AlfvenEnE0expanded}
\end{equation}
which shows highly anisotropic power distribution approaching $\delta$-function behaviour that concentrates power in the modes, orthogonal to the projection of the mean field. This limit gives rise to all multipoles in the power spectrum being of the same magnitude with alternating signs.

\begin{figure*}
\centering
\includegraphics[width=0.45\textwidth]{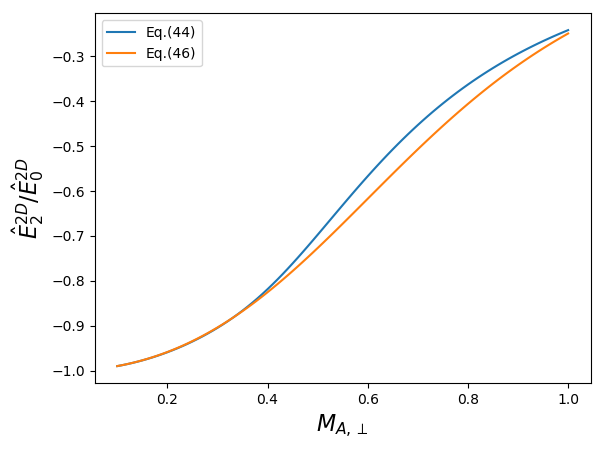}
\includegraphics[width=0.45\textwidth]{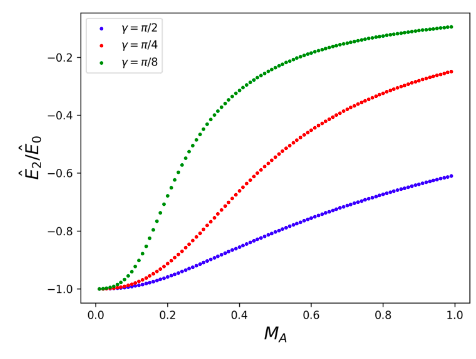}

\caption{\label{fig:E2E0}  (Left) Justification on our approximation formula (Eq.\ref{eq:AlfvenE2E0}) as compared to the actual analytical formula (Eq.\ref{eq:44}). the approximation is almost the same when $M_{A,\perp}<0.4$. (Right) Variation of $\widehat{E}^{2D}_2/\widehat{E}^{2D}_0$ as a function of $M_A$ according to Eq.~\ref{eq:AlfvenE2E0}.}
\end{figure*}

As $M_{A\perp}$ increases, the ever lower multipoles become subdominant. For $M_{A\perp} > 0.3$ one can limit consideration to the monopole and quadrupole. The following expression is a reasonable fit for the quadrupole even for $M_{A\perp}$ exceeding unity 
\begin{equation}
\frac{\widehat{E}^{2D}_2}{\widehat{E}^{2D}_0} \approx -1 +  M_{A\perp}^2 \frac{ 1 + 2 M_{A\perp}^2}{ 1 + M_{A\perp}^2 + 2 M_{A\perp}^4}
\label{eq:AlfvenE2E0}
\end{equation}
which we draw both Eq.\ref{eq:44} and  Eq.\ref{eq:AlfvenE2E0} on the left of Fig.\ref{fig:E2E0}.  On the right of Fig.~\ref{fig:E2E0} shows this ratio of projected power multipoles as the function of $M_A$ for three values of $\gamma$ as given by the exact form of Eq.~(\ref{eq:AlfvenE2E0}).

\subsection{Transforming Alfv\'en Modes from Local to Global Frame of Reference}
\label{subsec:alfven_wandering}

Changing our focus to physical turbulent Alfv\'en mode defined locally, we follow LP12 model that accounting for the large scale magnetic field direction can be considered as wandering of $\widehat{\lambda}$ direction along LOS, level of which is determined
by the {\torefereetwo Alfv\'en} Mach number $M_A$. LP12 suggests that the result of wandering (See also \citealt{leakage})can be described by replacing the anisotropic geometrical function by the weighted combination with the isotropic term as
\begin{eqnarray}
        G_{n-p}^{(A)} &\to& W_I \left( \frac{1}{2} \delta_{np}+\frac{1}{4} \left(\delta_{n-2,p} + \delta_{n+2,p}\right)\right) + W_L G_{n-p}^{(A)}  \nonumber \\
        \mathcal{W}_{n-p}^{(A)} &\to&  W_I \delta_{np} + W_L \mathcal{W}_{n-p}^{(A)}
        \label{eq:geom_isotropized}
\end{eqnarray}
and the power spectrum $ \widehat{E}(\widehat{\mathbf{k}} \cdot \widehat{\lambda}) $
by the average $\overline{ \widehat{E}(\widehat{\mathbf{k}} \cdot \widehat{\lambda})} $
according to Eq.~(\ref{eq:E_averaging}).

We adopt a simple model for the weights
\footnote{This model is more accurate for Alfv\'en modes than what was suggested
in LP12. We use this opportunity to note an inconsistency in LP12 where $W_I$ as used there in Eq.~(71) is twice the one introduced in  Eq.~(\ref{eq:WIWL}). See also \cite{leakage}.}
\begin{equation}
W_I \approx \frac{\sfrac{1}{2} M_A^2}{1 +  M_A^2}, \quad
W_L + W_I \approx \frac{1+  M_A^2}{1+ 2 M_A^2}~.
\label{eq:WIWL}
\end{equation}
together with the angle distribution of the power in Alfv\'en spectrum given by Eq.~(\ref{eq:P_lambda}).
This model reflects, on one hand, that in isotropic, $M_A \to \infty$, limit  $W_I \to \sfrac{1}{2}$ since Alfv\'en mode  has one degree of freedom whereas general isotropic solenoidal field would have two, and on the other hand, that in $ M_A \to 0$ limit
$W_I \sim \sfrac{1}{2} M_A^2$ since Eq.~(\ref{eq:P_lambda}) gives $\langle \sin^2(\theta_{\widehat{\lambda}}) \rangle \sim M_A^2$.

A useful parameter $\alpha(M_A)$
\begin{equation}
\alpha(M_A) = \frac{W_L}{W_I + W_L}
\approx \frac{1 + \sfrac{3}{2} M_A^2}{(1+M_A^2)^2}
, \quad 0 \le \alpha \le 1
\label{eq:alpha_wandering}
\end{equation} 
describes the level of isotropization arising due to change in the mean magnetic field direction along the LOS. The $\alpha=1$ case can be realized when there is an idealized {\torefereetwo Alfv\'en} mode on the background of a fixed mean field for small $M_A$. While $\alpha=0$ corresponds to completely isotropic tensor structure of the turbulence when $ M_A > 1$. The resulting geometrical structure of the local Alfv\'enic turbulent perturbations is that of a mix of $A$ and $F$ modes,  for our purpose here represented as a mix of $A$ and the isotropic $I=A+F$ combination of the correlation tensors. 

\subsection{Asymptotic Expressions for $f_0$ Factor}
\label{subsec:alfven_f0}

Let us start with explicit form of Eq.~(\ref{eq:fn_def})  for the monopole
coefficient $f_0$ when the geometrical functions are given by Eq.~(\ref{eq:geom_isotropized})
that takes into account the wandering of a local direction of the magnetic field along the LOS.
Using $\alpha$ defined in Eq.~(\ref{eq:alpha_wandering}) as the measure of wandering level, one
obtains the following expression
\begin{equation}
f_0^2 = \frac{1}{2} \frac{ (1-\alpha) \left(1 + \frac{\widehat{E}^{2D}_2}{\widehat{E}^{2D}_0} \right)
+ 2 \alpha \sum_p \frac{ \widehat{E}^{2D}_p}{\widehat{E}^{2D}_0} G_p^A}
{ 1 - \alpha + \alpha \sum_p \frac{\widehat{E}^{2D}_p}{\widehat{E}^{2D}_0} \mathcal{W}_p^A }
\label{eq:f0_alfven_firststep}
\end{equation}
Remarkably, for small $M_{A\perp}$, when the power spectrum multipoles are given by approximation Eq.~(\ref{eq:AlfvenEnE0expanded}), one can perform the summations in 
Eq.~(\ref{eq:f0_alfven_firststep}) completely, by noting that
\begin{equation}
\begin{aligned}
& \sum_p (-1)^{p/2} G^A_p = 0 \\
&\sum_p (-1)^{p/2} \mathcal{W}^A_p = \sin^2\gamma \\
&\sum_p (-1)^{p/2} (p/2)^2 G^A_p = - \frac{1}{2} \cos^2\gamma
\end{aligned}
\end{equation}
Substitution of these formulae into Eq.~(\ref{eq:f0_alfven_firststep}) gives a compact expression
for $f_0$ factor in Alfv\'enic turbulence
\begin{equation}
f_0^2 \sim  \frac{1}{2} \cdot \frac{1-\alpha(M_A) \sin^2 \gamma}{1-\alpha(M_A)\cos^2\gamma} M_{A\perp}^2
\label{eq:f_alfen}
\end{equation}
This shows that magnetic field wandering eliminates degeneracies {\it due to projections} that arise due to idealized structure of formal global A-mode when $\gamma=\pi/2$. However, some suppression of Alfv\'enic perturbation in line-of-sight projection for perpendicular field remains to be a real effect. 
From Eq.~(\ref{eq:f_alfen}) at $\gamma=\pi/2$ we find
\begin{equation}
f_0^2(M_A,\pi/2) = \frac{1}{2} (1-\alpha) M_{A\perp}^2
\approx \frac{1}{4} M_A^4 \frac{1 + 2 M_A^2}{(1+ M_A^2)^2}
\label{eq:f0Alfvenpi2}
\end{equation}
which shows that at perpendicular orientation of the mean field to LOS in the strongly sub-{\torefereetwo Alfv\'en}ic regime one has
$\bar{f}_0^2 \sim M_A^4$ dependence that comes equally from a high anisotropy of the power spectrum
and a small amount of mean field wandering, each effect contributing one $M_A^2$ factor. Though
this configuration is rather special, the case often appears in numerical simulations and one should
be aware of additional suppression of angle fluctuations in this case. 

Although the power spectrum treatment in Eq.~(\ref{eq:f_alfen}) has been truncated to $M_A^2$ order,
for $\gamma\smallrightarrow \pi/2$ expanding Eq.~(\ref{eq:f_alfen}) to $M_A^4$ gives the correct asymptotic behaviour for $f_0$
\begin{equation}
f_0^2 \sim \frac{1}{4} M_A^2 \left(M_A^2+2\cos^2\gamma \right),
    \label{eq:f0A_smallMAandgamma}
\end{equation}
since the omitted corrections are proportional to the product $M_A^2 \cos\gamma^2 $ and small.
Eq.~(\ref{eq:f0A_smallMAandgamma}) demonstrates that the dependence $f_0\sim M_A$ arises when $\cos\gamma> M_A / \sqrt{2}$ while $f_0\sim M_A^2$ otherwise. 

In the limiting case of sufficiently small $M_A \ll \sqrt{2} \cos\gamma$, Eq.~(\ref{eq:f_alfen}) gives
\begin{equation}
    f_0
    \approx \frac{1}{\sqrt{2}}\frac{\cos\gamma}{\sin^2\gamma} M_A ~.
    \label{eq:alf_wand}
\end{equation}
which corresponds to neglected field wandering. Thus
we can interpret the transition from
$f_0 \sim M_A$ to $f_0 \sim M_A^2$ as a transition to the regime when field wandering becomes important.

\begin{figure}
\includegraphics[width=0.45\textwidth]{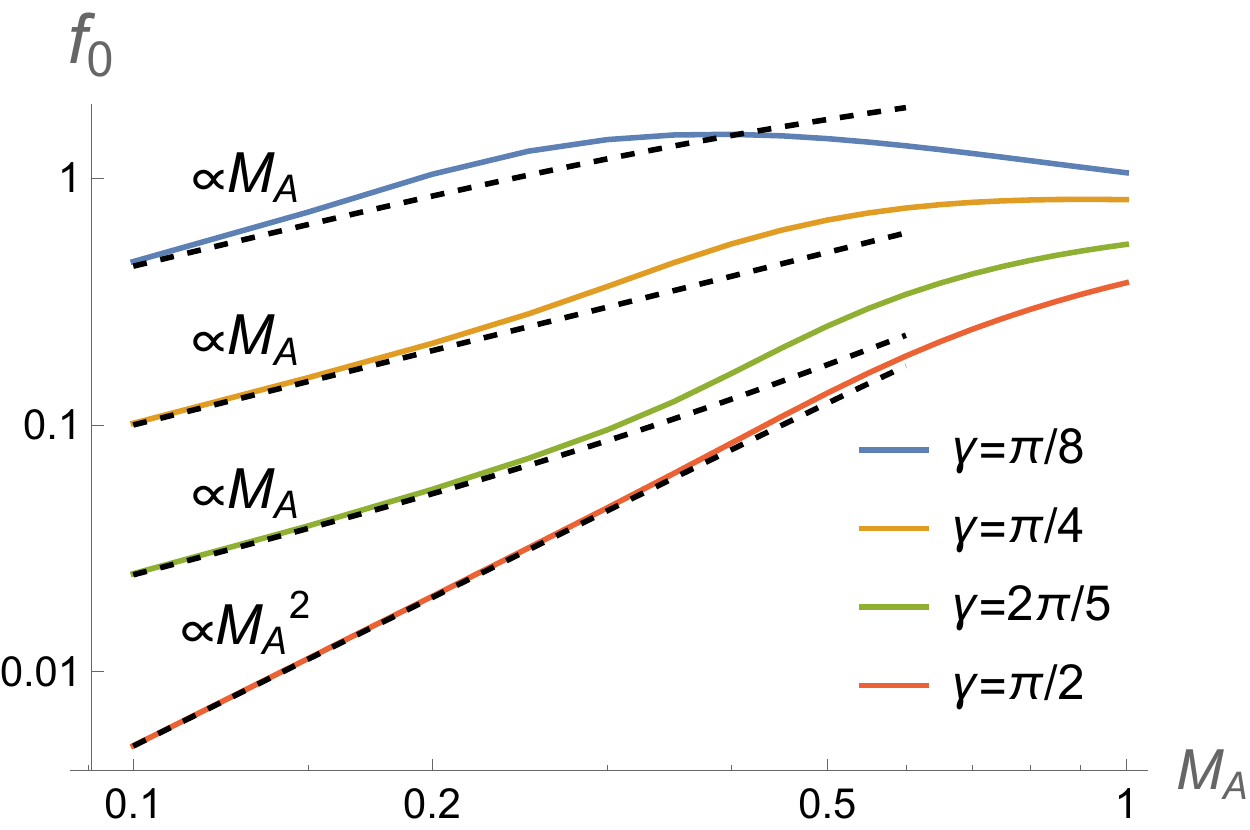}
\caption{$f_0$ factor for {\torefereetwo Alfv\'en}ic turbulence for several orientations of the mean magnetic field, $\gamma$. Dashed lines show asymptotic behaviour given by Eq.~(\ref{eq:f_alfen}).}
\label{fig:f02_pi2_Ma}
\end{figure}

The dependence of $f_0$ as a function of $M_{A}$ calculated numerically in our model is presented in Fig.~\ref{fig:f02_pi2_Ma} for several orientations of the magnetic field $\gamma$. The figure demonstrates the linear $f_0 \propto M_A$ behaviour at low $M_A$ for all orientations of the mean field 
not strictly perpendicular to LOS. The formula Eq.~(\ref{eq:f_alfen}) works well at least for $M_A \lesssim 0.3$ for all $\gamma$'s.
At higher $M_A \gtrsim 0.3$, the exact dependence becomes steeper than Eq.~(\ref{eq:f_alfen}), which is expected since
Eq.~(\ref{eq:f_alfen}) uses power spectrum expansion truncated to the quadratic $M_{A\perp}^2$ order.  At nearly
perpendicular configurations (e.g., $\gamma=2\pi/5$) we observe an almost quadratic $f_0 \propto M_A^2$ raise in $ 0.3 \lesssim M_A \lesssim 0.6$ range, before saturating at $M_A > 1$.  As $\gamma \to \pi/2$, the transition point from linear to quadratic dependence slides to ever smaller $M_A$ following $\cos\gamma$ in accordance to Eq.~(\ref{eq:f0A_smallMAandgamma}), revealing for $\gamma=\pi/2$ the quadratic $f_0 \propto M_A^2$ scaling across the whole range $ M_A \lesssim 0.6$, as accurately described
by Eq.~(\ref{eq:f0Alfvenpi2}).

\begin{figure}
\includegraphics[width=0.49\textwidth]{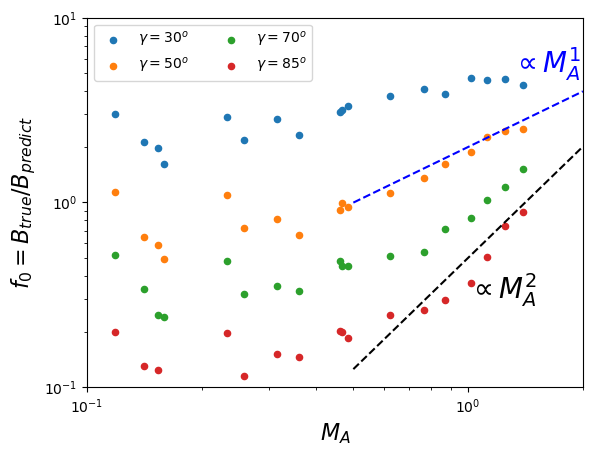}
\label{fig:fig5b_compare_new}
\caption{The numerical result of $f_0$ factor for {\torefereetwo Alfv\'en}ic turbulence for several orientations of the mean magnetic field, $\gamma$ at $30^o,50^o,70^o \& 85^o$. Dashed lines show asymptotic behaviour as indicated in Fig.\ref{fig:f02_pi2_Ma}. }
\end{figure}

For configuration of the mean field more aligned with LOS, $f_0$ factor is close to unity for $0.1 \lesssim M_A \lesssim 1$, though still scaling linearly $f_0 \propto M_A$ at small $M_A$'s.  Worth noting is that Mach number dependence  may become non-monotonic as $\gamma=\pi/8$ case in Fig.~\ref   {fig:f02_pi2_Ma} attests to.  
To test the predictions from Fig.\ref{fig:f02_pi2_Ma},  we perform the same calculation in numerical simulations by extracting the Alfv\'en contributions using \cite{CL03}. We see from Fig.\ref{fig:fig5b_compare_new} that the trend is very similar as Fig.\ref{fig:f02_pi2_Ma} for $M_A>0.5$, including the change of dependence of $f_0$ as $\gamma$ changes from $85^o$ to $30^o$, and the transition from steep to shallow dependence at $M_A \approx 0.5$. Notice we believe the reason why the curve is flatter when $M_A<0.5$ is due to the strong turbulence transition is more and more restrictive in our numerical simulations. (See Appendix \ref{app:mhdturb}).

Clearly, in general, the determination of magnetic field strength depends on our knowledge of the angle between the mean magnetic field and the line of sight $\gamma$. The statistical determination of this angle is possible (see Paper II), but in this paper it acts as a parameter. Note, that this presents an uncertainty both for DMA and the traditional DCF techniques. In our case, the functional dependence on this parameter is explicitly given.

\subsection{High-$\beta$ Nearly Incompressible MHD Turbulence}
\label{sec:highb}

If the sonic speed $c_s$ is larger than Alfv\'en velocity $V_A$, the media is high-$\beta$, as $\beta\sim c_s^2/V_A^2$. Such media is widely presented in interstellar environments.

Gas in HII regions, warm interstellar and circumstellar gases can be approximated as weakly compressible high-$\beta$ gas (See Tab.\ref{tab:ISMtable}). In this case fast MHD modes are very similar to the sound waves and their contribution vanishes due to the integration along the line of sight provided that $R$ is significantly smaller than the integration scale $L$.\footnote{This cancellation is easy to understand by considering the decomposition of the fast waves with ${\bf k}\| {\bf v}$ into Fourier components. The decomposed velocity component after Doppler broadening is still parallel to the line of sight. The integration of the waves oscillating along the line of sight results in a negligible contribution if the wavelength of the wave is much smaller than the integration length.}

In such weakly compressible media, the slow and {\torefereetwo Alfv\'en} modes have very similar power spectra, being two "polarizations" of a solenoidal wave and they have equal contributions to the observed fluctuations. The tensor of "Slow + Alfv\'en" fluctuations has simply the isotropic solenoidal form. In this case $W_{n-p}^I = \delta_{np}$ while $ G_{n-p}^I = \frac{1}{2} \delta_{np}+\frac{1}{4} \left(\delta_{n-2,p} + \delta_{n+2,p}\right) $. 
As a result we find
\begin{equation}
\begin{aligned}
& \overline{B}_\perp^2 = 4 \pi \rho f^2_n \frac{\mathcal{D}_n^v}{D_n^\phi}\\
& \text{where} \quad f^2_n(M_A,\gamma) = \frac{1}{2}+\frac{\widehat{E}^{2D}_{n-2} + \widehat{E}^{2D}_{n+2}}{4 \widehat{E}^{2D}_n}
\end{aligned}
\label{eq:B_slow_alfen}
\end{equation}

\begin{figure}
\includegraphics[width=0.45\textwidth]{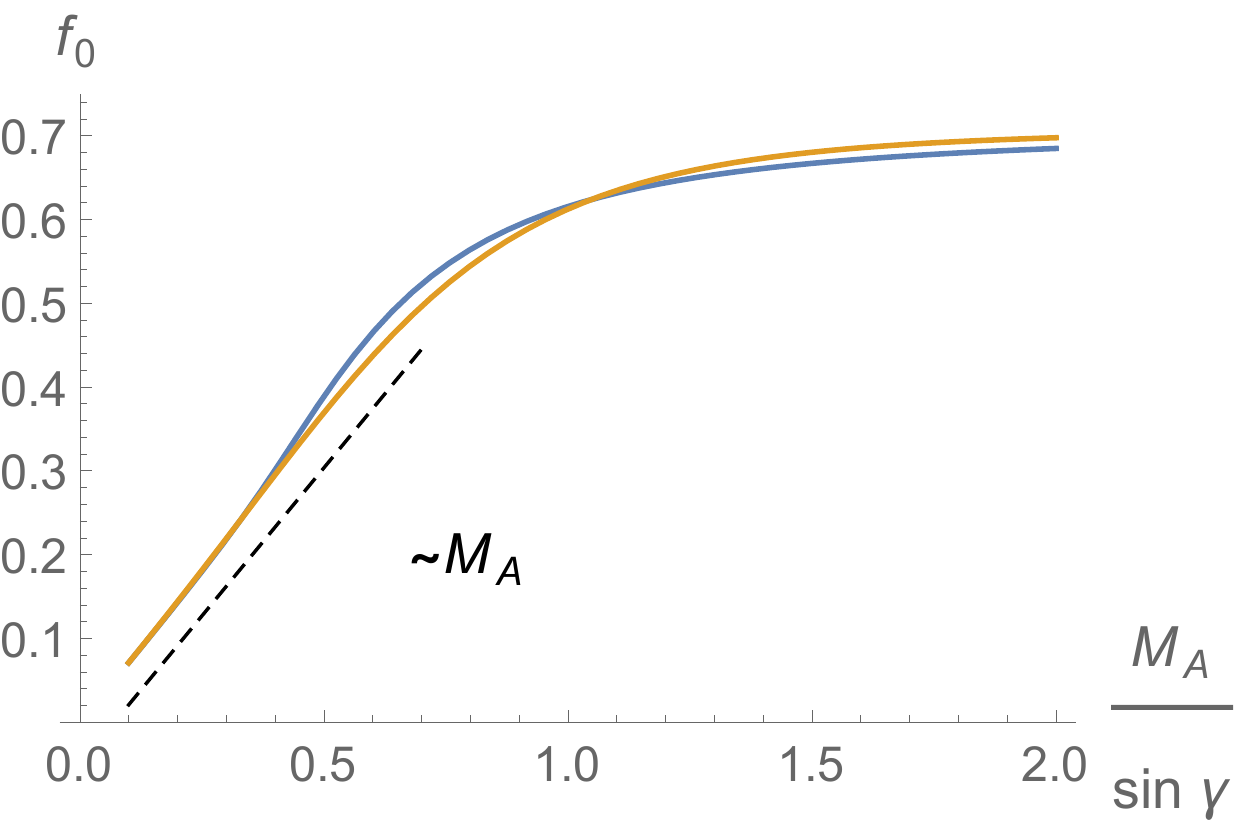}\vspace{2mm}\\
\includegraphics[width=0.45\textwidth]{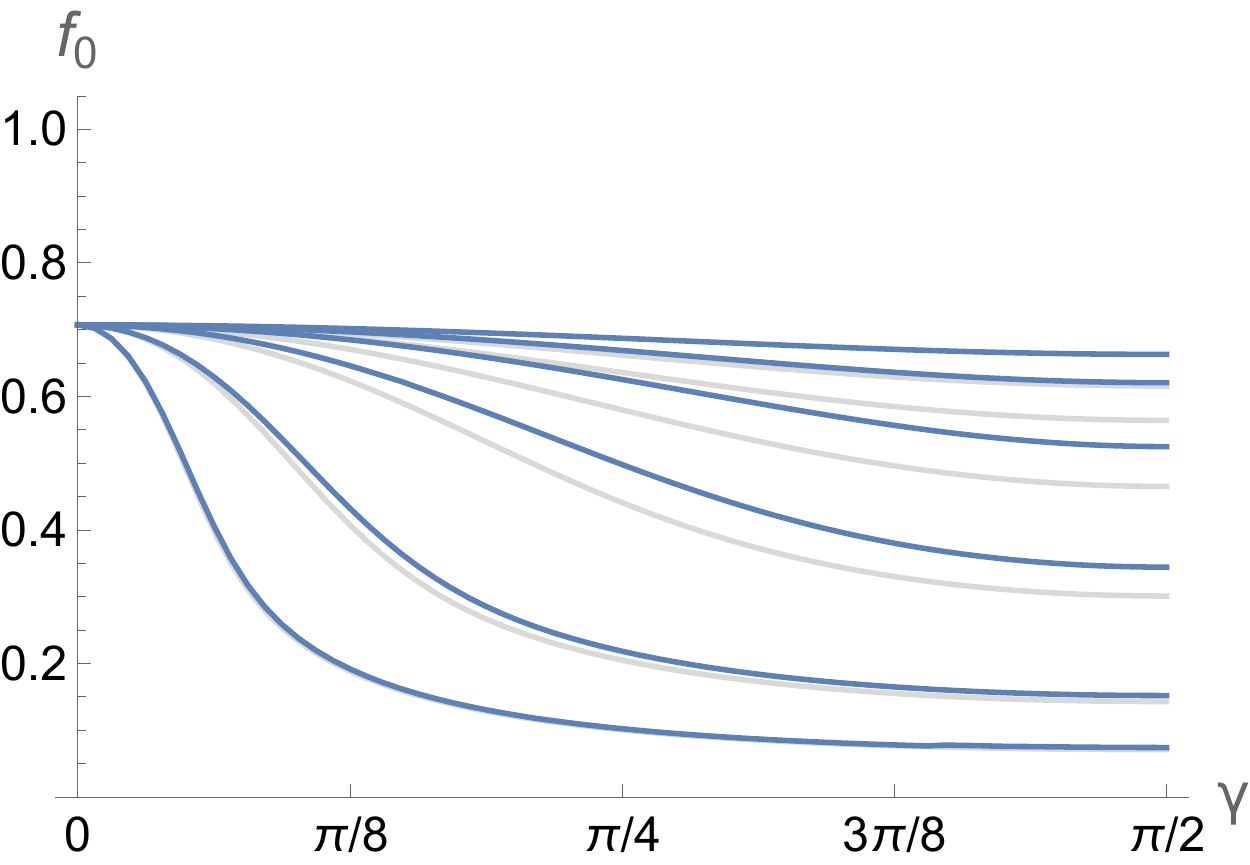}
\caption{Top: numerical factor $f_0$ in Eq.~(\ref{eq:B_slow_alfen}) for strong turbulence regime as a function of $M_{A\perp}$. Orange curve corresponds to approximation Eq.~(\ref{eq:f0bar_strong})
while the blue curve is the result with an accurate evaluation of spectral multipoles in Eq.~(\ref{eq:B_slow_alfen}).
 Bottom: factor $f_0$ according to a more accurate evaluation
of the field wandering via Eqs.~(\ref{eq:E_averaging}, \ref{eq:P_lambda})
as a function of 3D angle of the mean magnetic field $\gamma$ for, from top to bottom, $M_A=1, 0.8, 0.6, 0.4, 0.2, 0.1$. Grey lines are the corresponding curves obtained
via the approximation in the top panel. }
\label{fig:f02_I}
\end{figure}

Focusing on the monopole $f^2_0 = \frac{1}{2}( 1 + \widehat{E}^{2D}_2/\widehat{E}^{2D}_0 )$ as the simplest case to measure observationally, we notice that Eq.~(\ref{eq:B_slow_alfen}) is equivalent to {\torefereetwo Alfv\'en} case for $\alpha=0$, since the monopole case have
no effects to the anisotropic structure of the modes. In this case, however, 
it is an exact result that only the monopole and the quadrupole of the power distribution add contributions to $f_0^2$. Since  their ratio in high-$\beta$ plasma remains to be given by Eq.~(\ref{eq:AlfvenE2E0}), we obtain for sufficiently small $M_{A,\bot}$:
\begin{equation}
\bar{f_0}^2 
\approx \frac{1}{2} M_{A\perp}^2 \frac{ 1 + 2 M_{A\perp}^2}{
 1 + M_{A\perp}^2 + 2 M_{A\perp}^4}
\label{eq:f0bar_strong}
\end{equation}
for all orientations of the mean magnetic field, as shown in Fig.~\ref{fig:f02_I}. Eq.~(\ref{eq:f0bar_strong}) is more accurate in an intermediate sub-{\torefereetwo Alfv\'en}ic regime $M_{A\perp} > 0.3$ than Eqs.~(\ref{eq:f_alfen},\ref{eq:f0Alfvenpi2}) that we had for Alfv\'enic turbulence,  since the quadrupole modelled by Eq.~(\ref{eq:AlfvenE2E0}) is not truncated to the leading order in $M_{A\perp}$.

This result of high-$\beta$ weakly compressible turbulence should be contrasted with that of the purely Alfv\'enic turbulence,  especially in near perpendicular configuration at small $M_A$. The origin of small value of $f_0$ in sub-{\torefereetwo Alfv\'en}ic regime in both cases is the suppression in projection of  $\delta B_y$ that is responsible for angle variations of the observed magnetic field lines whereas velocity centroids fluctuations are not suppressed and dominated by the full power of Alfv\'en modes.
However, while high-$\beta$ result comes only from the anisotropy of
power distribution, in purely Alfv\'enic turbulence we have an additional geometrical 
suppression due to specific anisotropic structure of Alfv\'en perturbations.
Thus, ignoring projection effects 
in the patches of the sky where the mean field is perpendicular to LOS may lead to an overestimation
of the magnetic field strength by a significant factor for low $M_A$.

In the areas of the sky where the mean field direction is close to LOS, in weakly compressible turbulence {\torefereetwo Alfv\'en} mode dominates POS magnetic field fluctuation, while slow mode is responsible for the LOS velocity centroids. Both contribute similar power and $f_0^2 \sim \frac{1}{2}$. We remind however that
in the patches of the sky where $\gamma$ is below some critical angle $\gamma_{crit}$, DMA (and DCF) measurements can not be interpreted as a measure of the mean magnetic field.

\subsection{High-$\beta$ Turbulence with Isotropic Driving}

One may notice that in low $M_A$ regime the high-$\beta$ model with equipartition of energy between Alfv\'en and slow modes does not lead to isotropic distribution of velocities, due
to highly anistropic power distribution. Indeed, as $M_A \to 0$ we get 
$\langle v_\parallel^2 \rangle = \langle v_\perp^2 \rangle $, where parallel component
with respect to the magnetic field is dominated by the slow mode, while perpendicular
components are given by the Alfv\'en mode.  But this means that $\langle v_z^2 \rangle
= 2 \langle v_x^2 \rangle = 2 \langle v_y^2 \rangle$, if $z$ is along the magnetic field.
As anisotropy of the spectrum decreases with $M_A$ increase, velocity distribution becomes
more and more isotropic.

\begin{figure}
\includegraphics[width=0.45\textwidth]{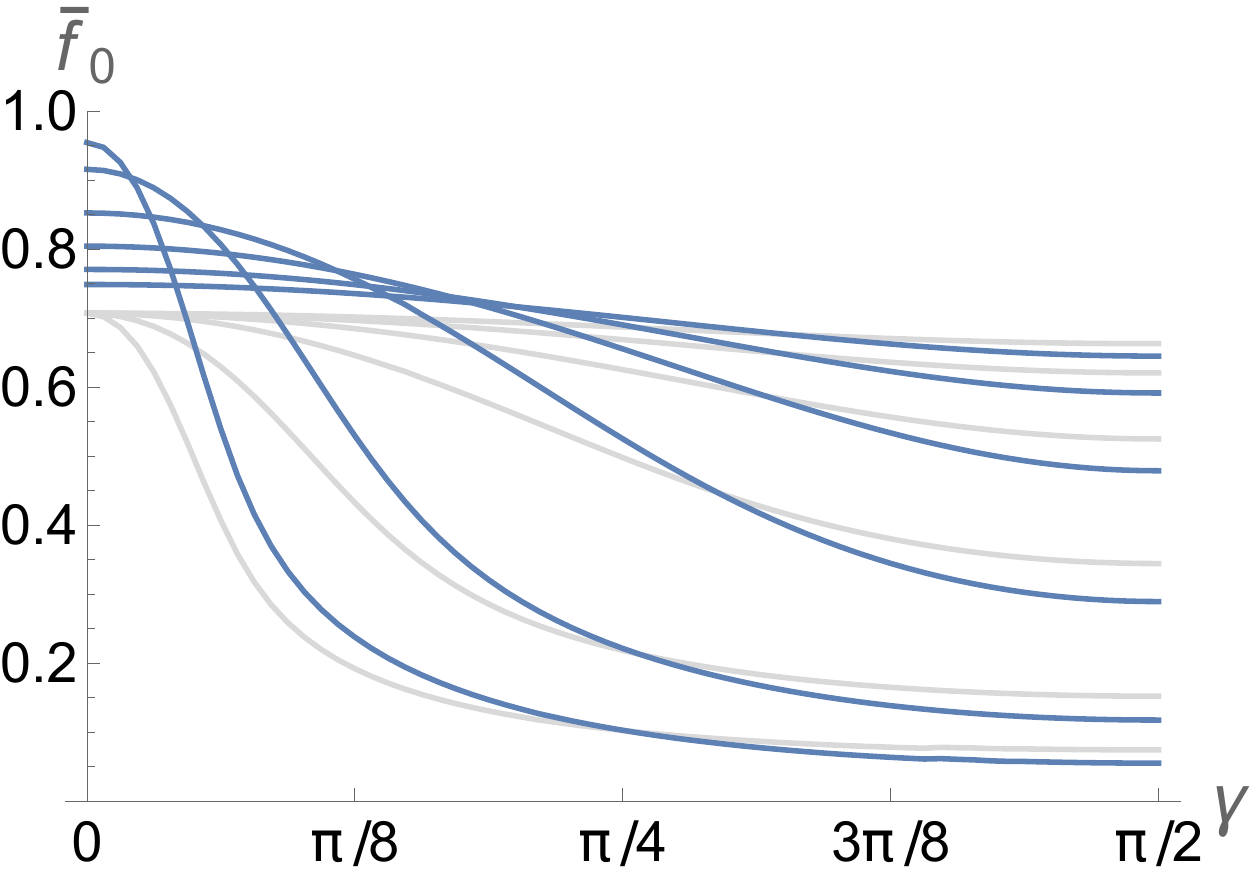}
\caption{The factor $f_0$ in high-$\beta$ turbulence with isotropic driving (blue) as compared with
 the case of equal energy in {\torefereetwo Alfv\'en} and slow modes (grey, same as in Figure~\ref{fig:f02_I}), shown as a function of 3D angle of the mean magnetic field $\gamma$ for, from top to bottom, $M_A=1, 0.8, 0.6, 0.4, 0.2, 0.1$. }
\label{fig:f0_isovel}
\end{figure}

If, however, the turbulence is driven isotropically at all Alfv\'en Mach numbers,
$\langle v_z^2 \rangle \approx \langle v_x^2 \rangle \approx \langle v_y^2 \rangle$, i.e., $\langle v_\parallel^2 \rangle = \onehalf \langle v_\perp^2 \rangle $, the energy in the slow mode will be less than that in the Alf\'en one, with the $M_A$ dependent fraction $\langle v_\mathcal{S}^2 \rangle / \langle v_\mathcal{A}^2 \rangle$, changing
from one half as $M_A \to 0$ to unity as $M_A > 1$.  We can approximate this transition as
$\langle v_\mathcal{S}^2 \rangle / \langle v_\mathcal{A}^2 \rangle \sim \frac{1}{2} + 
\frac{1}{\pi} \arctan{M_A}$ which gives $\langle v_\mathcal{S}^2 \rangle / \langle v_\mathcal{A}^2 \rangle=0.75$ at $M_A=1$, consistent with our simulations.  In Figure~\ref{fig:f0_isovel} we observe that the decrease of power in slow modes relative
to Alfv\'en ones, results, 
for small $M_A$, in an \textit{increase} of $f_0$ at small $\gamma$ by a factor up to $\sqrt{2}$,
and in a \textit{decrease} of $f_0$ at $\gamma \approx \pi/2$ by up to the same $\sqrt{2}$ (this limiting value corresponds to the case when slow mode contain half the energy of Alfv\'en ones). 

Dependence of the fraction of the slow modes on $M_A$ makes $f_0$ to depend separately on $M_A$ and $\gamma$ rather than just on $M_{A\perp}$ combination.  Nevertheless, the fit
\begin{equation}
f_0^2 = \frac{1}{2} \times \frac{1+\cos^2\gamma}{1+\sin^2\gamma } M_{A\perp}^2 \frac{ 1 + 4 (1+\sin^2\gamma)  M_{A\perp}^2 }{ 1 + 4 (1+\cos^2\gamma) M_{A\perp}^4}~,
\label{eq:highbeta2}
\end{equation}
which works very well for $\gamma \approx \pi/2$, continues to work for all $M_{A\perp} < 1$.
In this expression $f_0^2 \sim M_{A\perp}^2$ is a rigorous asymptotic at small $M_{A\perp}$, including the prefactor, that stems from Eqs.~(\ref{eq:AlfvenE2E0},\ref{eq:f_alfen}). The last term is a fit for the transition to $M_{A\perp} \sim 1$.

\section{Low-$\beta$ Medium Case}
\label{subsec:lowbeta}

Low-$\beta$ turbulence presents a complex case where all three turbulent modes play part in the estimation of the magnetic field.  We shall focus on the limit $\beta\to0$, in which slow modes are the magnetic compression propagating along magnetic field lines and therefore only marginally contribute to the changes of the magnetic field direction. Also, numerical simulations show that the contribution of the  fast mode is
subdominant to the velocity of the medium. In this situation, 
\begin{equation}
    f_n^2 \approx 
    \frac{\sum_{p} \left[ 
    \mathcal{\widehat{A}}_p^{2D} G_{n-p}^\mathcal{A}  + \frac{\langle \delta B_\mathcal{F}^2 \rangle \mathcal{I}_0^\mathcal{F}}{\langle \delta B_\mathcal{A}^2 \rangle \mathcal{I}_0^\mathcal{A}} \widehat{\mathcal{F}}_p^{2D} G_{n-p}^\mathcal{F} \right]}
    {\sum_{p}\left[ \mathcal{\widehat{A}}_p^{2D} \mathcal{W}_{n-p}^\mathcal{A} + \frac{\langle v_\mathcal{S}^2 \rangle \mathcal{I}_0^\mathcal{S}}{\langle v_\mathcal{A}^2 \rangle \mathcal{I}_0^\mathcal{A}} \widehat{\mathcal{S}}_p^{2D} \mathcal{W}_{n-p}^\mathcal{S}  \right]}
    \label{eq:fn_lowbeta_gen}
\end{equation}
Slow modes are expected to have similar spectrum as Alfv\'en modes with the same scaling $\mathcal{I}_0^\mathcal{S}(R)=\mathcal{I}_0^\mathcal{A}(R)$ and angular dependence $ \widehat{\mathcal{S}}_p^{2D}= \widehat{\mathcal{A}}_p^{2D} $
and the relative power in two modes is constant. 
On the other hand fast modes may have a different scaling $\mathcal{I}_0^\mathcal{F}(R)$ from Alfv\'en modes. 

We adopt as before the anisotropic model power spectrum for Alfv\'en mode 
given in Eq.~(\ref{eq:hatEmu_averaged})
and assume that the fast mode power is distributed
isotropically, $\mathcal{F}_p = \mathcal{F}_0 \delta_{0p}$.
Eq.~(\ref{eq:fn_lowbeta_gen}) can then be significantly simplified for $n=0$. Noting that $\mathcal{W}^S_{n-p} = \delta_{np} \cos^2\gamma $ (Eq.~\ref{eq:W_lowbeta}) and finding from Eq.~(\ref{eq:app_tildeDijIA}) that $G^F_0 = \onehalf \frac{1-\cos\gamma}{1+\cos\gamma}$ we obtain

\begin{equation}
    f_0^2(M_A,\gamma) \approx \frac{\sum_{p} 
    \mathcal{\widehat{A}}_p^{2D} G_{p}^\mathcal{A} + \onehalf \frac{\langle \delta B_\mathcal{F}^2 \rangle \mathcal{I}_0^\mathcal{F}}{\langle \delta B_\mathcal{A}^2 \rangle \mathcal{I}_0^\mathcal{A}} \frac{1-\cos\gamma}{1+\cos\gamma}}
    {\sum_{p} \mathcal{\widehat{A}}_p^{2D} \mathcal{W}_{p}^\mathcal{A} + \frac{\langle v_\mathcal{S}^2 \rangle}{\langle v_\mathcal{A}^2 \rangle }  \mathcal{\widehat{A}}_0^{2D} \cos^2\gamma }
    \label{eq:f0_lowbeta_gen}
\end{equation}

For practical applications the difference between the scaling $\mathcal{I}_0^\mathcal{F}(R)/\mathcal{I}_0^\mathcal{A}(R)$ of fast and Alfv\'en modes (see Cho \& Lazarian 2003, Kowal \& Lazarian 2010) is weak enough not to be of secondary importance for the range of $R$ involved in the DMA study. Thus, in the zeroth approximation, the difference in Alfv\'en and fast mode scaling may be disregarded.

For small $M_{A\perp} < 1$, the multipole summation for Alfv\'en modes that led to Eq.~(\ref{eq:f_alfen}) can be used again. We also need to additionally take into account the field wandering effect on the fast modes,
which is accomplished by replacing $G_0^F \to W_I + W_L G_0^F$, while the wandering effect on slow modes is accounted for by $G_0^\mathcal{S} \to W_I + W_L G_0^\mathcal{S}$ and using the same averaged spectrum, as for the Alfv\'en mode. The result is
\begin{eqnarray}
    \label{eq:f0_lowbeta_sum}
    f_0^2 && (M_A, \gamma) \approx \\
    \approx&& \frac{1}{2} \; \frac{ 
    \widehat{\mathcal{A}}_0^{2D} (1 - \alpha \sin^2\gamma) M_{A\perp}^2  + \frac{\langle \delta B_\mathcal{F}^2 \rangle }{\langle \delta B_\mathcal{A}^2 \rangle } \left(1-2 \alpha \frac{\cos\gamma}{1+\cos\gamma} \right) }
    {\widehat{\mathcal{A}}_0^{2D} \left( 1 - \alpha \cos^2\gamma + \frac{\langle v_\mathcal{S}^2 \rangle}{\langle v_\mathcal{A}^2 \rangle } (1-\alpha\sin^2\gamma) \right) }
    \nonumber
\end{eqnarray} 
where 2D monopole of the projected {\torefereetwo Alfv\'en} angular spectrum $\widehat{\mathcal{A}}_0^{2D}(M_A,\gamma)$ is given by  Eq.~(\ref{eq:A2D_p}) to be
\begin{equation}
\widehat{\mathcal{A}}_0^{2D} = \frac{2 \exp\left(-\onehalf M_{A\perp}^{-2}\right) I_0\left(\onehalf M_{A\perp}^{-2}\right)}{\sqrt{\pi} M_A \mathrm{erf}(1/M_A)} \sim \frac{2}{\pi \sin\gamma} ~.
\end{equation}
The last expression is the leading term of expansion at small $M_{A\perp}$. Provided that $M_{A\bot}\ll \frac{\langle \delta B_\mathcal{F}^2 \rangle }{\langle \delta B_\mathcal{A}^2 \rangle } $ and $\alpha\approx 1$ when $M_A$ is small,
\begin{equation}
    f_0^2 (M_A, \gamma) \approx\frac{\pi \sin\gamma}{4} \frac{ 
     \frac{\langle \delta B_\mathcal{F}^2 \rangle }{\langle \delta B_\mathcal{A}^2 \rangle } \left(\frac{1-\cos\gamma}{1+\cos\gamma} \right) }
    { 1 + \cos^2\gamma\left(\frac{\langle v_\mathcal{S}^2 \rangle}{\langle v_\mathcal{A}^2 \rangle } -1\right) },
    \label{eq:f0_lowbeta_sum1}
\end{equation}
and the dependence of $f_0$ on $M_A$ is lost due to the fact that the variations of angle $\phi$ arise in this case from isotropic fast modes. 

The above requirement on the relation between $M_A$
and $\frac{\langle \delta B_\mathcal{F}^2 \rangle }{\langle \delta B_\mathcal{A}^2 \rangle }$ can be somewhat relaxed for $\gamma \rightarrow \pi/2$. The factor $f_0$ in this limit exhibit the dependence only on the ratio of energies in fast and Alfv\'en modes:
\begin{equation}
   f_0^2 (M_A, \gamma)\approx \frac{\pi}{4}  \frac{\langle \delta B_\mathcal{F}^2 \rangle }{\langle \delta B_\mathcal{A}^2 \rangle }
    \label{eq:f0_lowbeta_sum2},
\end{equation}
which corresponds to the notion that slow modes in low $\beta$ regime are similar to sound waves that cannot contribute neither to the variations of $\phi$ nor to the variations of $\delta v$ for $\gamma=\pi/2$.

For $\gamma \to 0$ and $M_A < 1$, $\widehat{\mathcal{A}}_0^{2D} \approx 2/(\sqrt{\pi} M_A)$. Assuming also that $M_{A\bot}\ll 1$, Eq. (\ref{eq:f0_lowbeta_sum}) transforms to
\begin{equation}
    f_0^2 (M_A, \gamma)\approx\frac{1}{2} M_{A\bot}^2\frac{\langle v_\mathcal{A}^2 \rangle}{\langle v_\mathcal{S}^2 \rangle},
    \label{eq:f0_lowbeta_smallgamma}
\end{equation}
which demonstrate that in this regime the variations of $\phi$ arises from anisotropic Alfv\'enic modes and therefore $f_0$ gets $\sim M_{A\bot}$. However, the requirement of $M_{A\bot}\ll 1$ is very restrictive in this case. 

In the opposite limit of negligible fast mode contribution, i.e. $\frac{\langle \delta B_\mathcal{F}^2 \rangle }{\langle \delta B_\mathcal{A}^2 \rangle } \ll M_{A\bot}^2<1$, Eq.~(\ref{eq:f0_lowbeta_sum}) gives
\begin{equation}
     f_0^2 (M_A, \gamma) \approx \frac{1}{2} \; \frac{ 
     (1 - \alpha \sin^2\gamma) M_{A\perp}^2 }
    { 1 - \alpha \cos^2\gamma + \frac{\langle v_\mathcal{S}^2 \rangle}{\langle v_\mathcal{A}^2 \rangle } (1-\alpha\sin^2\gamma) },
    \label{eq:f0_lowbeta_sum4}
\end{equation}
which transfers to Eq.~(\ref{eq:f_alfen}) if $\frac{\langle v_\mathcal{S}^2 \rangle}{\langle v_\mathcal{A}^2 \rangle }\rightarrow 0$, i.e. for pure Alfv\'enic turbulence. Otherwise, slow modes do not contribute to the fluctuations of $\delta \phi$, but increase the velocity fluctuations, decreasing the value of $f_0$. 

\begin{figure*}[ht]
\centering
\includegraphics[width=0.45\textwidth]{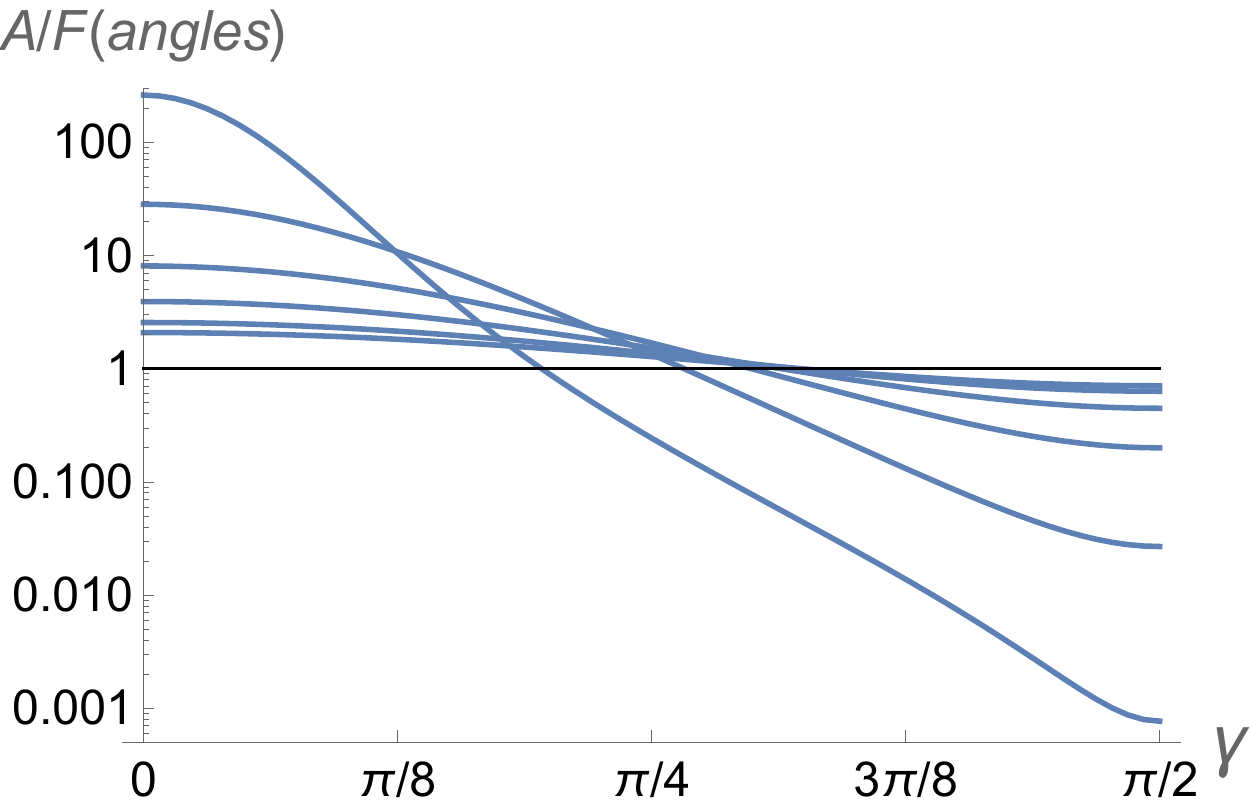}
\includegraphics[width=0.45\textwidth]{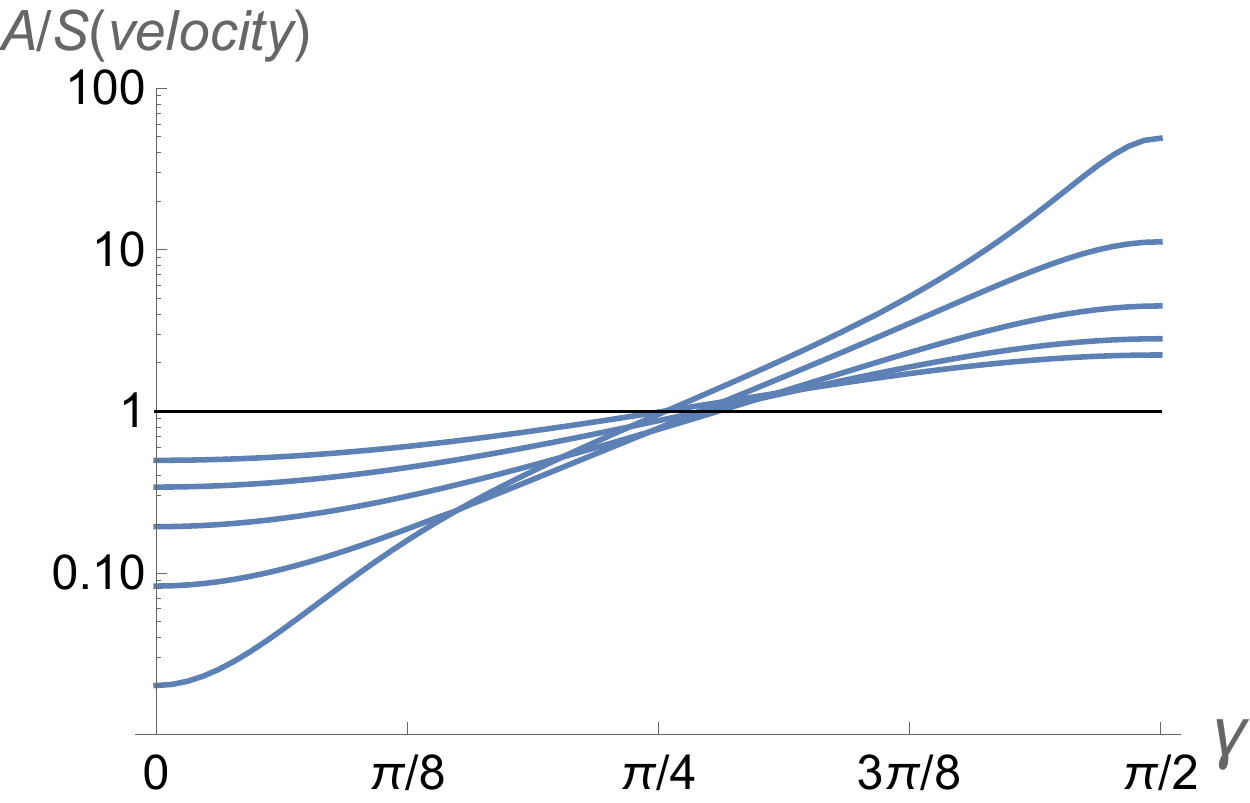}
\caption{The ratio of the variance normalized contributions of Alfv\'en to fast modes to the projected angle (left) and {\torefereetwo Alfv\'en} to slow contributions to velocity (right) structure functions,  as a function of $\gamma$ for low-$\beta$ turbulence. Different curves corresponds to, from the most constant to the most variable curve, $M_A=1, 0.8, 0.6, 0.4, 0.2, 0.1$. }

\label{fig:AFcomp}
\end{figure*}

Let us first look at the balance of contributions to angle fluctuations. 
In Fig.~\ref{fig:AFcomp} (left panel) 
we show the ratio of Alfv\'en to Fast modes to the projected angle structure function at equal power in both modes (i.e $\langle \delta B_\mathcal{F}^2 \rangle \approx \langle \delta B_\mathcal{A}^2 \rangle$). 
The curves are, thus, variance normalized, and reflect only the geometrical structure of the modes. They are to be multiplied by the ratio of powers when it is known at
the scale of study.  

Fig.~\ref{fig:AFcomp} focuses on the dependence of contributions as a function of orientation $\gamma$ and for different {\torefereetwo Alfv\'en} Mach numbers. Fast modes depend on $M_A$ due to field wandering only, while {\torefereetwo Alfv\'en}
modes depend on $M_A$ primary due to anisotropic power distribution, while the effect of field wandering dilutes this effect.  

We observe that when the magnetic field is predominantly aligned with POS, the fast mode plays an important role in POS angle fluctuations of the field. The reason behind is based on the earlier discovered suppression of Alfv\'en contribution for such orientations due to the projection effects, as discussed in \S~\ref{sec:projections}. The effect is stronger when $M_A$ is smaller. At the same time, fast mode magnetic perturbations in POS projection, given by $G^{\mathcal F}=G^F$ (see Eq.~(\ref{eq:app_tildeDijIA})),  do not vanish at $\gamma=\pi/2$.
Having isotropic power spectrum, low-$\beta$ fast modes will perturb equally the magnetic 
field components parallel and perpendicular to the mean field, thus perpendicular POS
perturbations will not be suppressed. Hence, for $\gamma\approx \pi/2$ the observed variations
of magnetic field  and, correspondingly,  the angle fluctuations will be  dominated by the
fast modes over the {\torefereetwo Alfv\'en} as well as slow one. When $\gamma$ is less than $\pi/2$, Alfv\'en modes begin contributing more to the projected angle and become dominant. Transition angle
depends on $M_A$ and how much power in fast modes there is.  

The balance of Alfv\'en and slow contributions
to the velocity centroids sets the denominator in Eq.~(\ref{eq:fn_lowbeta_gen}).   With similar spectra between these modes, the balance is set by
the scale independent power ratio $\frac{\langle v_\mathcal{S}^2 \rangle}{\langle v_\mathcal{A}^2 \rangle }$. In the right panel of Fig.~\ref{fig:AFcomp} we study $M_A$ and $\gamma$ dependence for equal power between two modes. In this case we find that Alfv\'en mode dominates velocity fluctuations at $\gamma > \pi/4$ while slow mode takes over at $\gamma < \pi/4$, in full accordance with simple model Eq.~(\ref{eq:f0_lowbeta_sum}). 

\begin{figure}
\includegraphics[width=0.42\textwidth]{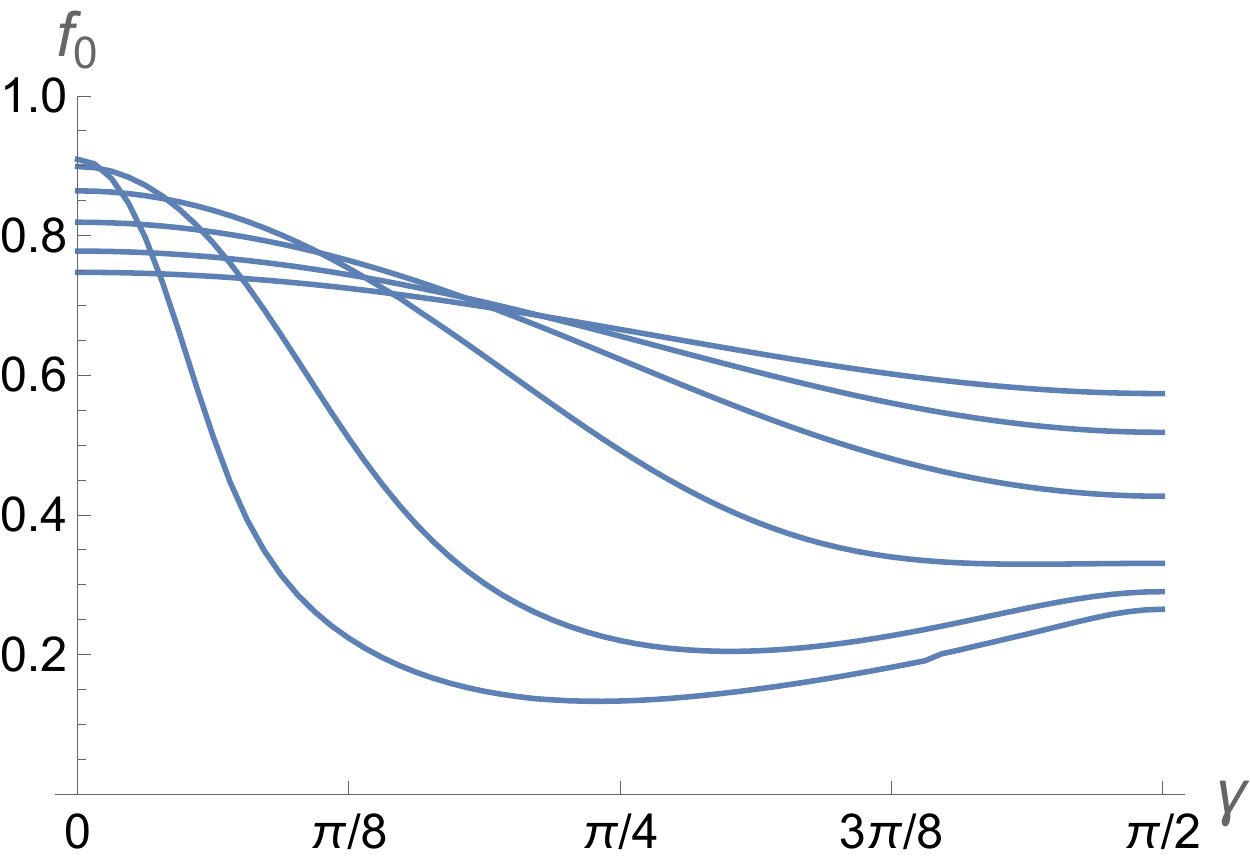}
\caption{Numerical factor $f_0$ as a function of $\gamma$ for low-$\beta$ turbulence with $\langle v_\mathcal{S}^2 \rangle / \langle v_\mathcal{A}^2 \rangle=0.6$ and $\langle \delta B_\mathcal{F}^2 \rangle / \langle \delta B_\mathcal{A}^2 \rangle=0.2$. Different curves corresponds to, from the most constant to the most variable curve, $M_A=1, 0.8, 0.6, 0.4, 0.2, 0.1$.  }
\label{fig:f0_lowbeta}
\end{figure}

The relative importance of slow and fast modes depends on driving. Our analysis of simulations in \S\ref{sec:numerical}  indicate a relatively low fraction of energy in fast modes at the level of 20\% relative to Alfv\'en ones and roughly equal contribution of slow and Alfv\'en modes.
In Fig. \ref{fig:f0_lowbeta} we show the modelled $f_0$ taking $\langle \delta B_\mathcal{F}^2 \rangle / \langle \delta B_\mathcal{A}^2 \rangle=0.3 M_A^{1/2}$ 
and $\langle v_\mathcal{S}^2 \rangle / \langle v_\mathcal{A}^2 \rangle=0.6$. This is a special case and we see that as $M_A$ increases the contributions of the modes to angles and velocities compensate each other to make $f_0$ nearly independent from $\gamma$. The most significant variations with $\gamma$ we observe for low $M_A$.

Fig \ref{fig:f0_lowbeta} illustrates that obtaining the magnetic field if both $\gamma$ and the composition of turbulence are unknown may include significant uncertainties. Those can be decreased, however, both through numerical and observational studies. For instance, numerical simulations of molecular cloud formation can help in getting an insight of what is expected in terms of mode composition. Observational decomposition of fluctuations into modes (see \citealt{leakage}) provides a synergistic insight into the redistribution of turbulent energy between the modes. An angle $\gamma$  molecular clouds can be roughly evaluated from the general spiral galactic structure or via combining Zeeman and polarization observations. A statistical evaluation of $\gamma$ is also possible (Chen et al. 2019, Hu et al. 2022). 

\section{Low $\beta$ case: Input from Numerical Simulations}
\label{sec:numerical}

\subsection{Two special cases validating our theoretical argument in \S \ref{subsec:lowbeta}}

\subsubsection{Synergistic use of numerical simulations}

Our analysis above provides crucial insight on how the statistics of velocity and angle $\phi$ behave according to the statistical theory of MHD turbulence. Our study in \S \ref{subsec:lowbeta} demonstrates that the use of DMA expressions, and also the magnetic field estimation method via the DCF-like formalism, requires accurate knowledge of the media provided that magnetized turbulence is present and dominant. For instance, the required initial knowledge is different in the case of low and high $\beta$ media. For instance, only slow and Alfv\'en modes are important for high $\beta$ case. This makes the study rather robust for separations $R$ that are smaller than $LM_A^2$. In particular, for high beta case we expect that the energy of Alfv\'en and slow modes are similar in order of magnitude or even equi-partitioned, which is expected true for sufficiently extended inertial range of incompressible turbulence. In this case,  the dependence of $f_0$ to $\gamma$ is much simpler to analyze according to theory, as we can see in \S 6 in detail. 

The low $\beta$ case is more challenging for practical studies. This, however, a very important case as it includes studies of turbulence in molecular clouds. The relative proportion of the modes can change and their contribution to the fluctuations of angle and velocity changes with angle $\gamma$.  This is reflected in \S \ref{subsec:lowbeta}. The main idea of our analysis in \S \ref{subsec:lowbeta} is that, when $\beta<1$, the velocity fluctuations mostly depend on the Alfv\'en and slow modes, while that of polarization angles depend on the Alfv\'en and fast modes. Therefore both the composition of turbulence and also the line of sight angle $\gamma$ is crucial in estimating the magnetic field strength.

Both angle $\gamma$ and composition of turbulence in terms of modes are parameters that can potentially be obtained from observations (see Yuen et al. 2022, Hu et al. 2022). However, in what follows we will numerically explore the output of numerical simulations to evaluate to what extend the DMA can be applied if these aforementioned parameters are not determined prior to the application of the technique.

\begin{figure*}[th]
    \centering
    \includegraphics[width=0.95\textwidth]{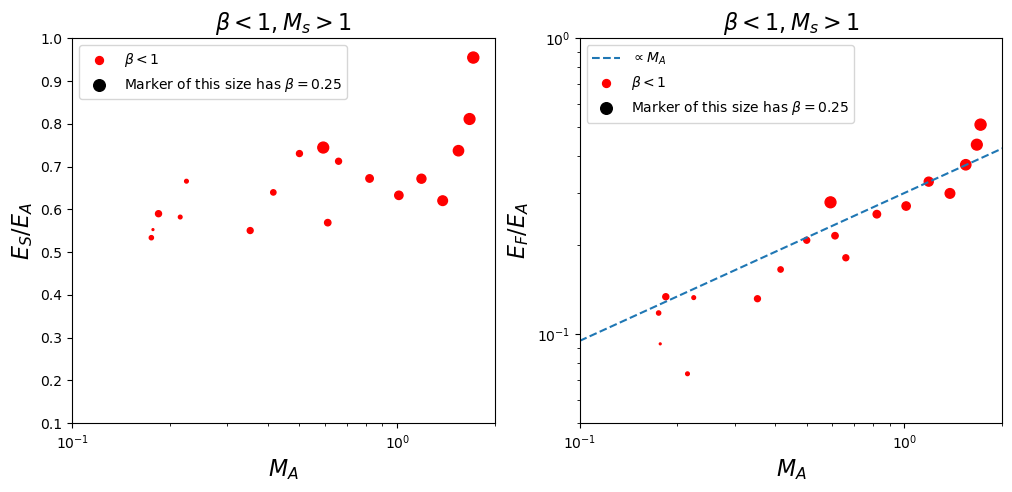}
    \caption{ The energy fraction of slow to Alfv\'en mode (left) and that of fast mode relative to the Alfv\'en mode (right) in our simulations with $\beta<1$ and $M_>1$ (Table \ref{tab:sim}) as a function of $M_A$. Here the size of the points are proportional to their respective $\beta$ values. The dashed line corresponds to $E_F/E_A=0.3 M_A^{1/2}$ over the range $M_A < 1$.
    }
    \label{fig:fraction}
\end{figure*}

For our suite of low-$\beta$ models, the distribution of
power in slow, $E_S$, and fast, $E_F$, modes relative to the power $E_A$ in Alfv\'en modes is
shown in Fig.~\ref{fig:fraction}. We find that the slow mode fraction is relatively
constant at $E_S/E_A \approx 0.6$ level, while the fraction of fast modes shows a weak 
dependence on $M_A$, $E_F/E_A \sim 0.3 M_A^{1/2}$. In \cite{CL02} it was found that when
the fast modes are generated through the conversion of energy from Alfv\'en modes starting from initial perturbations that are purely Alfv\'enic, the relative energy in the generated fast modes is proportional to $M_A$. In our simulations, the setting is different, the fast modes are generated by the solenoidal driving and  the dependence of the energy of fast modes on $M_A$ is found to be more shallow, i.e. $\sim M_A^{1/2}$ as seen in Fig.~\ref{fig:fraction}.

\subsubsection{$\pi/2$ Case}

\begin{figure}[th]
    \centering
    \includegraphics[width=0.45\textwidth]{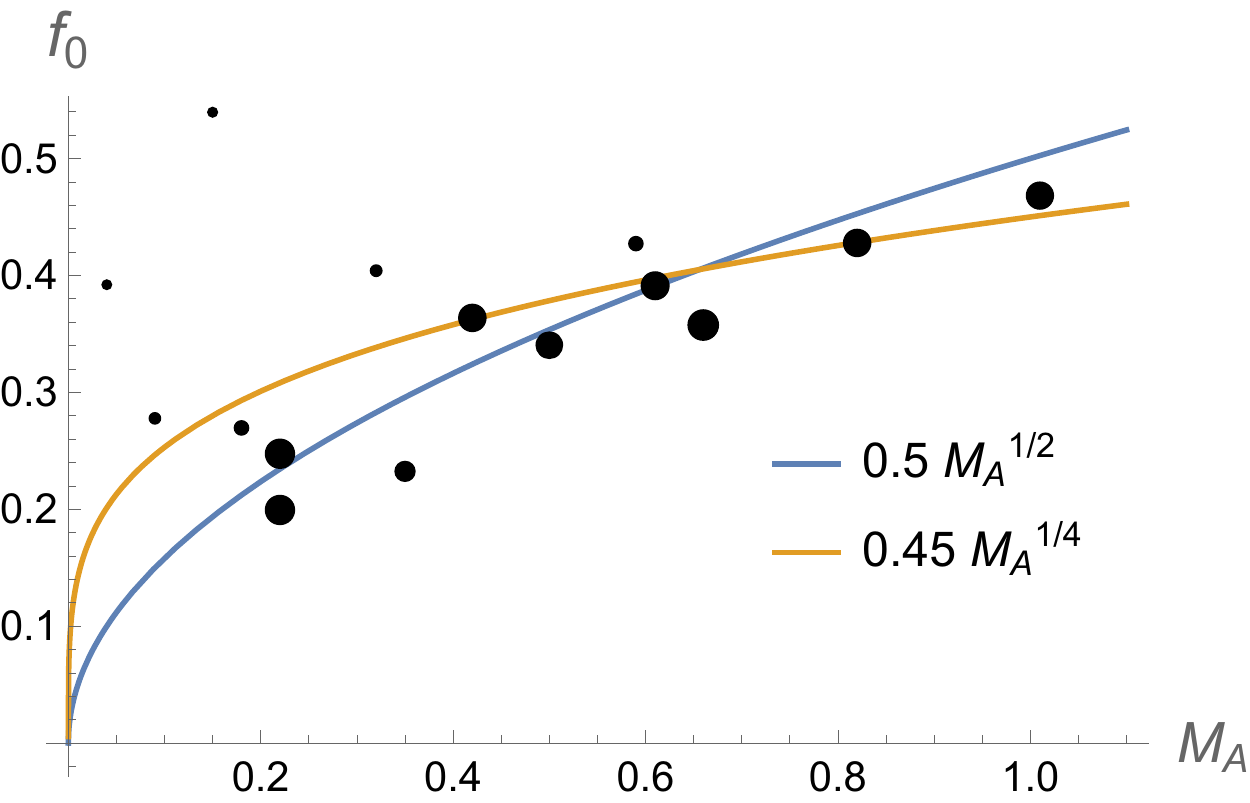}
    \caption{ $f_0$ in a series of low-$\beta$ simulations with $M_A$ ranging 
    from $0.04$ to $1.1$. The size of the data points reflects the sonic Mach number, $M_S$.
    The smallest point has $M_S =0.41$, while the largest one has $M_S=7.14$ 
    }
    \label{fig:f0_lowbeta_sims}
\end{figure}

In Fig.~\ref{fig:f0_lowbeta_sims} we show the numerically evaluated correction factor $f_0$
for a set of low-$\beta$ sub and trans-Alfv\'enic simulations for $\gamma=\pi/2$. The latter choice is motivated by the fact, that all the numerical studies of the DCF and alternative models (see e.g. \S 10) are done for this choice of angle. 

We observe that, while a significant scatter is present, the supersonic models
which have sonic Mach number $M_s > 1$ follow
\begin{equation}
f_0 \approx 0.5 M_A^{1/2}
\label{eq:f0_lowbeta_supersonic}
\end{equation}
dependence, while the outliers with significantly higher $f_0$ are
all the subsonic models with low $M_s < 1$. Namely, the two largest outliers have the lowest $M_S=0.4-0.5$ 
(models b11 and b21) in our list, while the models with the next lowest $M_S=0.92, 0.98$
(b12 and b22) are on the upper envelope of the quoted dependence.

If our reported dependence on the difference in $M_s$ is confirmed, this is a good news for practical magnetic field studies. There have been suggested a number of ways of obtaining the sonic Mach number $M_s$ observationally (e.g. Esquivel \& Lazarian 2005, Chepurnov et al 2009, Burkhart et al 2013, \citealt{GA} ).

\begin{figure*}[th]
    \centering
    \includegraphics[width=0.45\textwidth]{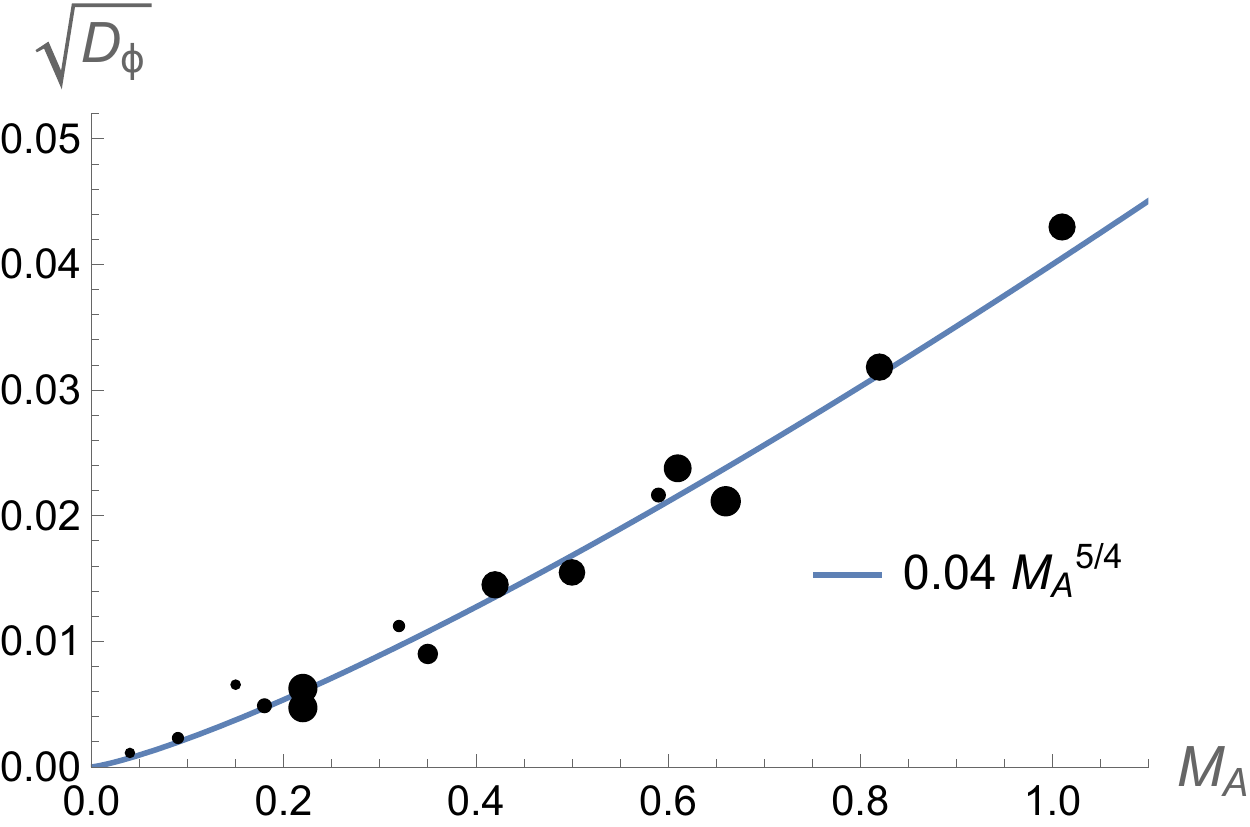}
    \includegraphics[width=0.45\textwidth]{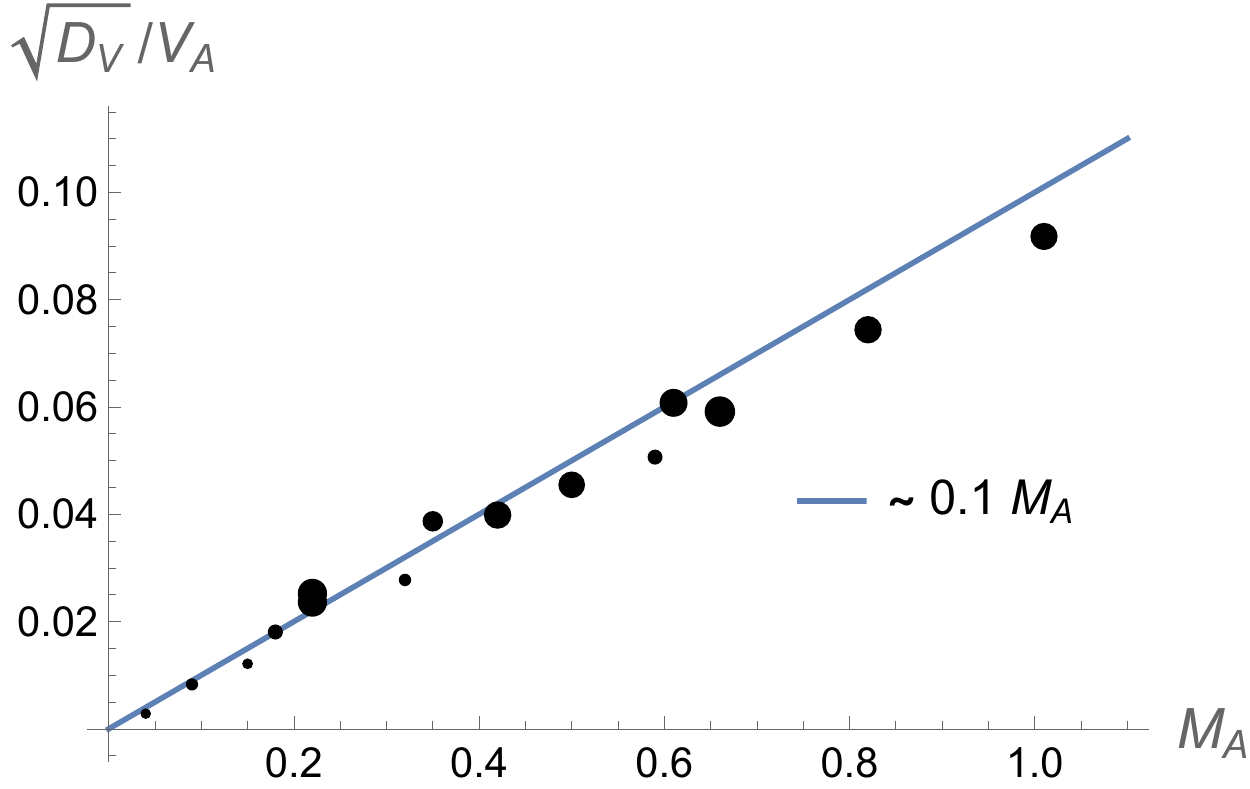}
    \caption{ Angle structure function $\sqrt{D^\phi}$ (left) and velocity structure function  $\sqrt{D_v}/V_A$ in the units of Alfv\'en velocity $V_A = \overline{B}/\sqrt{4 \pi \overline{\rho}}$ (right) evaluated at $R=L_{inj}/64$ lag in a series of low-$\beta$ simulations with $M_A$ ranging 
    from $0.04$ to $1.1$. The size of the data points reflects the sonic Mach number, $M_S$.
    The smallest point have $M_S =0.41$, while the largest one has $M_S=7.14$ 
    }
    \label{fig:fDvDphi_lowbeta}
\end{figure*}

To explore further the nature of the $M_A$ dependence for $f_0$, in Fig.~\ref{fig:fDvDphi_lowbeta} we plot
separately the (square root of the monopole of) the angle structure function $\sqrt{D^\phi}$ and the velocity
structure function $\sqrt{D_v}$, measured in the units of Alfv\'en velocity $V_A = \overline{B}/\sqrt{4 \pi \overline{\rho}}$.  The ratio of these two functions give $f_0$ in 
Fig.~\ref{fig:f0_lowbeta_sims}. 

The velocity structure function shows linear dependence on $M_A$, $\sqrt{D_v} \approx 0.1 M_A$ which just reflects overall proportionality of the level of fluctuations in Alfv\'en modes to $M_A$ as we discussed in \S 6 and Appendix \ref{app:mhdturb}.  The angle structure function, however, has a steeper
dependence,  that is fit quite well, at least for supersonic case low-$\beta$ turbulence, by  
$\sqrt{D^\phi} \approx 0.04 M_A ^{5/4}$. To understand what the extra $M_A^{1/4}$ dependence reflects and to compare these measurements with our theoretical picture in the previous section, we decompose the MHD turbulence into modes following the algorithm in \cite{CL03} and
look at the contributions of individual modes in the result.

\begin{figure*}[th]
    \centering
    \includegraphics[width=0.45\textwidth]{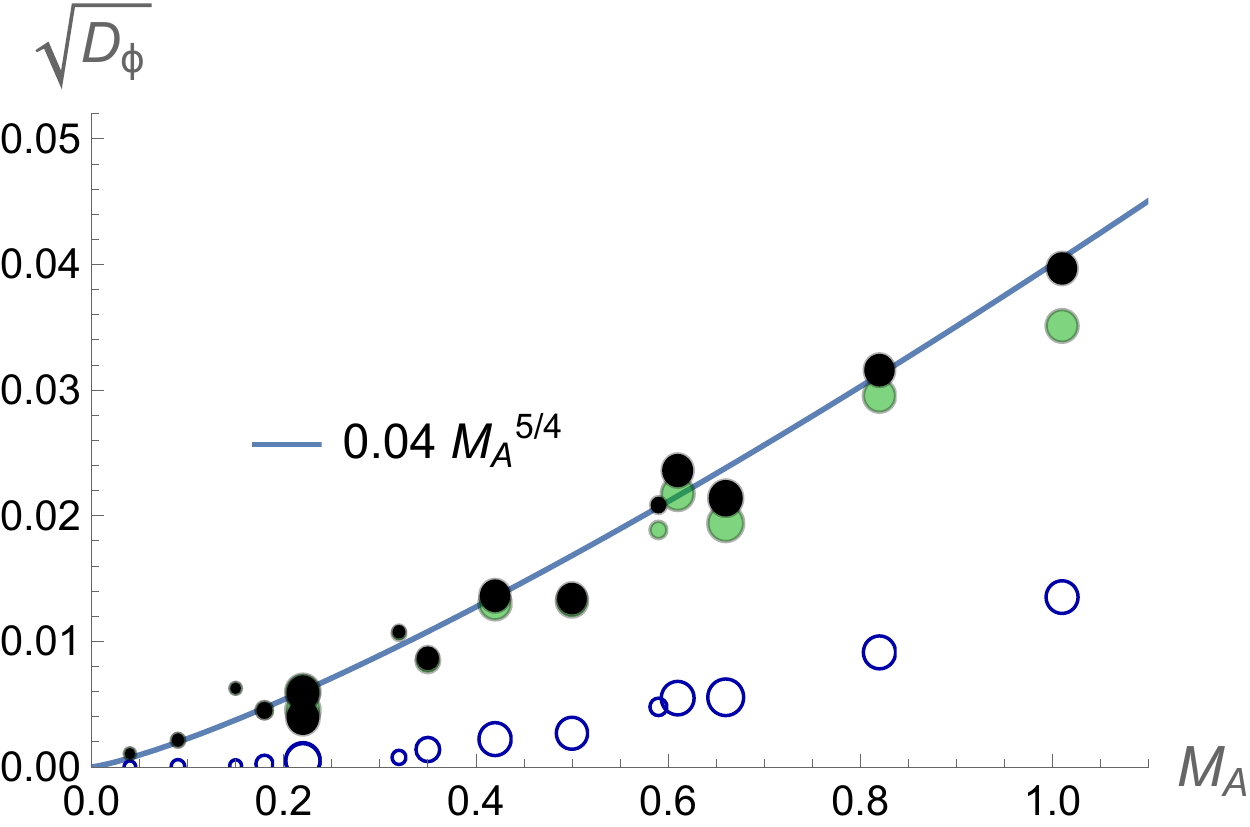}
    \includegraphics[width=0.45\textwidth]{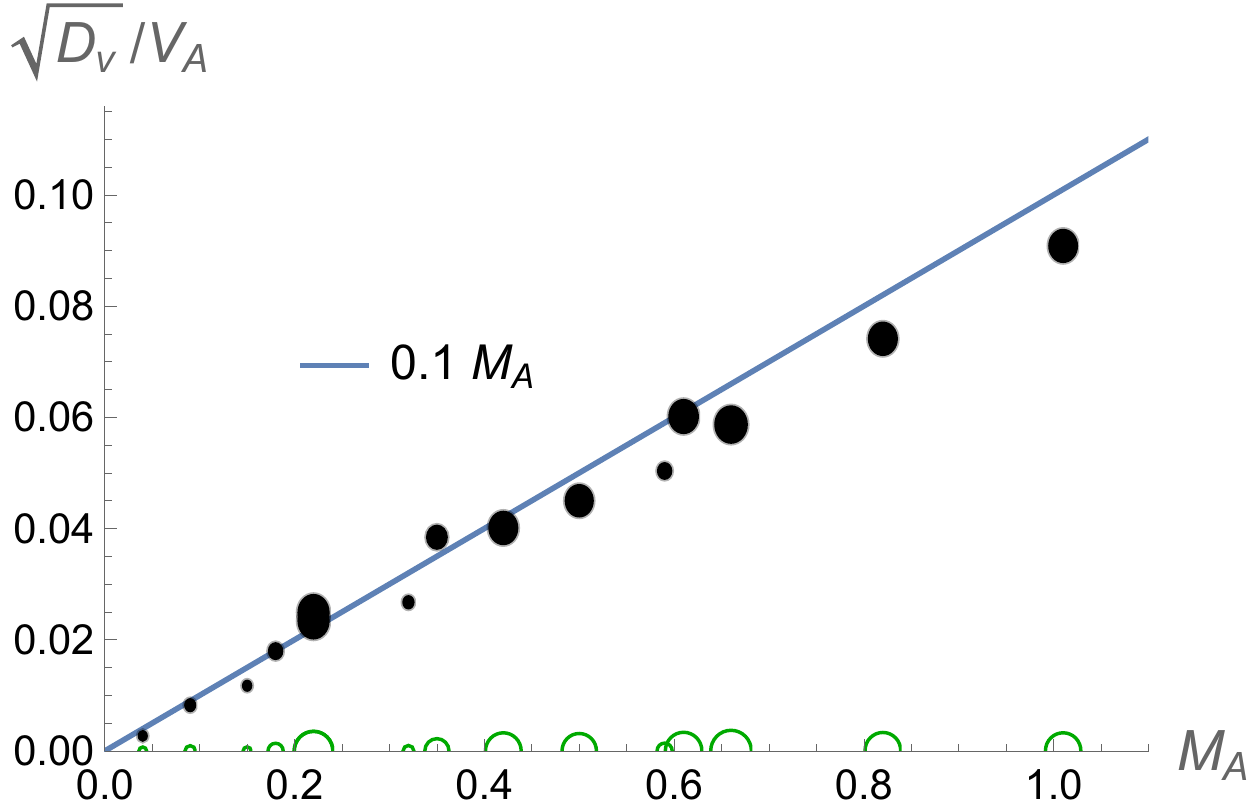}
    \caption{ Structure functions for $\gamma=\pi/2$. Left: Angle structure function $\sqrt{D^\phi}$  is dominated by fast modes (black dots) over Alfv\'en modes (blue circles). Right: velocity structure function  $\sqrt{D_v}/V_A$ in the units of Alfv\'en velocity $V_A = \overline{B}/\sqrt{4 \pi \overline{\rho}}$ is dominated by {\torefereetwo Alfv\'en} modes (black dots) versus fast modes (green circles)
    Structure functions are evaluated at $R=L_{inj}/64$ lag in a series of low-$\beta$ simulations with $M_A$ ranging 
    from $0.04$ to $1.1$. The size of the data points reflects the sonic Mach number, $M_S$.
    The smallest point has $M_S =0.41$, while the largest one has $M_S=7.14$ 
    }
    \label{fig:fDvDphi_modes_lowbeta}
\end{figure*}

In Fig.~\ref{fig:fDvDphi_modes_lowbeta} we see that at $\pi/2$ orientation, the contribution of the Alfv\'en modes
to the angle fluctuations have been suppressed  
while fast modes do not contribute to velocity
fluctuations, as expected in theoretical model (See \S 3, Tab.\ref{tab:proj}).  
We do not show a separate contribution of
slow modes, as on its own it is subdominant in amplitude at $\pi/2$ to both angle and velocity measurements (See \S 7).

Velocity structure function is completely determined by the Alfv\'en mode, there is virtually
no difference between right panels of Fig.~\ref{fig:fDvDphi_modes_lowbeta} and Fig.~\ref{fig:fDvDphi_lowbeta}.  For angle fluctuations fast modes is the dominant source.
Non-zero contribution of Alfv\'en modes that grows with a $M_A$ is the result of field 
wandering effect, and is absent if fluctuations of $\sin\gamma$ along LOS are not included in the projection. Thus, our theoretical model, resulting at $\gamma=\pi/2$ in Eq.~(\ref{eq:f0_lowbeta_sum2}), is on a firm basis.   

Eq.~(\ref{eq:f0_lowbeta_sum2})
predicts that any observed dependence of $f_0$ on $M_A$ is due to the $M_A$ dependence
of fast-to-Alfv\'en magnetic perturbation ratio $\langle \delta B_{\cal F}^2 \rangle/\langle \delta B_{\cal A}^2\rangle$. In our simulations (see Fig.~\ref{fig:fraction}), we found that
the relative energy in the fast modes in low $\beta$ sub-Alfv\'enic plasma roughly scales
as $0.3 M_A^{1/2}$. This dependence translates into 
$\frac{\langle \delta B_\mathcal{F}^2 \rangle }{\langle \delta B_\mathcal{A}^2 \rangle } \propto M_A^{1/2}$.  Thus the total amplitude of fluctuations from fast modes, which in low $\beta$ medium are compressions of magnetic field,  is
$\propto M_A \times M_A^{1/4}$, which exactly explains 
the best fit for angle structure function from the dominant fast modes in Fig.~\ref{fig:fDvDphi_modes_lowbeta}.  Note that Eq.~(\ref{eq:f0_lowbeta_sum2}) predicts
the magnitude $f_0(M_A=1) \approx 0.49$, in agreement with measurements.

We notice, however, that taking the ratio $\sqrt{D^\phi/D_v} V_A=f_0$ would result in an expected dependence $f_0 \sim 0.4 M_a^{1/4}$. At the same time, this ratio is rather noisy,
and we have found that the better fit to $f_0$ is given by Eq.~(\ref{eq:f0_lowbeta_supersonic}).  We feel that determining the exact scaling of $f_0 (M_A)$ for $\gamma=\pi/2$ may require more detailed numerical studies. Here we suggest that for low beta and $\gamma=\pi/2$ the dependence of $f_0$ is $\sim M_A^{\alpha}$, where $\alpha$ is between $1/2$ and $1/4$.

\subsubsection{$\pi/4$ case}

The contribution of Alfv\'en modes to projected magnetic angle fluctuations increases with $\gamma$ and therefore we consider a case of intermediate angles, i.e. $\gamma=\pi/4$. The intermediate angles are usually not discussed within the DCF study where the composition of the turbulence in terms of modes is ignored together with the effects of Alfv\'en and slow mode anisotropy.

For $\gamma=\pi/4$ we know both theoretically (see \S 3) and numerically (Fig.~\ref{fig:ST21}) that the strong projection suppression of Alfv\'en mode  is not present and as the result 
field wandering does not qualitatively change the results.
At $\gamma=\pi/4$  Eq.~(\ref{eq:f0_lowbeta_sum}) gives
\begin{equation}
    \label{eq:f0_lowbeta_sum_pi4}
    f_0^2 (M_A) \approx 
    \frac{ 
    M_{A}^2  + \frac{\kappa(\alpha)}{\widehat{\mathcal{A}}_0^{2D}}  \frac{\langle \delta B_\mathcal{F}^2 \rangle }{ \langle \delta B_\mathcal{A}^2 \rangle } }
    {1 + \frac{\langle v_\mathcal{S}^2 \rangle}{\langle v_\mathcal{A}^2 \rangle } }
\end{equation} 
where $\kappa(\alpha)=\frac{1+\sqrt{2}-2\alpha}{(1+\sqrt{2})(2-\alpha)}$ varies from
$\kappa(1)\approx 0.17$ to $\kappa(0)=0.5$ while $\widehat{\mathcal{A}}_0^{2D}(M_A) \approx 1$,
varying just by 15\%  from $M_A=0.1$ to $M_A=1$.

\begin{figure*}[th]
    \centering
    \includegraphics[width=0.45\textwidth]{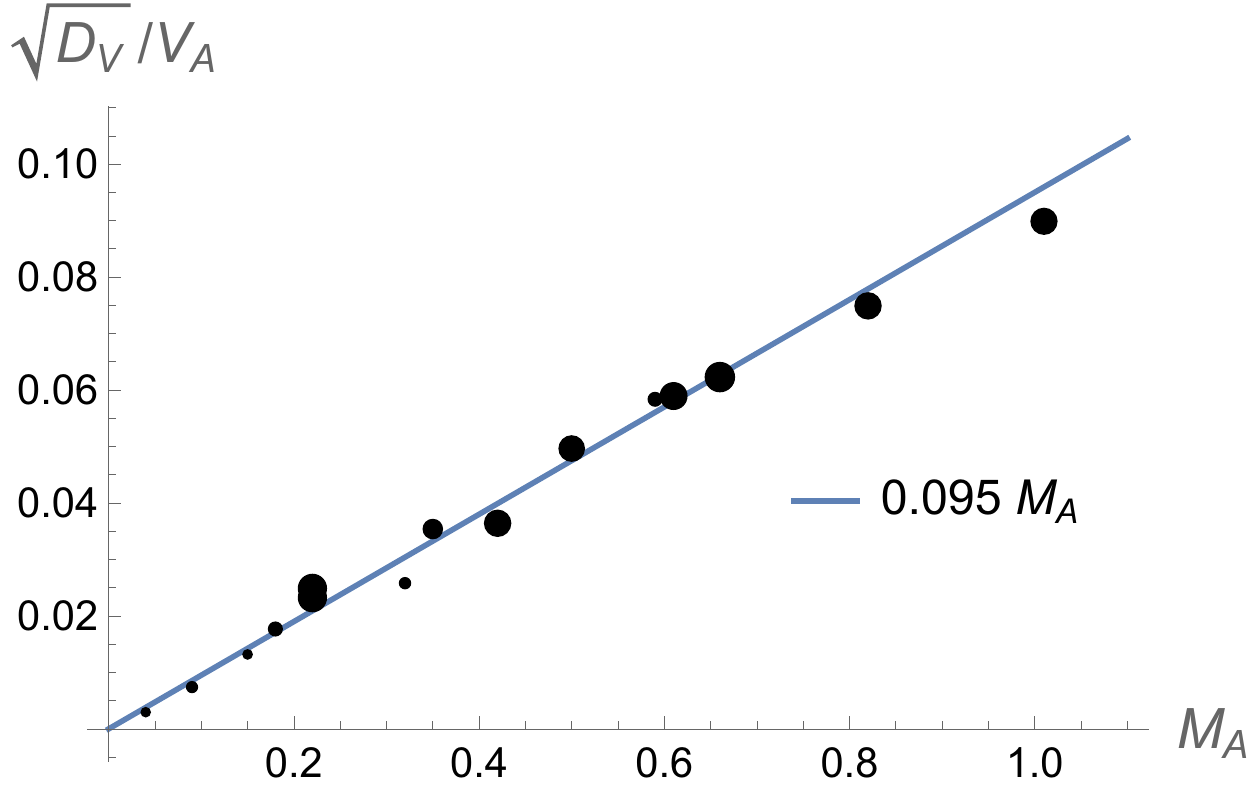}
    \includegraphics[width=0.45\textwidth]{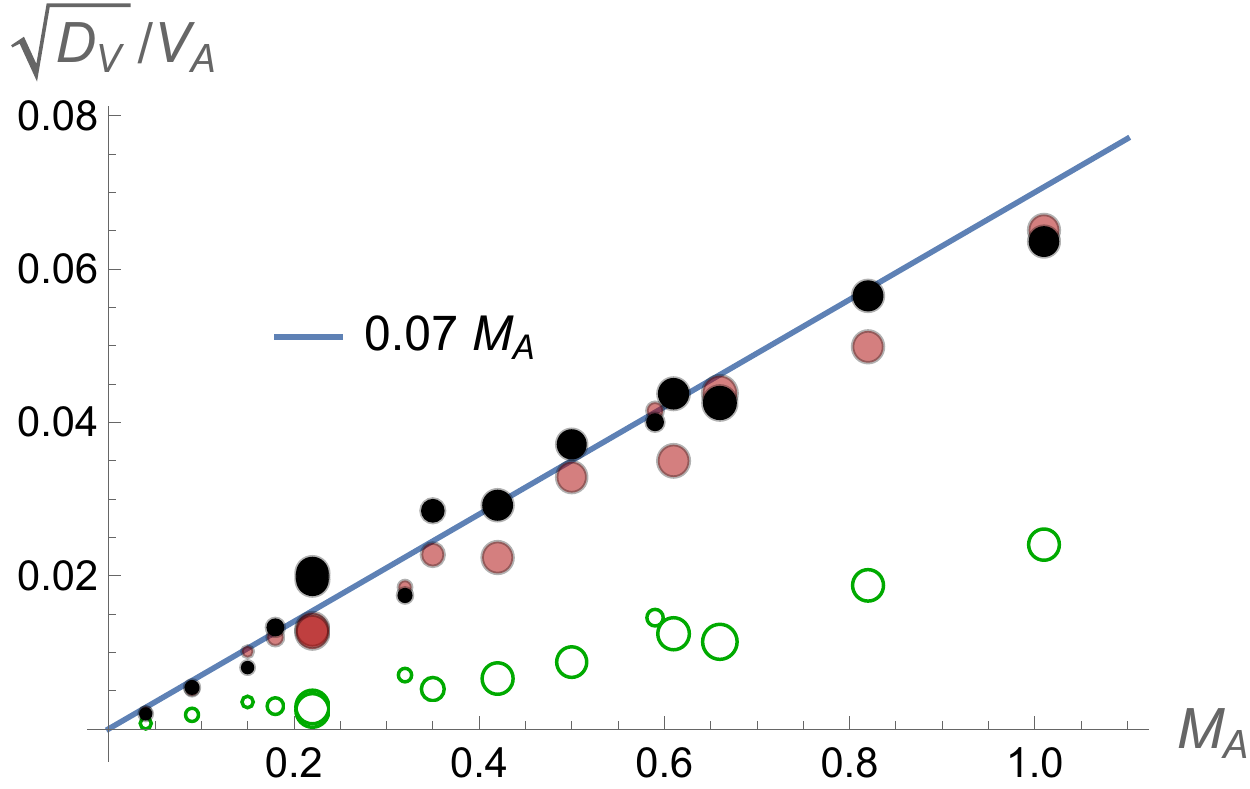}
    \caption{Right: velocity centroid structure function  $\sqrt{D_v}/V_A$ in the units of Alfv\'en velocity $V_A = \overline{B}/\sqrt{4 \pi \overline{\rho}}$  is determined (left)
    by almost equal contributions of {\torefereetwo Alfv\'en} (black) and "compressible" modes (red) when
    the mean magnetic field angle is $\pi/4$ to the LOS. Contribution of the fast modes (green circles) is subdominant part in the "compressible" mix.
    Structure functions are evaluated at $R=L_{inj}/64$ lag in a series of low-$\beta$ simulations with $M_A$ ranging 
    from $0.04$ to $1.1$. The size of the data points reflects the sonic Mach number, $M_S$.
    The smallest point have $M_S =0.41$, while the largest one has $M_S=7.14$ 
    }
    \label{fig:fDv_modes_lowbeta_pi4}
\end{figure*}

For velocity fluctuations, in Fig.~\ref{fig:fDv_modes_lowbeta_pi4} we observe, as expected, that $\sqrt{D_v}$ scales as $M_A$ and is determined almost in equal fraction by Alfv\'en and slow modes, as was predicted in Fig.~\ref{fig:AFcomp}, given that we have a 
similar energy content of two modes seen in Fig.~\ref{fig:fraction}. 

\begin{figure*}[th]
    \centering
    \includegraphics[width=0.45\textwidth]{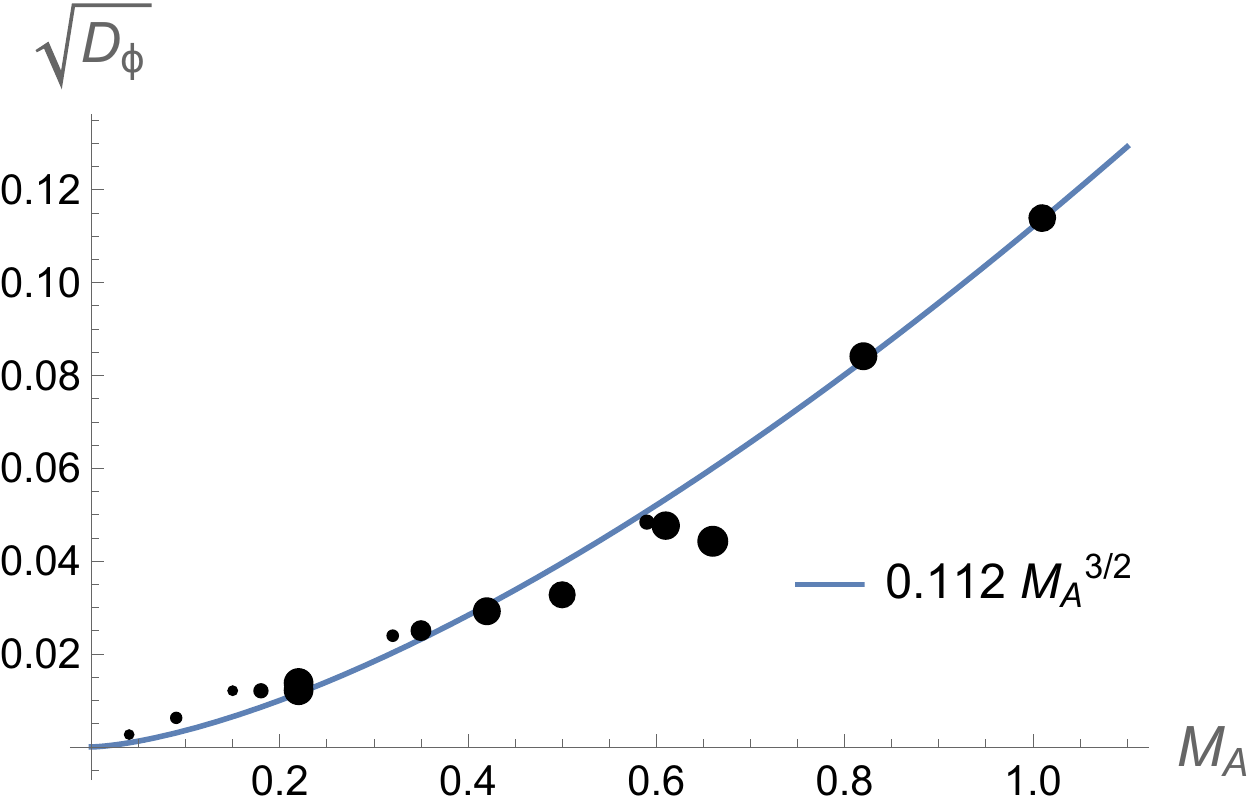}
    \includegraphics[width=0.45\textwidth]{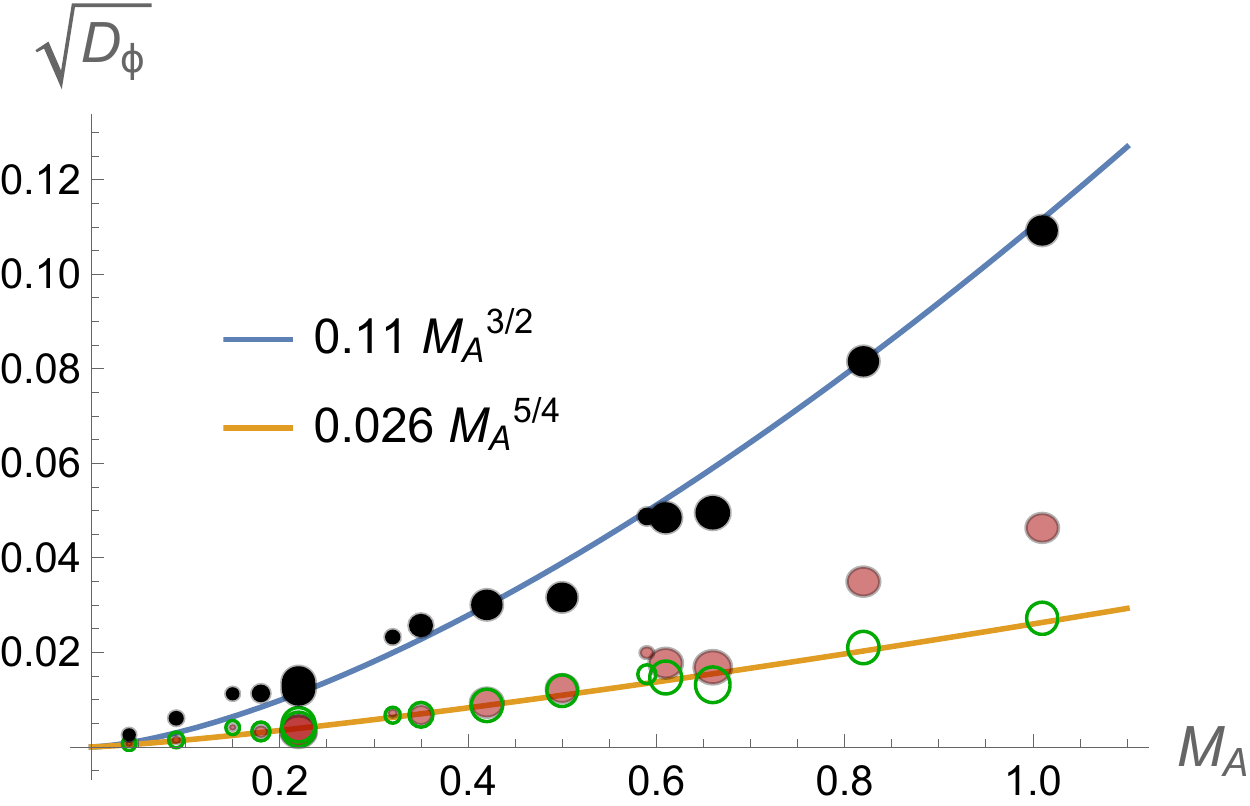}
    \caption{Right: Angle structure function  $\sqrt{D^\phi}$. Left: individual contributions
    from {\torefereetwo Alfv\'en} (black) and subdominant "compressible" modes (red) among which the fast mode contribution (green circle) still follows $M_A^{5/4}$ scaling. 
    The mean magnetic field angle is $\pi/4$ to the LOS.
    Structure functions are evaluated at $R=L_{inj}/64$ lag in a series of low-$\beta$ simulations with $M_A$ ranging 
    from $0.04$ to $1.1$. The size of the data points reflects the sonic Mach number, $M_S$.
    The smallest point have $M_S =0.41$, while the largest one has $M_S=7.14$ 
    }
    \label{fig:fDphi_modes_lowbeta_pi4}
\end{figure*}

On the other hand, for polarization angles, we see from Fig.~\ref{fig:fDphi_modes_lowbeta_pi4} that the Alfv\'en mode contribution is larger than that of the fast modes. It is expected
at large $M_A$, as Eq.~(\ref{eq:f0_lowbeta_sum_pi4}) predicts. However, according to Eq.~(\ref{eq:f0_lowbeta_sum_pi4}), 
fast modes ere expected to take over at small $M_A^2 \lesssim  \kappa \langle \delta B_\mathcal{F}^2 \rangle /\langle \delta B_\mathcal{A}^2 \rangle$ which, if we take $\kappa\approx 0.2$ and the fast mode fraction to be $0.3 M_A^{1/2}$, resolves to $ M_A \lesssim 0.15$. This is not observed
in $D^\phi$ in Fig.~\ref{fig:fDphi_modes_lowbeta_pi4} which shows Alfv\'en modes prevailing
over the fast ones even at smaller $M_A$.  This originates in Alfv\'en contribution falling 
as, approximately, $\sqrt{D^\phi} \sim M_A^{3/2}$, which is slower than $\sim M_A^2$ scaling expected in the theory ( recall that $\sqrt{D^\phi} \sim M_A f_0$).

\begin{figure}[th]
    \centering
    \includegraphics[width=0.45\textwidth]{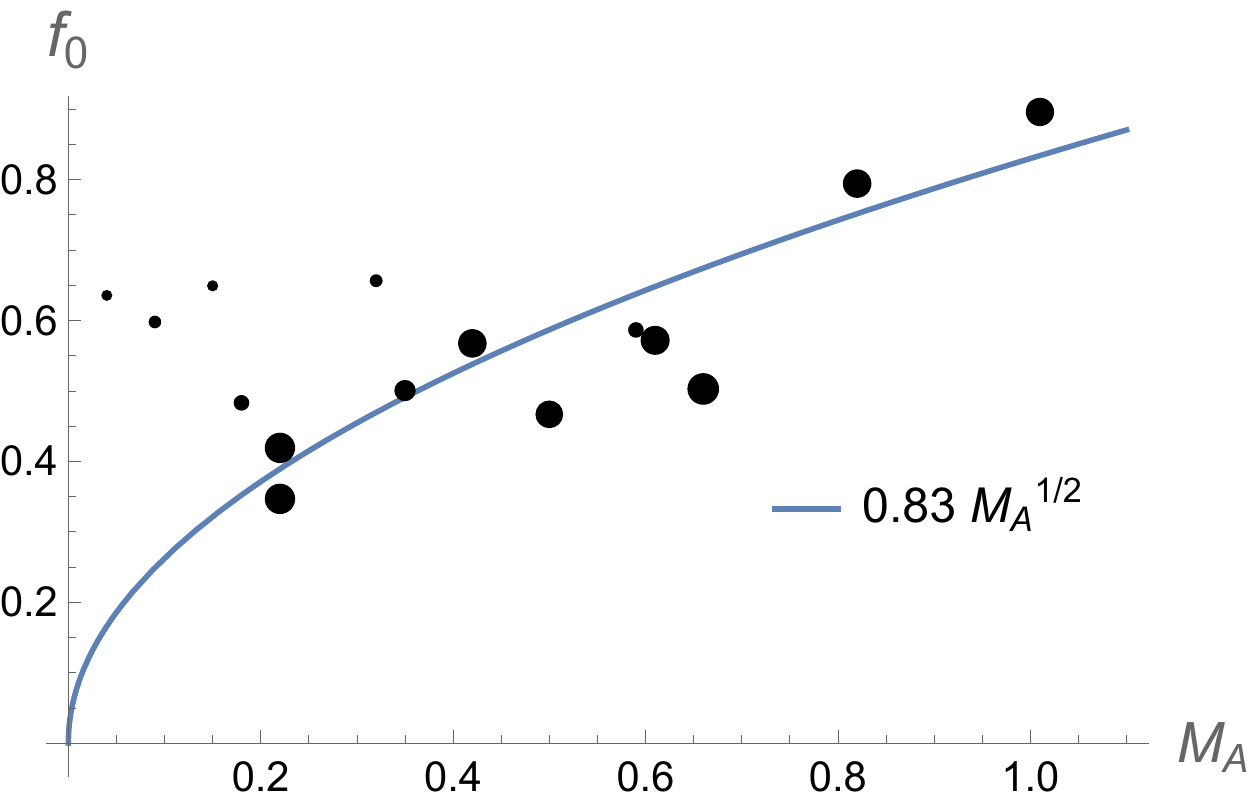}
    \caption{$f_0$ factor in a series of low-$\beta$ simulations with $M_A$ ranging 
    from $0.04$ to $1.1$ when mean magnetic field angle with LOS is $\pi/4$.
    The size of the data points reflects the sonic Mach number, $M_S$.
    The smallest point have $M_S =0.41$, while the largest one has $M_S=7.14$ 
    }
    \label{fig:f0_total_lowbeta_pi4}
\end{figure}

Correspondingly, instead of $f_0\sim M_A$ expected for the Alfv\'enic component, in Fig. \ref{fig:f0_total_lowbeta_pi4} we observe $f_0\sim M_A^{1/2}$ for supersonic simulations and $f_0\sim \text{const}$ for subsonic ones.  

We believe that here we may face a restriction related to the resolution of our numerical simulations. Let us remind the reader that the origin of $f_0 \propto M_A$ result for
Alfv\'enic mode comes from the anisotropy of the power spectrum, namely Eq.~(\ref{eq:AlfvenEnE0expanded}), that reflected  the properties of the strong turbulence,
see Eq.~(\ref{eq:hatEmu_averaged}).
As we discuss in Appendix \ref{app:mhdturb}, in sub-Alfv\'enic regime the Alfv\'en modes exhibit two different regimes of turbulence, the weak turbulence  from the injection scale down to the scale $LM_A^2$ and the strong one for scales less then the transition scale $LM_A^2$. All the calculations of the DMA are performed for the strong turbulence regime. However, for sufficiently small $M_A$ this regime cannot develop if $L M_A^2<d_{diss}$, where $d_{diss}$ is the scale of numerical dissipation. Therefore for small $M_A$ we deal with the pure weak Alfv\'enic turbulence which has different anisotropies from our model of strong turbulence that we employ in DMA. 
For our current testing, we can only perform analysis of the strong Alfv\'enic regime up to $M_A\ge 0.2$.
\footnote{In our highest resolution $L=1200$ simulations with $L_{inj}=L/2$, assuming a rough estimate for the dissipation scale $L_{disp}=12$ pixels, the minimal $M_A$ at which we can have a realistic strong regime is $M_A = \sqrt{L/(2 l_{diss})} \approx 0.14 $. However the lowest 
$M_A$ we have at this resolutions is $0.22$. Our lowest $M_A$ low-$\beta$ simulations
are at $L=480$, but at this resolution the strong turbulence regime require $M_A > 0.22$ } Future studies with higher numerical resolution should test our prediction.
For fast modes,  there exist only one regime of turbulence and therefore where fast modes are dominant, we see the scaling of $D_\phi$ that corresponds to the change of the fast mode energy with $M_A$, unaffected by resolution. 

\subsection{Dispersion and structure functions}
\label{sssec:phi}

At large lags the structure functions saturate at the twice the dispersion level. In this limit we transfer to the case similar to that the DCF deals with. Our analytical study does not cover this limiting case, as the DMA is limited to sufficiently small scales for which the MHD turbulence is in the strong regime. To have a more complete picture of the evolution of $\phi$ fluctuations below we explore this case numerically.

\begin{figure*}[th]
\centering
\includegraphics[width=0.45\textwidth]{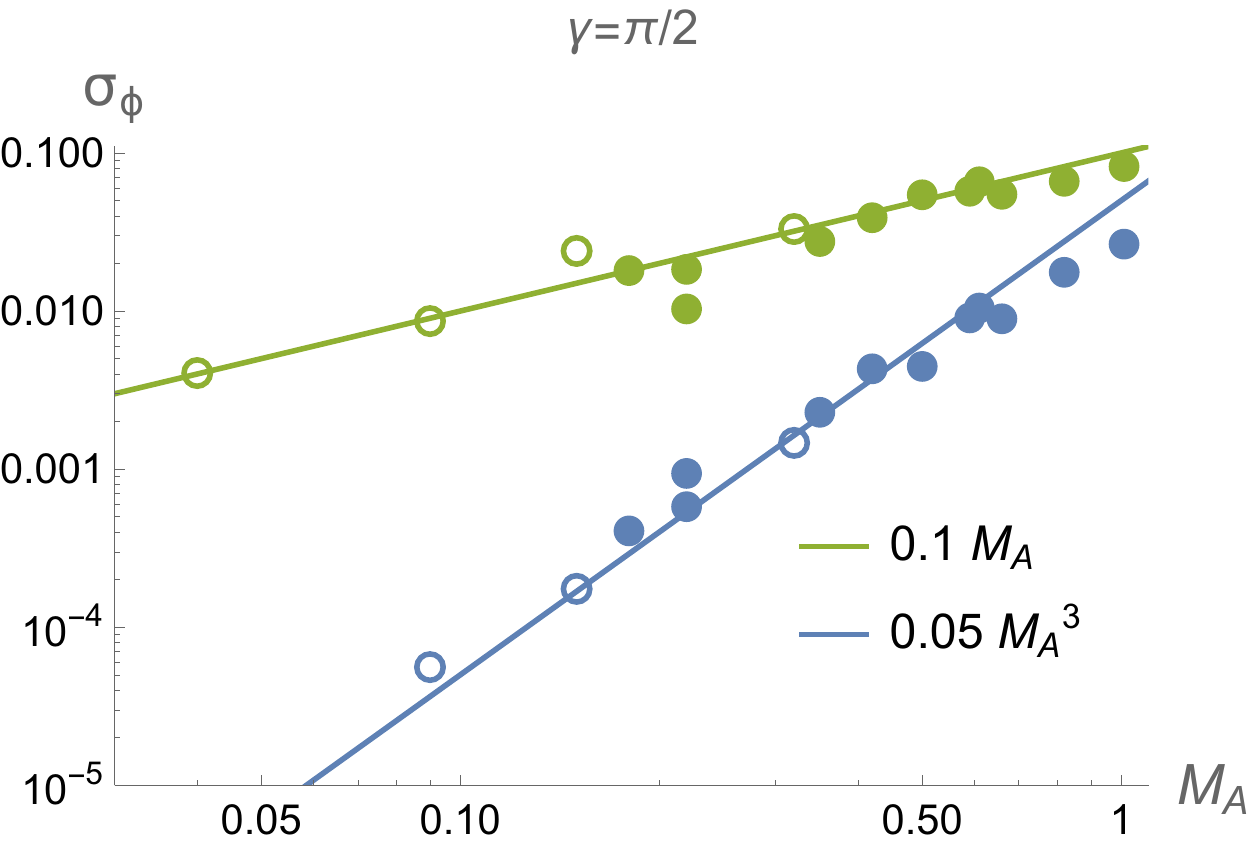}\hfill
\includegraphics[width=0.45\textwidth]{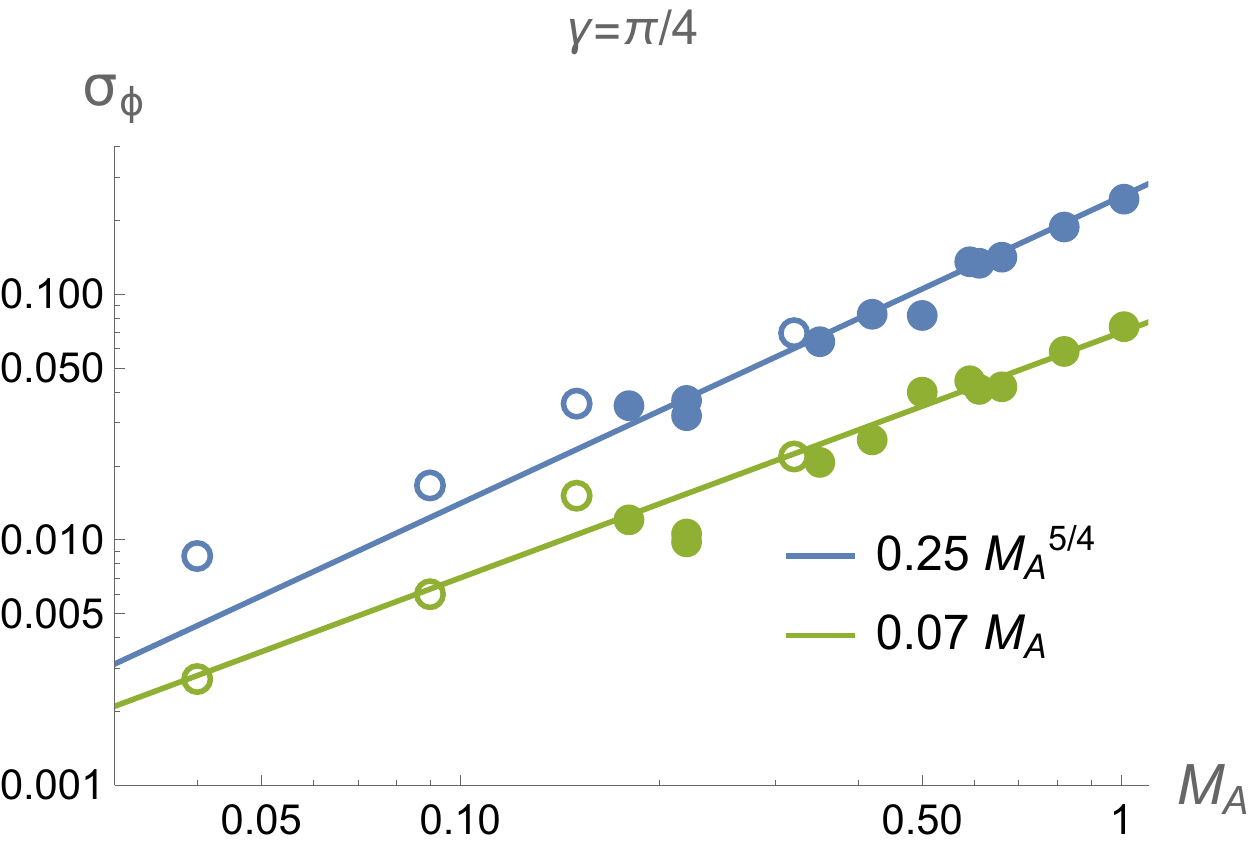}
\caption{\label{fig:ST21} Left: circular dispersions of the polarization angle $\sigma_\phi$ as functions of $M_A$ for all three fundamental modes modes in our numerical simulations {\it that has $\beta<1$} (See Table \ref{tab:sim}) at $\gamma=90^o$. Right: $\gamma=45^o$.
}
\end{figure*}

Fig.~\ref{fig:ST21} illustrates the dispersion of angles $\sigma_\phi$ arising from  different MHD turbulence modes in low $\beta$ turbulence. For simplicity we did not include the contributions of slow modes since in theory they have infinitesimal contribution to $\sigma_\phi$ when $\beta <1$ (See \S \ref{subsec:lowbeta}) As expected, for $\gamma=90$ (see left panel) the fast mode dominates over the Alfv\'en modes . In this scenario, the total $\sigma_\phi$ scales approximately as $\approx 0.1\; M_A$.  It is important to note that the magnitude of polarization angle fluctuations projected onto the sky is significantly less than $M_A$, i.e. $\sigma_\phi \ll M_A$. This adds to the accuracy of approximations that were made in our theoretical approach. The contribution of Alfv\'en  modes to $\sigma_\phi$ fits the $M_A^3$ scaling, becoming more shallow as $M_A \to 1$, in accordance with our description of ``field wandering'' (see \S~\ref{subsec:alfven_wandering}-\ref{subsec:alfven_f0}).

It would be interesting to compare the dependencies of $\sigma_\phi$ and $\sqrt{D_\phi}$ (left of Fig.~\ref{fig:fDvDphi_modes_lowbeta}) to $M_A$ as the former is simply the "large scale" result of the latter. From the left of Fig.~\ref{fig:fDvDphi_modes_lowbeta} it is indicative that the change of the fast modes energy with $M_A$ may be different at small and large scales, although it is difficult with the existing noisy data to definitively distinguish the change of power from $M_A^{5/4}$ to $M_A^1$.

\begin{figure}[th]
\includegraphics[height=0.35\textwidth,width=0.45\textwidth]{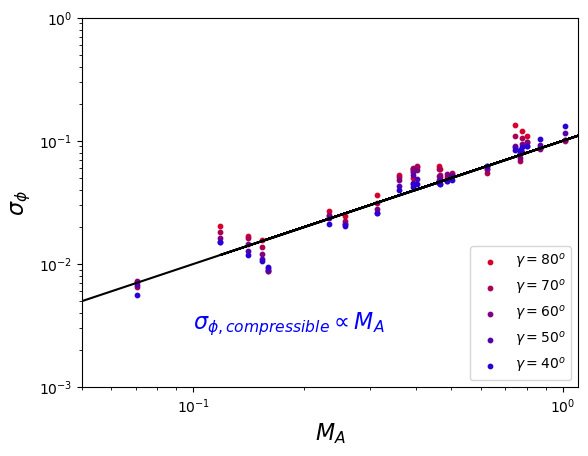}
\caption{\label{fig:comp} A figure showing how the variation of the inclination angle $\gamma$ will change the dispersion on the circular dispersion $\sigma_\phi$ in the selected "huge" simulations from Table \ref{tab:sim}. }
\end{figure}

The contribution of Alfv\'en modes becomes much more important when $\gamma$ switches from $\pi/2$ to $\pi/4$ both for $\sigma_\phi$ and $\sqrt{D_\phi}$, which is expected as in \S \ref{subsec:lowbeta}. For the case of $\sigma_\phi$ to $M_A$, which we can see from the right panel of Fig.\ref{fig:ST21}, we see that Alfv\'en mode takes over fast modes in terms of the amplitude. Furthermore, we notice that the dispersion $\sigma_\phi$ arising from Alfv\'enic modes scaling is best described as $M_A^{5/4}$, though the slope $M_A^{3/2}$ as was $\sqrt{D_\phi}$ is also possible (see right panel of Fig.~\ref{fig:fDphi_modes_lowbeta_pi4} ). This is again differs from $M_A^2$, which is the scaling for
$\sqrt{D_\phi}$ expected theoretically for Alfv\'en modes (see \S 3).  The difference between the two scaling laws is significant and we interpret the difference as resulting from the transfer of the Alfv\'enic turbulence to the weak regime at large scales. We do not have a theoretical prediction for this regime, but $\sigma_\phi\sim M_A$ seems plausible. Furthermore, we note that the fast modes contribution to $\sigma_\phi$ remains $\propto M_A$, but at $\pi/4$ its relative role decreases compared to the case of $\pi/2$.

In terms of practical separation into modes, a better accuracy can be achieved for the separation into Alfv\'en and the mixture of slow and fast modes. The latter is frequently called "compressible", although this does not really reflect the nature of the modes, which have both compressible and incompressible parts. Fig.~\ref{fig:comp} shows that the contribution of these "compressible" modes to $\sigma_\phi$ $\propto M_A$. Notice that the scaling of $\sigma_\phi$ to $M_A$ does not change as $\gamma$ changes from $80^o$ to $40^o$, however there is a slight amplitude change as $\gamma$ changes. In the case of low $\beta$, this constancy of the slope is related to the isotropy of magnetic perturbations induced by fast modes. Therefore, at the range of fast mode domination, the $\sigma_\phi$ part arising from "compressible" modes is independent of $\gamma$. For this case if the velocity field has typical scaling $\sigma_v, \sqrt{D_v} \propto M_A$, then it is very probable to arrive to $f_0 \sim \text{const}$, recovering the classical DCF expression given by Eq. (\ref{eq:BBsin}). Notice also that in the case of Alfv\'enic modes dominance to $\sigma_\phi$ it is {\torefereetwo troublesome} to ignore the dependence of $f_{DCF}$ on $\gamma$. Our discussion above shows that the error can be significant for sufficiently small $\gamma$.

\section{PRACTICAL APPLICATION OF DMA} 
\label{sec:practical}

In the previous sections we have studied a variety of cases involving turbulence with different admixtures of Alfv\'en, slow and fast modes for both low and high $\beta$ medium. This provides a valuable toolbox for obtaining magnetic field strength in a variety of astrophysical situations. 

\subsection{How Do We Apply DMA to Observations?}

From \S \ref{sssec:phi} we can understand how $f_0$ are related to the statistics of velocity and polarization angle. 

We shall discuss in depth (\S \ref{sec:practical}) on how to utilize the extra dependence of $M_A$ will alter the formula of magnetic field estimation that we used before (Eq.~\ref{eq:DMA_eq}). However, before we proceed with the exact magnetic field estimation, let us discuss what we expect for the statistics of $f_0$.

Due to the complex dependencies of $\hat{D}$ and $D_\phi$, the statistics of $f_0$ obviously depends on the relative contribution of modes, the Alfv\'enic Mach number $M_A$, the compressibility $\beta$ and the line of sight angle $\gamma$. From what we discussed in \S \ref{sssec:phi}   the contributions of Alfv\'en mode does not act the same in velocity and polarization angle as we discussed in Tab.\ref{tab:proj}. In particular, for the low $\beta$ case, the relative contribution of Alfv\'en mode  increases as the angle $\gamma$ decreases from $\pi/2$ to zero. Therefore, initially at $\gamma=\pi/2$ we expect the fluctuations of $\phi$ to be dominated by fast modes. The transition to Alfv\'enic-dominated variations of $\phi$ happens at some intermediate angle $\gamma$, the value of which depends on the ratio of the energy of modes that is affected by the properties of driving. Therefore, for strongly compressible driving the transitional angle $\gamma$ is expected to be smaller than for the incompressible driving that we employ in this paper. In addition, for very small $M_A$ fast modes can dominate for a wider range of angles due to the dependence of the contribution the Alfv\'en modes to $\phi$ on $M_A$ (see Eq. \ref{eq:f0_alfven_firststep}). 

In the situations that fast modes dominate both variations of velocity and $\phi$, no dependence of $f_0$ on $M_A$ is expected. However, what we observed in Fig.~\ref{fig:ST21} that while for $\gamma=\pi/2$ fast modes are dominant for the $\phi$ variations, the variations of velocity can be dominated by Alfv\'enic component. Thus we get the observed dependence of $f_0$ on $M_A$.

The main takeaway of the current work is that the DCF scaling factor $f_0$ is a function of $M_A, \gamma$ and the relative composition of modes. The non-trivial dependence of $f_0$ and also its lack of systematic testing over the decades for all three items contribute to the reasons on why not until our work do we observe that the $f_0$ factor is not a constant. More importantly, since the contributions of different mode plays dramatically different role in terms of magnetic field estimations, whether a particular mode is important depends on the plasma compressibility $\beta$. For instance, in the case of molecular clouds with $\beta \ll 1$, the scaling factor $f_0$ depends on the mode composition in the form of $\langle \delta v_{\cal A}^2/\delta v_{\cal S}^2\rangle$ and $\langle \delta B_{\cal F}^2/\delta B_{\cal A}^2\rangle$. This discovery is non-trivial and has not been discussed before in literature.

Notice that the mode ratio is also a function of the line of sight angle $\gamma$. As we explained in \S \ref{sec:pure_alfven} the structure functions of velocities $D_v(R)$ are expected to scale with $M_A^2$ for all MHD modes, irrespectively of their anisotropy. At the same time, in \S \ref{sec:pure_alfven}. we showed that the turbulence anisotropy significantly modifies the behavior of the angle structure functions $D_\phi (R)$ compared to the naive expectations employed in DCF. The effect is most prominent for Alfv\'en modes and it is also present for slow modes. For equal energy in Alfv\'en and slow modes, the fluctuations of angles induced by slow modes are important in high $\beta$ regime and subdominant in low $\beta$ regime.

To utilize our results from \S 6-8, we start with Eq. (\ref{eq:Dc_over_Dtheta_rho_alt}):
\begin{equation}
   {B_\perp^2} \approx 4 \pi \bar{\rho} f_0^2 \frac{D_v (R)}{D_\phi (R)},
   \label{eq:pedest}
\end{equation}
where the structure functions of velocities $D_v(R)$ and angle $D_\phi (R)$ should be calculated at sufficiently small scale $R<L_{inj}M_A^2$ in order to insure that the MHD turbulence is in the strong regime (see Eq.~\ref{trans}). Here the factor $f_0 = f_0(M_s,M_A,\gamma, \text{mode fraction})$ should be computed according to the analysis that we provided in \S\ref{sec:pure_alfven}, \ref{subsec:lowbeta}. To proceed, we recognize that $f_0$ is generally in the form of $f_0 \propto C M_A^n$ for some $C=C(\gamma)$ and some constant $n$.  The idea of reformulation is to express $M_A$ back via the mean magnetic field
\begin{equation}
M_A \equiv \frac{V_L}{V_A} = V_L \frac{\sqrt{4 \pi \rho}}{\overline{B}}
~,
\label{eq:MaviaBVL}
\end{equation}
substitute it into $f_0$ dependence and resolve now implicit expressions for $\overline{B}$ via  the ratio of structure functions wrt $\overline{B}$.

In this scenario, if $f_0 \propto C M_{A\perp}^n$, $M_{A,\perp} = M_A/\sin\gamma$
 \footnote{Here we are assuming that the $\gamma$ value is not too small so that our expressions in \S\ref{sec:pure_alfven}, \ref{subsec:lowbeta} do not fall into a regime that $M_{A,\perp}>1$ despite $M_{A,tot}<1$.},
one can write the square of the {\it perpendicular} magnetic field strength as 
\begin{equation}
    B_{\perp}^2 \approx \left(C^{\frac{2}{n+1}} 4\pi\right) \rho V_L^2 \left(\frac{\widehat{D_v}}{D_\phi} \right)^{\frac{1}{n+1}}
    \label{eq:DMA_eq}
\end{equation}
where readers should be careful that $V_L$ is the injection velocity and  $\widehat{D_v} = D_v/V_L^2$.

It follows from Eq. (\ref{eq:DMA_eq}) that our empirical study in \S \ref{sec:numerical} suggests that for low $\beta$ medium $ f_0\approx 0.8 M_A^{1/2}$ for supersonic turbulence and $\gamma=\pi/2$ and therefore 
\begin{equation} 
B_{\bot}^2\approx 0.4 \rho V_L^2 \left(\frac{\widehat{D_v}}{D_\phi} \right)^{2/3}
\label{eq:approx_num}
\end{equation}

Naturally, the dependence for $\gamma=\pi/2$ does not persist for other $\gamma$ as the contribution from different modes to $\phi$ and velocity fluctuations changes with $\gamma$. One may argue that the contribution of the modes may get to the universal equipartition between them for sufficiently extended cascade. However, this has not been proven and, even if true, it is not clear that for a given observation this is satisfied. At least near the injection scale, the energy in the modes depends on the properties of turbulence driving. For our choice of incompressible driving, we may see that Eq. (\ref{eq:approx_num}) is applicable also to $\gamma=\pi/4$, provided that the 
corresponding prefactor changes from $0.4$ to $ 0.8$. 

\begin{figure*}[th]
\label{fig:merged_comparison}
\centering
\includegraphics[width=0.45\textwidth]{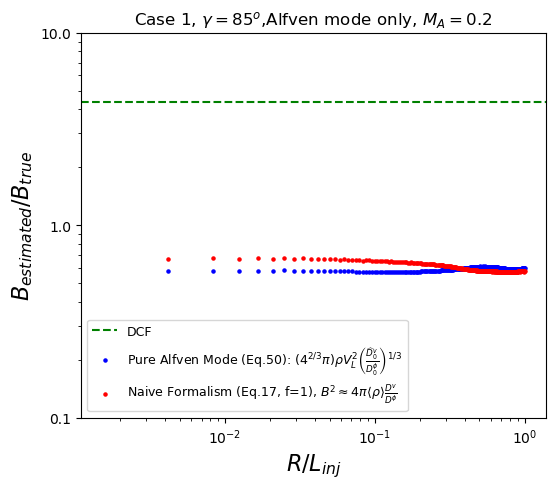}
\includegraphics[width=0.45\textwidth]{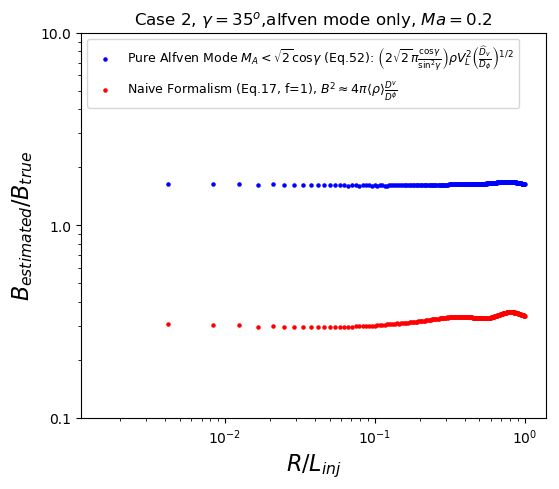}
\includegraphics[width=0.45\textwidth]{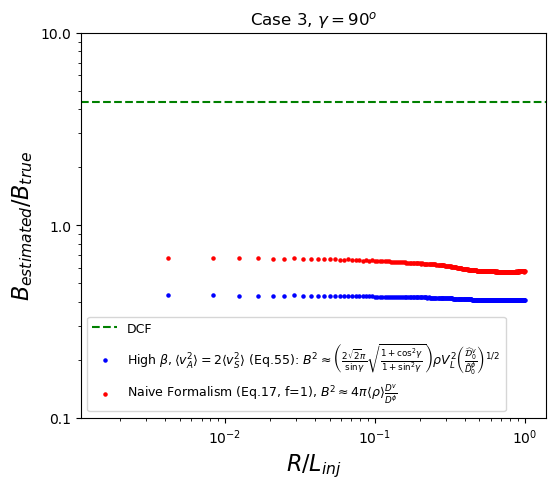}
\includegraphics[width=0.45\textwidth]{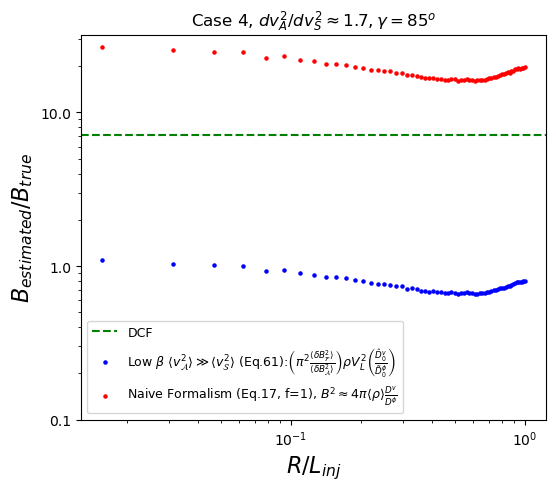}
\includegraphics[width=0.45\textwidth]{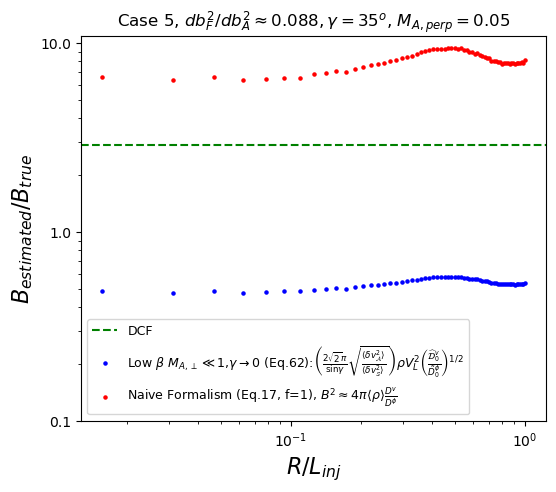}
\includegraphics[width=0.45\textwidth]{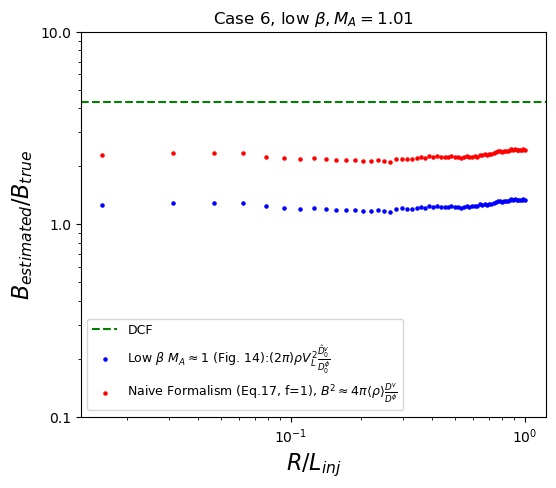}
\caption{The ratio between the estimated magnetic field over the true magnetic field $B_{estimated}/B_{true}$ for different cases list in Table. \ref{tab:result_summary} . The optimal method would have this ratio $=1$ The figures right now are in ascending order as in Table \ref{tab:result_summary}. From top left: case 1, top right: case 2, middle left: case 3, middle right: case 4, lower left: case 5, lower right: case 8.  Here we take $f_{DCF}=1/2$. For Case 2, the ratio between the DCF result to the true B-field value went over 10.}
\end{figure*}

In Table.~\ref{tab:result_summary} we collect the limiting cases that we discussed in \S \ref{sec:pure_alfven} and \S\ref{subsec:lowbeta}, and in Fig.~\ref{fig:merged_comparison} we provide the numerical comparison on our formulae in Table \ref{tab:result_summary} as compared to both the "naive" formulation assuming $f_0=\text{const}$  in Eq.~\ref{mean_B} and also the traditional DCF calculation. For reader's reference, if the curve is closer to $1$, that means the estimation of magnetic field from the specific formula is closer to the actual magnetic field strength .Notice that if the DCF result is not plotted, that means it is significantly deviated from the ranges that we are plotting (i.e. only $0.1<B_{DMA}/B_{true}<10$) is plotted. We can see that our complex formalism (blue curves on each panel in Fig.~\ref{fig:merged_comparison}) performs significantly better than both the "naive" formalism and also the DCF method.

\begin{deluxetable*}{c c c c c c }
\tablecaption{Selected DMA theoretical asymptotic predictions and relevant regimes \label{tab:result_summary} using Eq.~(\ref{eq:DMA_eq}).  }
\tablehead{
Case
& Mode composition  
& Conditions 
& $f_0$ dependence 
& $f_0$ equation 
& Dependence of $\langle {B^2_{\perp}}\rangle$ on $\widehat{D}^v_0,D^\phi_0$
}
% $\widehat{\delta v} = \delta v/V_L$
\startdata \hline \\
% pure alfven mode , gamma = pi/2
(1) & Pure Alfv\'en Mode  & $M_A \ll 1, \gamma \approx \pi/2$ & $\frac{1}{2}M_A^2$ & Eq.\eqref{eq:f0Alfvenpi2} & $2^{-2/3} \left( 4 \pi \rho V_L^2 \right) \left(\frac{\widehat{D}_0^v}{D_0^{\phi}}\right)^{1/3}$\\
% pure alf Ma <sqrt(2)\cos\gamma
(2) & Pure Alfv\'en Mode & $M_{A\perp} \ll \mathrm{min}\left(\sqrt{2} \frac{\cos\gamma}{\sin\gamma},1\right)$ & $\frac{1}{\sqrt{2}}
\frac{\cos\gamma}{\sin^2\gamma} M_{A}$ & Eq.\eqref{eq:alf_wand} & $ \frac{1}{\sqrt{2}}\; \frac{\cos\gamma}{\sin\gamma}
\left( 4 \pi  \rho V_L^2 \right)  \left(\frac{\widehat{D}_v}{D_\phi}\right)^{1/2}$\\
%high \beta incompressible, i.e. dv_A=dv_S 
 & High $\beta $, $\langle v_A^2 \rangle =\langle v_S^2 \rangle $,
& $M_{A,\perp} \ll 1$
& $\frac{1}{\sqrt{2}\sin\gamma}M_A$
& Eq.\eqref{eq:f0bar_strong}
& $\frac{1}{\sqrt{2}} \left( 4 \pi  \rho V_L^2 \right)
\left(\frac{\mathcal{\widehat{D}}_0^v}{\tilde{D}_0^\phi}\right)^{1/2}$\\
%high \beta, Ma\gg1, dv_A^2 = 2dv_S^2
(3) & High $\beta$, $\langle v_A^2 \rangle =2\langle v_S^2 \rangle $& $M_{A,\perp}\ll 1$ & $ \frac{1}{\sqrt{2}\sin\gamma}\sqrt{\frac{1+\cos^2\gamma}{1+\sin^2\gamma}} M_{A}$ & Eq.\eqref{eq:highbeta2} & $\frac{1}{\sqrt{2}}\sqrt{\frac{1+\cos^2\gamma}{1+\sin^2\gamma}} \left(4 \pi \rho V_L^2 \right)
\left(\frac{\mathcal{\widehat{D}}_0^v}{\tilde{D}_0^\phi}\right)^{1/2}$\\
% low beta. first do eq 62
(4) & Low $\beta$& $\gamma\rightarrow \pi/2$ & $\sqrt{\frac{\pi}{4}\frac{\langle \delta B_{\cal F}^2\rangle}{\langle \delta B_{\cal A}^2\rangle}}$ & Eq.\eqref{eq:f0_lowbeta_sum2} & $\frac{\pi}{4} \frac{\langle \delta B_{\cal F}^2\rangle}{\langle \delta B_{\cal A}^2\rangle}\left(4 \pi \rho V_L^2 \right)
\left(\frac{\mathcal{\widehat{D}}_0^v}{\tilde{D}_0^\phi}\right)$\\
% low beta. first do eq 63
(5) & Low $\beta$ & $M_{A,\perp} \ll 1$,$\gamma\rightarrow 0$ & $\frac{1}{\sqrt{2}\sin\gamma}\sqrt{\frac{\langle \delta v_{\cal A}^2\rangle}{\langle \delta v_{\cal S}^2\rangle}} M_A$& Eq.\eqref{eq:f0_lowbeta_smallgamma} &  $\frac{1}{\sqrt{2}}\sqrt{\frac{\langle \delta v_{\cal A}^2\rangle}{\langle \delta v_{\cal S}^2\rangle}}\left( 4 \pi \rho V_L^2 \right) \left(\frac{\mathcal{\widehat{D}}_0^v}{\tilde{D}_0^\phi}\right)^{1/2}$\\
% case 6
(6) & Low $\beta$ & $M_A\approx 1$ & $\sim \frac{1}{\sqrt{2}}$ & Fig.\eqref{fig:f0_lowbeta}& $ 2\pi  \rho V_L^2 \frac{\mathcal{\widehat{D}}_{0}^v}{D_{0}^\phi}$ \\ 
\enddata
\end{deluxetable*}

\subsection{Effect of Density Fluctuations}

In the present work so far we disregarded the effect of density fluctuations, assuming that we can use the mean density instead. The contribution of density fluctuations can be significantly reduced using the VDA approach (\citealt{VDA}, Appendix \ref{polar_VDA}), which is particularly efficient for subsonic turbulence for both velocity and polarization angle fluctuations.  Below, however, we do not apply any of these techniques and deal with the data as is given by synthetic observations where both the fluctuations of $\phi$ and velocity centroids calculated in the presence of turbulent density.

\begin{figure}
\includegraphics[width=0.49\textwidth]{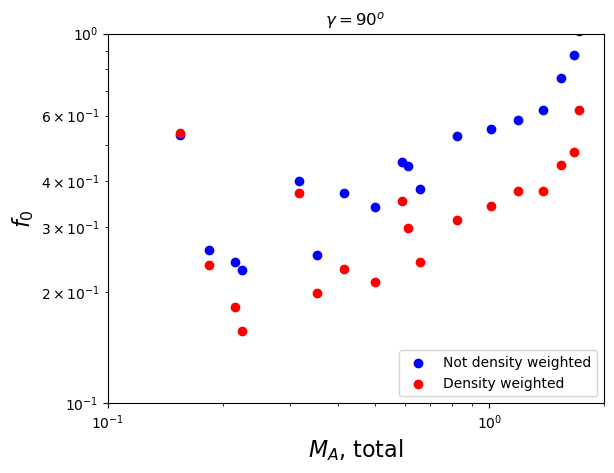}
\caption{A comparison of function $f_0$ before (blue) and after (red) the density fluctuations are accounted for a selected ranges of low $\beta$ simulations 
with $0.2<M_A<1.3$ 
}
\label{fig:den_eff}
\end{figure}

Fig.\ref{fig:den_eff} shows a comparison of $f_0$ obtained with synthetic observations with constant $\rho$ and with actual turbulent $\rho$. The overall behavior of $f_0$ is not changed. Withing the DMA the density over the averaging block should be employed for the calculation of the B-strength.

\subsection{Obtaining $V_L$}

The velocity at the injection scale enters the our expressions for $B$-strength (see Eq. \ref{eq:DMA_eq}, Table.\ref{tab:result_summary}) One way to obtain this velocity is via measuring of the total line width of the Doppler-shifted lines.

This approach is not applicable, however, in the presence of galactic shear. In this situation, $V_L$ can be obtained from the turbulence sonic Mach number $M_s=V_L/c_s$, where $c_s$ is the sound velocity. There are a number of ways of obtaining $M_s$ from the statistical analysis of data (see Chepurnov \& Lazarian 2009, Burkhart et al. 2014, \citealt{GA}). Therefore, $V_L$ can be obtained as
\begin{equation}
    V_L=M_s c_s
\end{equation}
where $c_s$ is known if the temperature of the gas is known, or by kernel estimation method proposed in \cite{GA}. More discussion of how to obtain $V_L$ will be provided in Lazarian, Yuen \& Pogosyan (2022, Paper II).

\subsection{Sub-Alfv\'enic and Super-Alfv\'enic Turbulence}
\label{sec:superAl}

Within this paper we consider the application of DMA to sub-Alfv\'enic turbulence. One might wonder how do we extend our analysis to super-Alfv\'enic turbulence? This is a totally valid question that we plan to address in an upcoming publication but first we should discuss how super-Alfv\'enic turbulence is different from that of sub-Alfv\'enic turbulence from the theoretical point of view.

The main concept that we would like to repeatedly emphasize in this paper about the turbulence system is that, the statistics of full 3D turbulence is {\it dramatically different} from that of the projected observables. In 3D turbulence, it is well known from theory (See, e.g. \cite{CL02}, or a recent review from Beresnyak \& Lazarian 2019) that the seemingly super-Alfv\'enic turbulence will behave like a sub-Alfv\'enic turbulence as long as the scale of consideration is smaller than the cut-off scale $L_{inj} M_A^{-3}$. If we measure the statistics of a super-Alfv\'enic turbulence and a sub-Alfv\'enic turbulence in sufficiently small scale, they will both exhibit the Alfv\'enic behavior as predicted in theory and also verified in numerical simulations. This also suggests a main way of distinguishing the turbulence behavior as the sub-Alfv\'enic turbulence at large scale only reverts to the weak cascade, but that for super-Alfv\'enic turbulence reverts to hydrodynamic cascade. The differences of the statistics of these cascades can be measured via structure functions at large scale. The understanding of the correlation scale $L_{inj} M_A^{-3}$ could also be put in this way: in 3D statistics, for scales smaller than $L_{inj} M_A^{-3}$ there will be Alfv\'enic cascade, and larger than that it should be hydrodynamic $L_{inj} M_A^{-3}$. Notice that $M_A$ is much more easier to obtain than the actual magnetic field strength, e.g. by measuring the anisotropy of turbulence (see Esquivel \& Lazarian 2005, Lazarian \& Pogosyan 2012, Kandel et al. 2016, 2017a). A powerful way of finding $M_A$ is based on exploring the distribution (Lazarian et al. 2018) or curvature of velocity gradients \citep{curvature}. 

\begin{figure*}
\includegraphics[width=0.49\textwidth]{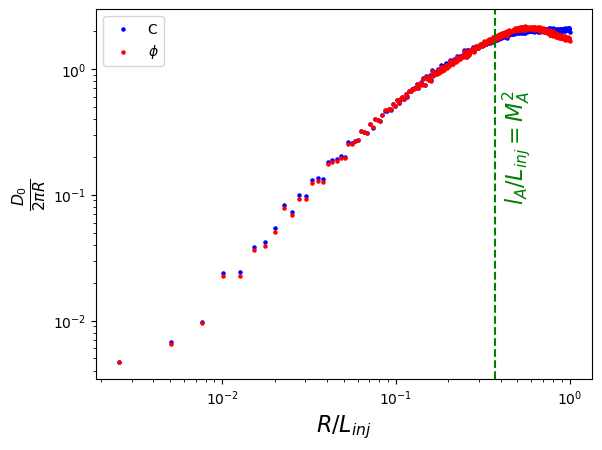}
\includegraphics[width=0.49\textwidth]{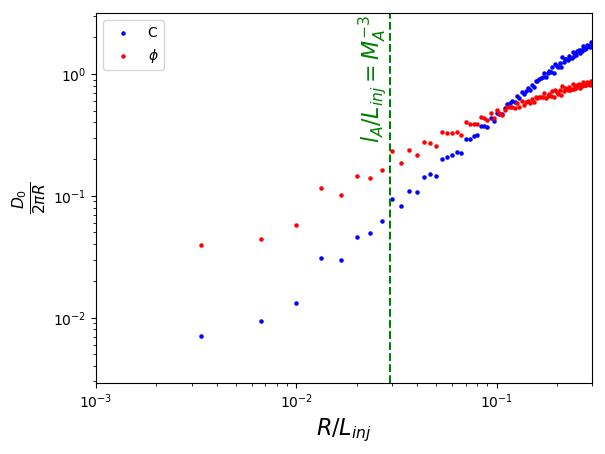}
\caption{(Left) Structure functions for velocity and the positional angle for sub-Alfv\'enic turbulence with $M_A\approx 0.61$ for $\gamma = \pi/2$. The green line corresponds to the scale $l_A=L_{inj} M_A^{2}.$  (Right) Structure functions for velocity and the positional angle for super-Alfv\'enic turbulence with $M_A\approx 3.24$ ("e7r3" from Table.\ref{tab:sim}) for $\gamma = \pi/2$. The green line corresponds to the scale $l_A=L_{inj} M_A^{-3}.$ }
\label{fig:scaling}
\end{figure*}

\begin{figure}
\includegraphics[width=0.45\textwidth]{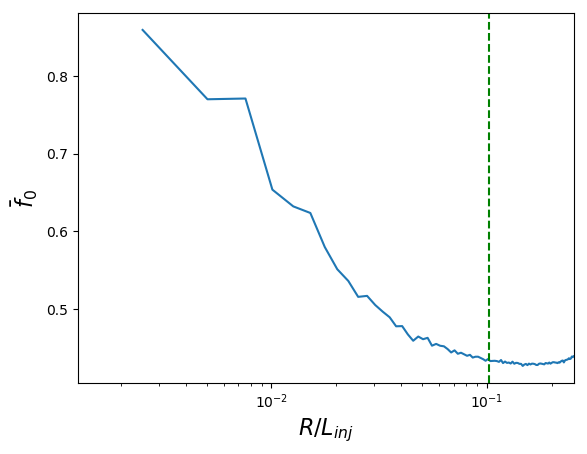}
\caption{Change of ${f}_0$  as a function of length scale $R$ (in numerical units) for the pure Alfv\'enic modes from a super-Alfv\'enic simulation (\S 4.1).  $l_A=LM_A^{-3}$ is marked by the green vertical line. }
\label{fig:f_vs_R}
\end{figure}

However, could we interpolate the scaling argument about into 2D observables? If it could be, we can definitely apply our theory to observations as long as sufficient filtering is applied to both centroid and polarization map.  Fig.\ref{fig:scaling} shows how the structure functions of velocity and polarization angle behave in two selective simulations with sub-Alfv\'enic and super-Alfv\'enic turbulence, respectively. We can see from the LHS of Fig.\ref{fig:scaling} that, in the case of sub-sonic turbulence, the structure functions for both velocity and polarization angle behave very similarly, as predicted in theory carrying the same $R^{1+m}$ factor, and saturates at the exact same transitional scale of $l_{tr} = L_{inj}M_A^2$ which defines the transition from weak to strong turbulence. However, for the case of super-Alfv\'enic turbulence, we can see from Fig. \ref{fig:scaling} that the slope of the polarization angle structure function is significantly flatter than that of the velocity structure function {\it until} $R<L_{inj} M_A^{-3}$.The difference between the two scales allows to estimate $M_A$ at least to the level to distinguish the case of sub-Alfv\'enic turbulence to which the DMA in its present form is applicable.  Furthermore, in Fig. \ref{fig:f_vs_R} it shows that the effective $f$ is changing at the $L_{inj} M_A^{-3}$ scale. An additional indication of the super-Alfv\'enic turbulence is that the slope of the spectral functions for $\phi$ gets more shallow compared to the slope velocity. Note, that the two slopes are similar for the sub-Alfv\'enic turbulence.  

\subsection{Obtaining the mode fraction from independent measurement}

Some of our limiting case formulae (Table~\ref{tab:result_summary}) depend on the mode fractions $\delta B_{\cal F}/\delta B_{\cal A}$ or $\delta v_{\cal A}/\delta v_{\cal S}$. Observers might question how do we obtain the mode composition from observations? While we show in Fig.\ref{fig:fraction} that these two ratios are simple functions of $M_A$, it would be synergistic if we obtain these two parameters independently, instead of a derived variable of $M_A$, so as to minimize the uncertainty of the magnetic field strength estimation. 

While there is no widely accepted method in obtaining the actual ratios between three modes, recent progresses have been made via the newly developed Synchrotron Polarization Analysis method (SPA, \citealt{2020NatAs.tmp..174Z,leakage}) allowing the retrieval of the ``compressible''-to-Alfv\'en ratio from direct observations of polarization maps. Notice that the ``compressible modes'' that SPA method is discussing about is the F-type tensor we discuss in the current work, i.e. including the contributions of slow and fast mode. Therefore the SPA technique provides a upper bound for the slow or fast to Alfv\'en mode ratio. This value is exceptionally important when we are in the cases where the $\beta$ is either very high (warm, diffuse ISM, see \citealt{HO21} for the reason why colder ISM, including both the unstable and the cold phases, are not ideal candidates) or very low (molecular clouds). In either of these cases, we have the corresponding predictions outlined in the previous sections, e.g. high $\beta$ case in \S 6 and low $\beta$ case in \S 7 and 8. The additional knowledge of mode ratio from SPA will enable us to obtain a more accurate estimation of magnetic field through on our results in Table~\ref{tab:result_summary}). 

\section{Comparison with Recent Modifications of DCF}
\label{sec:compare}

Due to the importance of the DCF technique, there were numerous attempts to improve the technique. Below we discuss a few selected studies that have some overlap with the present work.

\subsection{\cite{CY16}: Accounting for Multiple Injection Scales along the Line of Sight}
\label{app_CF}

\subsubsection{Motivation and Approach}

\cite{CY16}, henceforth CY16, considered the case of the injection scale of turbulence $L_{inj}$ being less than the extend of the line of sight ${\cal L}$ within the emitting turbulent volume. The authors noted that while the total line width employed as $\delta v$ in Eq. (\ref{eq:BB}) represents the full dispersion of velocity in the volume, the variation $\delta \phi$ in Eq. (\ref{eq:BB}) is a result of a random walk of $\phi$ from one turbulent injection scale to another. As a result, the DCF technique overestimates the magnetic field unless ${\cal L}$ is less or equal to $L_{inj}$. To remedy this problem CY16 proposed to measure $\delta v$ using velocity centroids. 

To explain the problem addressed by CY16, consider a setting with mean magnetic field being along $x$-direction in the plane of the sky and the Alfv\'enic fluctuations $\delta B$ is along $y$-direction. If the 3D magnetic field is $b_{reg}$, it is adds up linearly along the line of sight and therefore the observed $B_x$ is $\int_{\cal L} b_{reg} dx\approx b_{reg}{\cal L}$. On the contrary, the fluctuating magnetic field $b_{turb}$ with correlation scale $L_{inj}$ is added up in the random walk fashion with $\delta B_y$ providing $\int_{\cal L}b_{turb} dx \approx b_{turb}\sqrt{L_{inj} {\cal L}}$.\footnote{The authors have not considered the effect of Alfv\'enic mode extreme suppression for $\gamma=\pi/2$ that we discussed in \S \ref{sec:projections}.} As a result an additional factor enters the $\delta B_y/B_x$ ratio, namely, the observed fluctuation gets reduced by a factor $\approx \sqrt{L_{inj}/{\cal L}}$, which corresponds to a random walk suppression.

\subsubsection{CY16 Expression}

To account for this factor, CY16 considered the ratio of the line of sight velocity and the centroid velocity. The latter is given by Eq. (\ref{centroid}), while the former is the usual $\delta v_{los}$ arising from the velocity dispersion at the scale $L_{inj}$. The velocity measured by centroids is, on the contrary $\delta C =\int_{\cal L}\delta v_{los} dx/{\cal L}\approx \delta v_{los}\sqrt{L_{inj} /{\cal L}}$.  As a result, if $\delta v_{los}$ is substituted by the dispersion of Velocity Centroid $\delta C$ that can be used {\it both} for the case of ${\cal L}\sim L_{inj}$ and ${\cal L}\gg L_{inj}$. In other words, the expression 
\begin{equation}
\label{eq:CY}
B_{POS} \approx f_{CY16}\sqrt{4\pi\rho} \frac{\delta C}{\delta \phi}
\end{equation}
with constant $f_{CY16}$ that can be obtained from numerical simulations. Eq.\eqref{eq:CY} has a wider range of applications than the original DCF expression as they show in the series of numerical works \citep{CY16,YC19,Cho19}. In particular, the magnetic field strength computed based on Eq.\eqref{eq:CY} would not depend on $L_{inj}/{\cal L}$ ratio.

However, as we see from the applications of the DCF technique, it is very rare that the objects that observers are considering are having the size comparable to the injection scale. For instances, common molecular clouds like Taurus, Perseus have their size at the order of $1-10pc$, but the turbulent injection scale is believed to be at the order of $100pc$ (Chepurnov et.al 2015, \citealt{spectrum}). 
If we only consider the signals from the cloud along the line of sight, the ratio $L_{inj}/L$ is unlikely to be smaller than 1. In other words, the suppression of the dispersion $\delta \phi$ due to random walk arising from the addition of contributions from independent turbulent regions along the line of sight is rather unlikely. Naturally, as all the modifications of the techniques, \cite{CY16} assumes that the equipartition of the magnetic energy and the kinetic energy, which is known to be violated for the sub-Alfv\'enic turbulence
\citep{2004PhRvE..70c6408H}.  

\subsubsection{Comparison with DMA}

The similarity of the DMA with CY16 is that both techniques employ centroids of velocities. The difference is that the DMA employs the structure functions rather than dispersion. This provides the differences between our approaches at the level of our Eq. (\ref{mean_B}), which is the starting equation for the derivation of the DMA formalism. As a result, our measurements can be performed locally providing the distribution of magnetic field strength in a molecular cloud rather than a single value. In addition, structure functions are less affected by the large scale non-turbulent distortions. 

It is very important that the DMA treatment did not stop at Eq. (\ref{mean_B}) but, on the basis of MHD turbulence theory, provides analytical predictions for the functional dependence of $f_0$ on angle between the line of sight and mean magnetic field $\gamma$, as well as the Alfv\'en Mach number $M_A$. In contrast, CY16 considers the constant $f_0$ from the perspective of number of turbulent eddies along the line of sight, irrespective to the properties and orientations of the eddies according to the theory of MHD turbulence theory.

An important question arises of whether the expressions for $f_0$ that we obtained within the DMA can be used with the CY16 approach. Indeed, the structure functions of the velocities and the angle saturate at the values equal to twice of dispersion of the corresponding observables. However, this is not true, as the properties of sub-Alfv\'enic MHD turbulence change along the cascade from strong regime at small scales to weak regime at larger scales. The DMA deals with the robust scaling of small scale strong turbulence, while at larger scales the turbulence gets into the weak regime that we repeated described in the main text and also Appendix \ref{app:mhdturb}. The weak regime of turbulence is beyond the existing DMA description and therefore there is no way of direct utilizing the analytical expressions obtained in this study for improving the predictive abilities of CY16.  

{\torefereetwo
\subsection{\cite{2009ApJ...696..567H}: Model of uncorrelated turbulence}
\label{sec:comparison}

The Angle Dispersion Function (ADF) method was introduced in \cite{2009ApJ...696..567H} and \cite{2009ApJ...706.1504H}. Its similarity with the DMA is related to employing structure functions of $\phi$. However, while the DMA employs the structure functions at scales $R<L_{inj}$, \cite{2009ApJ...696..567H},
as we will argue, is only applicable when $R > L_{inj}$. 
}

\subsubsection{Model of Turbulence Adopted}

\cite{2009ApJ...696..567H} employs structure functions of the polarization angle, which is similar to the DMA, but uses the linewidth to measure the velocity dispersion.\footnote{In fact, the use of structure functions of polarization angles was suggested in \torefereeone {\cite{1994AJ....107.1433K} and numerically explored in \cite{Fal08}}. } 
In this work of the authors  replace $\delta \theta \rightarrow D^{1/2}_\phi$ and investigate its properties with the traditional DCF assumption that MHD turbulence is isotropic (cf \S \ref{app:mhdturb}), which is one of the differences from the DMA. 

Within these assumptions \cite{2009ApJ...696..567H} propose that the variations of the angle $\phi$ can be modelled by the Taylor expansions of structure
functions of polarization angles:
\begin{equation}
    D_\phi({\bf R}) \sim b^2 + m^2 R^2
    \label{eq:HH1}
\end{equation}
where $b,m$ are fitting factors. There $b$ describes the contribution from turbulence, which is applicable in the situation when the structure of turbulence is not important and only large scale dispersion of $\phi$ matters. The factor $m$ is related to the large-scale variations of mean magnetic field, which are assumed to be small making the Taylor expansion possible. 

Within the adopted model, it is then shown that the turbulent to regular magnetic field strength ratio is:
\begin{equation}
    \frac{\langle B_t^2\rangle}{\langle B^2\rangle} \sim \frac{b^2}{2-b^2}
    \label{eq:HH2}.
\end{equation}
The rest of the approach is based on  the DCF method, with  $\langle B_t^2\rangle/\langle B^2\rangle$ associated with $\delta \phi^2$ which finally gives the estimate
\begin{equation}
    \langle B^2\rangle \sim (2-b^2)4\pi \langle \rho \rangle \frac{\delta v^2}{b^2}.
    \label{eq:c-HH}
\end{equation}

As \cite{2009ApJ...696..567H} employs the velocity dispersion measured using the linewidth, the technique has to deal with the issue of estimating the number of independent fluctuations along the line of sight $\sim {\cal L}/L_{inj}$. This is elaborated further in \cite{2009ApJ...706.1504H} that estimate 
${\cal L}/L_{inj}$ in molecular clouds.  In CY16 and the DMA this issue is solved by construction by using  velocity centroids structure function.

\subsubsection{Applicability of the Model}

We shall point that the model of turbulence that amounts to a constant in the structure function (more accurately, a constant at all lags except $R=0$) is the model of negligible correlation length in the turbulent fluctuations. Basically, turbulent fluctuations are treated as a white noise. 

{\torefereetwo 
However, as we discussed in Appendix \ref{app:mhdturb}, the turbulent cascade is correlated on scales up to the energy injection scale $L_{inj}$ (see \cite{MY75}). For $R<L_{inj}$ where we sample turbulent motions over the inertial range, we expect 2D $SF_{tur}\propto R^{m+1}$ with fractional power index, e.g., for Kolmogorov turbulence $m=2/3$, that extends for multiple orders of magnitudes, as shown in the form of electron power spectra (Armstrong et.al 1995), H$\alpha$ data (Chepurnov \& Lazarian 2010) and also more recently with successive combinations of $HI$ and CO data \citep{spectrum}. In general, the expectations from turbulent fluctuations is that $m+1 < 2$. Thus \cite{2009ApJ...696..567H} assumption of the correlation scale of turbulence being
negligible makes their model not applicable at $R < L_{inj}$ to turbulence that we expect to encounter in the ISM and molecular clouds (see Table~\ref{tab:ISMtable}).  This is a significant difference with the DMA in which the turbulent
scaling at $R<L_{inj}$ is taken into account.   

It can be asked whether having observational or synthetic beam that exceeds the turbulence correlation length  and, therefore, suppresses the correlated inertial part of the spectrum, will restore the validity of Eq.~(\ref{eq:HH1}) if one samples the structure function
at much smaller scales than the beam width. This is not so. Computing explicitly
the structure function  filtered with Gaussian window of width $\sigma > L_{corr}$, one finds that $D(R) \sim 1 - \exp(-R^2/\sigma^2)$. Then at small lag $D(R) \sim R^2$, which is a general result, insensitive to the particular beam shape.\footnote{Smoothing creates a very steep spectrum 
at the scales $R < \sigma $, but for spectra steeper than $k^{-5}$ in 3D or $K^{-4}$ in 2D the structure function saturates with the quadratic behaviour. Note that $D(R) \propto R^2$ behaviour is regular, i.e. corresponds to the situation when it is possible to expand structure or correlation function in Taylor series at $R \to 0$. Thus, smoothing can be said to regularize the behaviour of the structure function at small, $R \gg \sigma $, lags to follow generic Taylor series.  In contrast, the turbulent $\propto R^{m+1}$ scaling is singular, in a sense that it is not captured by the Taylor series. }
No constant contribution $b$ appears, and Eq.~(\ref{eq:c-HH}) fails.

Thus, the model adopted in \cite{2009ApJ...696..567H}  can be applicable only when the correlation scale of magnetic turbulence is smaller than the lag $R$ between the points for which the structure function is calculated. For Kolmogorov-like turbulence this means $ R > L_{inj}$.

\cite{2009ApJ...706.1504H} have significantly extended the analysis of  \cite{2009ApJ...696..567H} with
a detailed treatment of the telescope beam and the depth of observations as applied to  molecular clouds, without using
a restrictive model of Eq.~(\ref{eq:HH1}). Nevertherless, their analytical progress relies on modelling the 
turbulence and polarization autocorrelation funcions in the specific form of a Gaussian, which width $\delta$ defines the correlation length. This model again does not reflect 
the actual scaling of the turbulence. In the language of structure functions, autocorrelation of a Gaussian form gives $D(R) \propto R^2$ again at $R < \delta$ and $D(R)\approx const$ at $R > \delta$, not the fractional power scaling that the real turbulence shows. 

The analysis in \cite{2009ApJ...696..567H} and \cite{2009ApJ...706.1504H} is potentially applicable to turbulence that has most of energy at small scales. These "shallow" spectra with 3D $k^{\alpha}$, $\alpha > -3 $ have not been observed for magnetic or velocity turbulence, although we cannot exclude their presence in some special circumstances. }

\subsubsection{Our Explanation of the Observational Data }

The approach described in \cite{2009ApJ...696..567H} \& \cite{2009ApJ...706.1504H} 
was applied to observational data to both molecular clouds \citep{2011ApJ...733..109H,2012ApJ...749...45C,2016ApJ...820...38H} and also galactic disks \citep{2013ApJ...766...49H}. In view of other observational studies of galactic turbulence (see Armstrong et al. 1995, Chepurnov \& Lazarian 2009, Chepurnov et al. 2010, Li et al. 2021, \citealt{spectrum,cattail}) it is difficult to accept the setting {\torefereetwo given our arguments in the previous subsections}. 

A possible explanation of some of the observational data in the aforementioned papers is that the small measured correlation scale of polarization angle fluctuations may correspond to the Alfv\'enic scale $L_{inj} M_A^{-3}$ of super-Alfv\'enic turbulence. In fact, we observe this type of behavior in Fig. \ref{fig:scaling}.  For instance,
in \cite{2012ApJ...749...45C} it was claimed that magnetic field fluctuations in OMC-1  have the correlation scales $\sim 9^{"}$ and $\sim 7^{"}$ for the Stokes Q and U parameters and $\sim 13^{"}$ for {\torefereeone unpolarized intensity} fluctuations. From the theory of super-Alfv\'enic turbulence (see Appendix A2), the first two numbers can be associated with the angular size associated with $l_A$, and the third number with the injection size $L_{inj}$. Using Eq. (\ref{la}) one can estimate $M_A$ for OMC-1 as $(13/9)^{1/3} \text{ to } (13/7)^{1/3}\approx 1.13 - 1.22$. This means that OMC-1 is a mildly super-Alfv\'enic object. {\torefereetwo The potential super-Alfv\'enic measurements suggest that the structure function analysis must be modified accordingly as the behavior of the structure function changes according to the theory of MHD turbulence. Note, in our paper we focus on the case of sub-Alfv\'enic turbulence. The super-Alfv\'enic case contains extra complication due to the non-trivial saturation on the slope of the polarization angle structure function. We shall discuss that case in later publications. }

\subsection{\cite{2021A&A...647A.186S}: Attempt to Account for Turbulence Compressibility}
\label{subsec:ST21}

\subsubsection{\torefereetwo Issues with the Starting Equation}

An attempt to take into account the effect of ``compressible modes'' within the DCF formalism was undertaken in \citeauthor{2021A&A...647A.186S} (\citeyear{2021A&A...647A.186S}, henceforth ST21). Following the DCF the ST21 assumed the equi-partition of kinetic and magnetic energies at the injection scale, but attempted to take into account the parallel to mean magnetic field ${\bf B_0}$ component of magnetic turbulent fluctuations ${\delta \bf B}_{\|,turb}$ present in the compressible media. Therefore the corresponding technique was termed in \cite{2021MNRAS.tmp.3119L} "parallel-$\delta B$ version of DCF". For the sake of brevity we shall refer to it as "parallel DCF" or Par-DCF". 

In the Par-DCF, Skalidis et al. (2021, henceforth SX21) further suggested that when considering energy
$E \propto B^2$, for $\delta B_{\|,turb}^2\ll B^2$ one can approximate
\begin{equation}
\begin{aligned}
    B_{total}^2&=B_{mean}^2+2B\delta B_{\|,turb}+\delta B^2 \\ 
    & \approx B^2_{mean}+2B_{mean}\delta B_{\|,turb}
\label{wrongB}
\end{aligned} 
\end{equation}
and therefore the term $2B\delta B_{\|,turb}$ was associated with the fluctuation of magnetic energy, i.e.
\begin{equation}
    \delta E_{mag}=\frac{B_{mean}\delta B_{\|,turb}}{4\pi}.
    \label{wrongE}
\end{equation}
However this is not right, since on average such $\delta E_{mag}$ will be zero as $\delta B_{\|,turb}$ is not a positively defined, fluctuating in sign and, by the definition of the fluctuation,  $\langle \delta B_{\|,turb}\rangle=0$. The multiplication of the fluctuation by $B_{mean}$ does not change the result and, naturally, $\langle B_{mean}\delta B_{\|,turb}\rangle=0$.

To address the averaging to zero, Skalidis et al. (2021) proposed that the mean $\delta B_{\|,turb}$ is to be substituted by root mean squared value $\sqrt{\langle \delta B^2 \rangle}$, which is {\torefereetwo mathematically not justified. It is obvious that the magnetic energy is $\sim \delta B^2$ and not the cross product of the mean field and the fluctuation.

A recent publication of \cite{2022arXiv220213020B} comes up with a suggestion that we interpret as an attempt not to deal with magnetic energies as the DCF does, but consider a second moment of magnetic energy fluctuation. The physical meaning of this quantity is not clear, but it serves the purpose of preserving the desired cross term $B \delta B_\|$. It is interesting that the presented numerical simulations are suggestive that the square root of the the second moment of the energy fluctuation is equal to the kinetic energy of the turbulence. This is a surprising result for which we do not see a physical justification. This result means that for subAlfvenic turbulence with $B\gg\delta B$, the energy of magnetic fluctuations $\sim \delta B^2$ are significantly lower than the energy of turbulent fluctuations $\sim \delta v^2$. This excludes the possibility that Alfvenic modes have significant contribution to turbulence and means that most of the energy must rest with the pure hydrodynamic motions that marginally involve magnetic field fluctuations. Such motions are compressions of fluid along magnetic field, which provides significant limitations on the type of media and driving that can result in such motions. This also at odds with earlier results e.g. in \cite{2004PhRvE..70c6408H}, Cho \& Lazarian (2002, 2003), Kowal \& Lazarian (2010), as well as the results in the present paper. Note, however, that the aforementioned numerical studies dealt with the incompressible driving of turbulence. 

Future research should clarify numerous numerical issues that potentially can distort the results of MHD simulations. For instance, results in \cite{2004PhRvE..70c6408H} indicate that the distribution between the energies of velocity and magnetic field fluctuations can be affected the ratio of the size of the numerical box to the injection scale. At low $M_A$ the effects of numerical box get more important and those can prevent establishing the regime of weak turbulence, as was demonstrated in Santos-Lima et al. (2021). If the relations used in Par-DCF can be justified from physical arguments, we do not believe that such arguments have been presented yet.}

\subsubsection{Par-DCF Expression for B-strength}

This problem with the SX21's starting expression being problematic was first noticed in \cite{2021MNRAS.tmp.3119L}, where the authors numerically calculated the value $B_{mean}\delta B_{\|,turb}$ for their simulations and found that that the value of this term is very small, i.e. significantly smaller than $\delta B_{\|,turb}^2$ term. In particular in Appendix A3 of \cite{2021MNRAS.tmp.3119L} 
it is written {\it "In view of the questionable assumptions underlying their method, more work is needed to understand the physical basis for the method (of \cite{{2021A&A...647A.186S}})..." }. This motivates us to further explore the SX21 expression for the magnetic field, even though, as we discussed earlier, we do not understand the physics behind this derivation.

Equating the kinetic energy with the expression of turbulent magnetic energy given by Eq. (\ref{wrongE}) ST21 obtained 
\begin{equation}
    B_0 = \sqrt{2\pi\rho}\frac{\sigma_v }{\sqrt{\sigma_\phi}}.
    \label{tassis}
\end{equation}
where, as in Eq. (\ref{eq:BB}) the fluctuations of velocities and angle are associated with the dispersions of these quantities that can be obtained from observations, i.e. $\sigma_v\equiv \delta v$ and $\sigma_\phi\equiv \delta \phi$.
Note that Eq. (\ref{tassis}) is different from the DCF expression given by Eq. (\ref{eq:BB}). The most important is the difference between the two equations is that the square root of dispersion of $\sigma_\phi$ enters in Eq. (\ref{tassis}).

\subsubsection{Does Eq. (\ref{tassis}) Describe Effects of Turbulence Compressibility?}

Our simulations employed for this paper are compressible MHD simulations with {\it solendoial} driving while SX21  employ both {\it solendoial} and {\it compressive} driving. However, SX21 claim that their finding do not depend on the difference in the compressibility of driving.

The study in SX21 is performed for low-$\beta$ simulations and $\gamma=\pi/2$. The latter, as we have shown in this paper is a very special case for numerical testing. Indeed, at this $\gamma$ due to the periodic boundary condition, there exist  the strong suppression of the contribution of Alfv\'en modes to $\delta \phi$. Unlike SX21, we perform the decomposition of the turbulence into Alfv\'enic, slow and fast modes. This allows us to quantify the effects of compressibility that the Eq (\ref{tassis}) is supposed to reflect. Our results in Fig. \ref{fig:fDvDphi_modes_lowbeta},\ref{fig:ST21} show that for low-$\beta$ simulations the dispersion of $\sigma_\phi$ is dominated by fast modes and scales as $M_A$ or $M_A^{5/4}$ depending on the choice of the scales $R$. However, neither these dependencies correspond to $\sigma_\phi\sim M_A^2$ reported in SX21.

Incidentally, SX21 claim that this difference of scaling of $\sigma_\phi$, i.e. being $\sim M_A$ and $\sim M_A^2$ is "based on the different scaling relation with the magnetic fluctuations in the incompressible (Goldreich \& Sridhar 1995) and compressible turbulence (Federrath 2016; Beattie et al. 2020)." Our numerical results contradict to this claim. It is Alfv\'enic and slow modes following Goldreich \& Sridhar scaling that can show in {\it some settings}, e.g. for scales $R\ll L_{inj}$, $\sim M_A^2$ scaling. In contrast, compressible modes do not show $\sigma_\phi\sim M_A^2$ scaling in our simulations (See lower right corner of Fig.\ref{fig:ST21}).  An additional discussion of our numerical results related to the applicability of Eq. (\ref{tassis}) is provided in Appendix \ref{sec:numerical_modes}.

We also can notice a resemblance of Eq. (\ref{tassis}) and some of our expressions obtained for {\it incompressible} turbulence (see Table \ref{tab:result_summary}). Indeed, following the DMA approach, it is easy to see that one can rewrite Eq. (\ref{tassis}) as
\begin{equation}
B_0 \sim f_{compr} \sqrt{4\pi\rho}\frac{\delta v}{\delta \phi}
\label{tassis2}
\end{equation}
where $f_{compr} \sim M_A$.  

However, this resemblance is {\it  coincidental}. The SX21 study is focused on molecular clouds, which are low $\beta$ medium. Therefore, the simulations that SX21 employ to test the correspondence with their formulae are compressible simulations in low $\beta$ medium. For this setting our Table \ref{tab:result_summary} provides $f_0\sim M_A$ for a very special case of $M_A\ll 1$ and $\gamma \rightarrow 0$. However, the numerical study in SX21 was done for $\gamma=\pi/2$. In this case, the contribution of both Alfv\'en and slow modes are expected to be subdominant. 

\begin{figure}[th]
\centering
\includegraphics[width=0.49\textwidth]{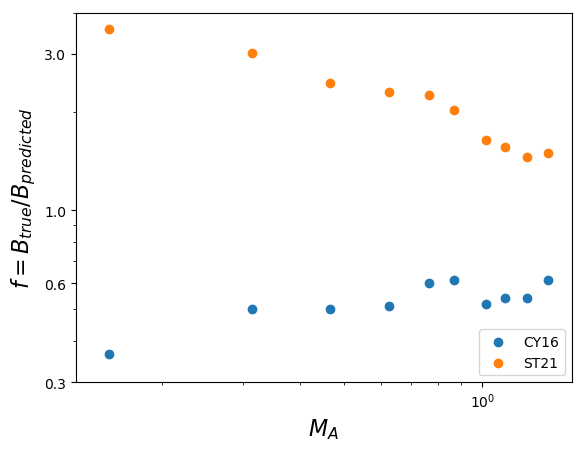}
\caption{\label{fig:CY_ST} The ratio $f = B_{true}/B_{predicted}$ that enters Eq. (\ref{tassis2}) as a function of $M_A$ for $\gamma=\pi/2$ for MHD simulations  with $M_s=6$.  The simulations are from Table 6 for low-$\beta$.}
\end{figure}

In Fig. \ref{fig:CY_ST} we provide the comparison of DCF results in the form of CY16 relations and Eq. (\ref{tassis}) for a range of $M_A$. Our results show that the CY16 show a more predictable behavior compared to Eq. (\ref{tassis}). Indeed, the flat dependence in CY16 can be corrected by choosing a constant pre-factor on the basis of numerical simulations, as it done frequently in many DCF studies. Eq. (\ref{tassis2}), on the contrary, shows a dependence on $M_A$ which is indicative of a problem in the choice of the function in Eq. (\ref{tassis2}). Thus our testing favors the traditional DCF rather than Eq. (\ref{tassis}). 

Our prediction of $f_0\sim M_A$ corresponds to the case of nearly {\it incompressible} turbulence in high $\beta$ medium (see Table \ref{tab:result_summary}). In this case the slow mode fluctuations induce changes of $\phi$ and, as a consequence, induce the $f_0\sim M_A$ dependence. This effect cannot be present in the SX21 setting. 

We would like to state that our main point is that we object to the physical interpretation of  Eq. (\ref{tassis}) given in the literature. At the same time, we have no reason to question this equation as the consequence of empirically established fact.

\section{Discussion}
\label{sec:discussion}
\subsection{DMA versus Davis-Chandrasekhar-Fermi Technique}
\label{sec:strength}

\subsubsection{Limitations of the DCF}

The DCF is an empirical technique that is widely used in spite of the serious problems related to its accuracy in obtaining the value of magnetic field strength. In the past there were many attempts in fixing the DCF technique (see Ostriker et.al. 2002, Houde et al. 2009, Cho \& Yoo 2016). However, these attempts were not founded based on the theory of MHD turbulence, in particular, the properties of the eigenmodes in MHD turbulence is usually disregarded or not properly considered.

The theoretical basis of the DCF is the equipartition of the kinetic energy and the energy of magnetic fluctuations, which is problematic in view of finding that the energy of magnetic fluctuations are significantly smaller than the kinetic energy in sub-Alfv\'enic turbulence (Haugen et al 2004). This equipartition is equal only at the regime of strong turbulence that we deal at small scales using the DMA.

Some of the problems of the DCF we addressed by introducing Eq. (\ref{mean_B}) that employs instead of global dispersion of $\delta \phi$ and $\delta v$ the differential measures in the form of structure functions that can be measured locally. However, this does not solve a major problem of  the DCF, namely, the DCF ignores the knowledge that we have about MHD turbulence, most importantly, the MHD turbulence anisotropy.

Reformulating the DCF on the basis of the modern MHD theory is difficult due to two different regimes of turbulence, weak and strong, that influence the dispersion of magnetic field and velocity. This problem is alleviated within the DMA that, by construction, deals only with strong turbulence sampling turbulent fluctuations at small scales.

\subsubsection{ DMA versus DCF }
\label{sec:ach}

In this paper, we analytically derive an important relation that relates the squared mean magnetic field strength to the structure functions of the velocity centroid and polarization angle, with an extra $f_0$ factor that is related to the geometric factor of the MHD modes. Our general expression given by (Eq.\eqref{eq:Dc_over_Dtheta_rho_alt}) describes obtaining magnetic field strength with any mixture of MHD modes. Later, we consider the limiting cases of turbulence corresponding to molecular clouds, i.e. low $\beta$ medium, and diffuse warm gas, i.e. high $\beta$ medium, analyze the effects of the change angle between the magnetic field and the line of sight.  {\kh Based on the formalism in \S \ref{sec:modes}, }We demonstrated that for the DMA can return accurate values of magnetic field strength (see Fig \ref{fig:merged_comparison}) for various astrophysical settings.

An additional advantage of the DMA compared to the DCF is that it can be successfully used for studying cases when line broadening is sub-thermal. The separation of the velocity components into the thermal and non-thermal part is rather complicated and causes additional errors \citep{2021arXiv210302219E}. This limits the accuracy at which this process can be performed withing the DCF. At the same time the DMA does not require such separation, as the structure function of centroids is not sensitive to the thermal part of the line (see \citealt{LE03,EL05,KLP17a}).  Moreover, The recently proposed Velocity Decomposition (VDA) approach \citep{VDA} allows the actual separation of non-thermal velocity fluctuations from channel maps. The fluctuations of densities induce changes to the statistics of the measured quantities, while the VDA allows to remove these interfering contributions. In Appendix \ref{polar_VDA} we provide an approach that allows decreasing the effects of density for the polarization studies.

In Paper II (Lazarian, Yuen, Pogosyan) we discuss another approach to finding magnetic field strength by employing Alfv\'en and sonic Mach numbers instead of the structure functions of magnetic field direction and the velocities that we use within the DMA. The two technique are new and the study of their synergy will be presented elsewhere.

\subsubsection{Important Role of Various MHD Modes}

The DCF technique disregards the fact that MHD turbulence consists of different MHD modes with slow and Alfv\'en modes being very anisotropic. To address this issue, in the DMA we study the turbulent velocity and magnetic field angle fluctuations at the small scales, i.e. at the scales at which we can apply our quantitative knowledge of compressible MHD turbulence  (GS95, LV99, Lithwick \& Goldreich 2001, Cho \& Lazarian 2002, 2003, Kowal \& Lazarian 2010, see also Beresnyak \& Lazarian 2019). We employ the statistics of MHD turbulence in the observer's frame mostly following the LP12 approach and somewhat extending it. 

In the DCF the accuracy of the magnetic strength determination is increased by the empirical adjusting the expression prefactor $f_0$ by using numerical simulations. On the contrary, the DMA provide expression of $f_0$ as a function of $M_A$, angle $\gamma$ and properties of the media at hand. This significantly increases the precision of the new technique and explains the reported uncertainties of the DCF.

\subsection{DMA and Recent Attempts to Modify DCF}

The well-known limitations of the DCF induce numerous efforts to modify the technique. We have reviewed a couple of recent ones applying in sequence the following criteria:
\begin{itemize}
    \item Solid physical model behind the derivation.  
    \item Model of turbulence applied for deriving the expression.
    \item Advantages of the derived expression compared to the original DCF.
\end{itemize}

For instance, we showed in \S 6.3 that the ST21 study does not satisfy the first criterion. Their approach if treated as based on the numerical finding should not be dismissed but subject to more numerical studies. The differences that we find interpolating our results can be due to the differences in the adopted driving. 

The first criterion is satisfied for Angle Dispersion Function (ADF) method in \cite{2009ApJ...696..567H,2009ApJ...706.1504H}. Incidentally, the ADF, similar to DMA, employs structure functions for the polarization angle, but stays within the DCF approach with both using the line width for evaluation of velocity fluctuations as well as returning to the use of dispersions of magnetic field and velocities in the final expressions. On the contrary, the DMA uses the small scale structure functions of both polarization angle and the velocity measures.  

At the same time the vital difference between the DMA and the ADF is that the latter implicitly assumed that the injection scale of the turbulent velocity $L_{inj}$ is much smaller than the separation of scales at which the structure functions are calculated. This assumption does not agree with what we know of turbulence in molecular clouds. For instance, according to McKee \& Stone (2021) the injection of turbulence scale is at least two times the size of molecular clouds, while in \cite{spectrum} it is shown that the turbulence spectrum of velocity fluctuations in Taurus smoothly transfers to much larger scales in diffuse media, making the turbulence injection scale significantly larger than the scale of the molecular clouds. As a result the applicability of the approach the applicability of the ADF is limited to special cases that should yet be identified by observations. 

The improvement of the DCF technique that corresponds to all three criteria above is given in \cite{CY16}.  Similar to the DMA it employs centroids of velocities. Formally, our Eq. (\ref{mean_B}) transfers into \cite{CY16} expression in the limiting case of $l=R\gg L_{inj}$. However, the analytical expressions derived in the DMA approach cannot be transferred to be used in \cite{CY16}, as the DMA deals with the well defined properties of MHD cascade at small scales, which are different from the properties of dispersion $\delta v$ and $\delta \phi$ at the injection scale that \cite{CY16} employs. In addition, the distribution of magnetic field strength over the observed object is available with the DMA.

\subsection{Obtaining Magnetic Field Strength in 3D}

\noindent{\bf Using Dust polarization}. GAIA data on the distances from the stars provide a valuable source of the 3D distribution of magnetic field in the galaxy. This information was used, for instance, in Gonsalvez-Casanova \& Lazarian (2019) to confirm that the 3D distribution of the POS magnetic field obtained with the an innovative Velocity Gradient Technique  (VGT) (see Yuen \& Lazarian 2017ab, Lazarian \& Yuen 2018, Hu et al. 2019, Yuen et al. 2021) is consistent with the star light polarization data. The direct application of the DCF to the distribution of polarization directions from the stars with known distances is prohibited by the fact that the velocity dispersion of the nearby gas is dominated by the galactic rotation. This is not a problem for the DCF, however. For instance, the DMA could be applicable to studies of magnetic field using the 21 cm line of atomic hydrogen. This line is broadened by both thermal motions and also galactic rotation, but one can still use structure functions of velocities using Reduced Velocity Centroids (RVCs) introduced in \cite{LY18a}. 

In fact, the DMA can be used directly with the stars without using any diffuse media spectroscopy. Indeed, it was recently found that the statistics of the velocity of young stars reflects the statistics of turbulence \citep{2021ApJ...907L..40H}. Therefore the structure functions of the velocities of stars can be obtained with GAIA and combined with the starlight polarization data. The advantage of the GAIA data, that the structure functions of velocity are available not only for the LOS component of velocity, but also for the POS velocity components. As a result, more detailed studies can be performed using the DMA formalism.\footnote{Incidentally, this will be complimentary to using the GAIA starlight data approach proposed in  Hu et al. (2021). There a new technique of finding media magnetization  by measuring the anisotropy of the distribution of star velocities was introduced.}\linebreak

\noindent{\bf Using Spectral line polarization}. Magnetic field direction can be obtained by measuring the polarization from spectral lines that arises from Goldreich-Kylafis effect and the Ground State Alignment (GSA) (see Appendix B). The polarization of spectral lines exhibit properties that makes them attractive from the point of view of 3D magnetic field strength measurements. First of all, one can ensure that the measurements of the magnetic field direction and the Doppler broadening arise from the same volumes of gas. Very importantly, however, that the emission of different spectral lines is localized to different regions. For instance, the GSA is being destroyed by collisions and thus it produces polarization in rarefied regions of interstellar medium with strong illumination by radiation sources. As a result, unlike dust polarization, the GSA is less affected by the line of sight averaging. In addition, spectral lines are affected by galactic rotation and this allows to get separate the magnetic field measurements at different distances along the line of sight. As we discussed above, the DMA can work successfully in the presence of Doppler broadening arising from galactic rotation.\linebreak

\noindent{\bf Using magnetic field directions from VGT}. Spectral lines can be used to obtain the magnetic field directions using the VGT. This does not require any polarization information and the same spectral line information can be used both for evaluating the structure functions of the magnetic field and the structure functions of velocity. The VGT, however, provides yet an alternative way of measuring magnetic field strength that is based on the statistical properties of gradients. We discuss this in Paper II (Lazarian, Yuen, Pogosyan). The DMA-VGT and the other technique are complementary in obtaining the 3D distribution of magnetic field strengths.

\subsection{Synergy of theory and numerical simulations}

In this paper we employed numerical simulations to both to test the theoretical expectations and provide some of the input parameters. For instance, our numerical studies showed how the energy is being distributed between different turbulent modes for  different settings.  

Being guided by theory, we expected the non-trivial change of the results with angle $\gamma$ between the line of sight and mean magnetic field. Therefore our numerical testing was performed for a variety of $M_A$ and $\gamma$. In contrast, the DCF numerical simulations that we are aware of, are limited to $\gamma=\pi/2$, which our study shows to be a very serious limitation. 

Theory provides important warnings for interpreting numerical simulations. We list of couple of them below.

For instance, numerical testing faces problems arising from the insufficient resolution of numerical simulations. This problem is very serious for Alfv\'en and slow modes. As the consequence of insufficient numerical range, these modes for sufficiently small $M_A$ can be simulated only in one regime, the weak MHD turbulence regime. In astrophysical reality, the inertial range is usually sufficiently extended and both regimes of weak and strong MHD turbulence are present. This entails a serious difference in the properties of actual and numerically simulated turbulence. In this situation the numerical testing of both DCF and DMA becomes poorly justified.

In addition, for $\gamma=\pi/2$ the periodic boundary conditions induce the effect of infinite integration that strongly suppresses the Alfv\'enic contribution. This is something that should be seriously considered during the DCF testing. At the same as $\gamma$ changes from $\pi/2$, the effects of periodicity along the line of sight is altered and the suppression of Alfv\'en waves is removed. This is a subtle but important effect that complicates the testing.
The periodicity is being disturbed for $\gamma\ne \pi/2$ and, as a result, the suppression of Alfv\'enic modes gets altered. Thus small changes in $\gamma$ around $\pi/2$ can result in significant changes of the observed contribution of Alfv\'en modes. These effects have never been discussed within the numerical studies of DCF where setting of $\gamma=\pi/2$ is employed

On the contrary, in the case when fast modes dominate the measured fluctuations of $\phi$, there exist only one regime of fast mode turbulence and therefore we see the scaling of $D_\phi$ that corresponds to the change of the fast mode energy with $M_A$, i.e. while the fast mode energy increases with $M_A$ in proportion to $M_A^{1/2}$, the increase of $\sqrt{D_\phi}$ goes as $M_A^{1+1/4}=M_A^{5/4}$. Indeed, we observed this dependence for $\gamma=\pi/2$.

There exist uncertainties related to the known properties MHD turbulence that still require studies. For instance, we are not certain to what extend the turbulence at small scales gets independent on the conditions of turbulent driving in terms of its mode composition. Naturally, at the injection scale, the distribution of energy withing compressible and incompressible modes is determined by the turbulence driving. However, one may expect that due to the partial coupling of modes (Cho \& Lazarian 2003), the energy can be redistributed between them as turbulence cascades to small scales. If so, for sufficiently extended turbulence range, one may get turbulence properties independent of the initial driving. 
This issue calls for further studies.

\subsection{Prospects of the Technique and a Broader Impact of the present study}

The expressions of the DMA allow to derive the $B$-strength with any required precision. However, they exhibit dependencies on on angle between the line of sight and magnetic field $\gamma$, medium magnetization $\beta$ and mode composition. 

Potentially, these parameters can be obtained from independent studies. For instance, in \cite{leakage} the procedures for obtaining the relative distribution of MHD modes are proposed and tested with synthetic observations. The plasma $\beta$ does not need to be obtained precisely and its evaluations can be improved via the iterative application of the DMA.
When these parameters are uncertain, the DMA expressions allow to evaluate the uncertainties of B-strength obtained with the technique. 

The new technique is really timely these days where both polarimetry and velocity measurements can done with high spacial resolution. This allows to measure more detailed statistics compared to the earlier days. Thus, with the DMA one can get detailed information about the magnetic field and its distribution over the turbulent astrophysical volume. 

The DMA samples motions at smaller scales at which the motions preserve its Alfv\'enic character. This it opens prospects for obtaining the magnetic field strength in super-Alfv\'enic turbulence that we will explore in our next publication (Yuen et al in prep).

The current study is based on the LP12 formalism describing MHD turbulence in the laboratory frame.
The first application was the analytical description of synchrotron intensity fluctuations. The part of it related to velocities was further advanced in the series of papers that followed (Kandel et al. 2017, 2018). 

In the present paper, the LP12 approach was elaborated for describing the observed statistics of polarization directions. The importance of this advance goes beyond the particular problem in hand, i.e. the problem of obtaining magnetic field strength by combining polarization and spectroscopic data. For instance, the developed formalism allows to address the studies of magnetic turbulence from observations (see Lazarian \& Pogosyan 2016). 

In addition, we noticed the limitation of LP12 study in terms of dealing with high $\beta$ media case. Indeed, the equipartition of energy in slow and Alfv\'en modes is assumed in our earlier studies. This is a reasonable assumption an idealized infinite inertial range, when the measurements of turbulence properties lose their dependence on turbulent driving. However,  for the measurements sufficiently close to the injection scale, the properties of turbulence are influenced by the driving. This is definitely the case for present day numerical simulations. For instance, for the case of the conventional isotropic turbulence driving, at low $M_A$, the energy of slow modes is expected to be $\approx 1/2$ of the energy of Alfv\'en modes. The difference between the two is expected to decrease with the increase of $M_A$, which we also confirmed by numerical simulations. Therefore, in the present study we extended the LP12 approach for the case of isotropic turbulence driving for high $\beta$ case. This extension is important for the applications that are different from the DMA.

\section{SUMMARY}
\label{sec:conclusion}

The paper introduces a new way to measure the strength of magnetic field by combining the spectroscopic data and the polarization measurements. The gist of our approach is to use the differential measures, i.e. structure functions at smallest available scales, to characterize the fluctuations of both the velocity and the polarization. A distinguishing feature of this study is the use the theory of MHD turbulence to describe the relevant small-scale fluctuations, at $R$ significantly smaller than the injection scale $L_{inj}$ This is in contrast to the traditional Davis-Chandrasekhar-Fermi (DCF) where the dispersion of the above two quantities were employed and the anisotropic properties of MHD turbulence are disregarded.  As a result, our Differential Measure Analysis (DMA) technique is different from the traditional DCF and its more recent modifications/improvements. We can briefly summarize our results in the following way:
\begin{enumerate}
    \item The DMA can be applied to small patches of observational data and can successfully deal with data inhomogeneity and interfering processes not related to the turbulent cascade. As a result, the DMA can provide a detailed distribution of the POS component of magnetic field.
     \item The anisotropic nature of MHD turbulence makes the DCF approach not accurate. Our study focuses on the small scale asymptotic behavior of basic modes of MHD turbulence and provides robust expressions for magnetic field. The generalization of our results to the dispersion that the DCF deals with is challenging, due to changes of the nature of MHD cascade at large scales. On the basis of our study we can state that the coefficient of proportionality between the magnetic field strength and the ratio of the velocity and polarization angle structure functions is not a constant, but, in general, a function of Alfv\'en Mach $M_A$, angle between the mean magnetic field and the line of sight $\gamma$, as well as of relative fraction of basic MHD modes in the turbulent volume.
     
     \item We obtained general expressions for the DMA both in interstellar medium with magnetic pressure larger than the gaseous pressure, i.e. low $\beta$ medium, and for gaseous pressure larger than the magnetic pressure, i.e. high $\beta$ medium. Starting with our general expressions, we derive a set of simplified expressions that are applicable to magnetic field studies in molecular clouds and diffuse media (see Table \ref{tab:result_summary}). 
     
     \item Our study of high $\beta$ medium provides robust expressions that can be applied to observational data with minimal assumptions. For instance, in the case of equipartition of Alfv\'en and slow modes, no additional information related to the value of the angle $\gamma$ is required.
     
     \item Our study of low $\beta$ medium demonstrates pronounced dependence on $\gamma$.
     Therefore the earlier numerical studies of the DCF case limited to $\gamma=\pi/2$ are not adequate. The composition of turbulence in terms of basic MHD turbulence modes can significantly alter the results.  
     
\end{enumerate}

Our study provides general expressions that can be used for obtaining magnetic field by combining polarization and spectroscopic observations. It testifies that a further increase of the accuracy of obtaining of magnetic field strength in molecular clouds can be achieved by employing additional information, e.g. the information on the composition of MHD turbulence in terms of Alfv\'en, slow and fast modes. This information can be obtained both from numerical simulations and observations.

\noindent{\bf Acknowledgment} We thank Jungyeon Cho for providing the set of incompressible MHD simulation data and the inspiring discussions. We thank Chris Mckee and Marijke Haverkorn for valuable discussions, providing comments and suggestions to our manuscript. Valuable discussion with Siyao Xu about the effects of generation of perpendicular magnetic field in sub-Alfv\'enic turbulence is acknowledged.  {\torefereeone We thank Martin `Houde for refereeing the paper and providing extensive suggestions and comments to our manuscript.} A.L. and K.H.Y. acknowledge the support the NASA ATP 80NSSC20K0542 and NASA TCAN 144AAG1967. The numerical part of the research used resources of both Center for High Throughput Computing (CHTC) at the University of Wisconsin and National Energy Research Scientific Computing Center (NERSC), a U.S. Department of Energy Office of Science User Facility operated under Contract No. DE-AC02-05CH11231, as allocated by TCAN 144AAG1967. K.H.Y also thanks Ka Wai Ho (UW Madison) for providing part of the XPU-parallelized codes (\url{https://www.github.com/doraemonho/LazRotationDev}). D.P. thanks Theoretical Group at Korea Astronomy and Space Science Institute (KASI) for hospitality. 

\appendix

\begin{deluxetable}{l c c}[h]
\tablecaption{\label{tab:notations}List of notations used in this work}
\tablehead{\emph{Parameter} & \emph{Meaning} & \emph{First appearance}}
\startdata
\hline\hline
${\bf r}$ & 3-D separation ${\bf x}_2-{\bf x}_1$ &  Eq. \eqref{d_theta} \\ 
${\bf R}$ & 2-D separation ${\bf X}_2-{\bf X}_1$ &  Eq. \eqref{centroid} \\ 
$z$ & Line of sight (LOS) variable &  Eq. \eqref{eq:tantheta} \\ \hline
${\bf x}$ & 3-D position vector & Eq. \eqref{struc_b}\\
${\bf X}$ & 2-D position vector & Eq. \eqref{d_theta}\\
$l$ & Distance of the 3d separation $|{\bf r}|$ &  Eq. \eqref{struc_b} \\ 
${\cal L}$ & Size of a turbulent cloud & Eq. \eqref{eq:sf2d3d}\\
${L_{inj}}$ & Turbulence injection scale & Eq. \eqref{eq:sf2d3d}\\ \hline
$A$ & A(lfven)-type vector component & Fig.\ref{fig:frame}\\
$F$ & F-type vector component, defined in \cite{LP12}, F-type contains both slow and fast mode contributions & Fig.\ref{fig:frame}\\
$P$ & P(otential)-type vector component &  Fig.\ref{fig:frame}\\
$\hat{\cal A}$ & Alfv\'en mode spectra & Eq. \eqref{eq:totalf}\\
$\hat{\cal S}$ & Slow mode spectra & Eq. \eqref{eq:totalf}\\
$\hat{\cal F}$ & Fast mode spectra & Eq. \eqref{eq:totalf}\\ \hline
$\rho({\bf r})$ & 3-D Density & Eq.\eqref{eq:alf}\\ 
$\rho({\bf X},v)$ & Emitters' intensity in the PPV space & Eq. \eqref{centroid} \\
$B$ & 3-D magnetic field  & Eq.\eqref{eq:alf}\\ 
$b_{turb}$ & Turbulent part of the magnetic field & Eq.\eqref{struc_b}\\
$B_\perp$ & Projected magnetic field   & Eq.\eqref{struc_main}\\
$B_{x,y}$ & The x \& y components of the magnetic field, perpendicular to LOS & Eq.\eqref{eq:stokes_dust}\\
$Q,U$ & Stokes Q \& U parameters & Eq.\eqref{eq:stokes1}\\
$v$ & 3-D velocity & Eq.\eqref{eq:alf}\\ 
$C$ & Velocity Centroid & Eq.\eqref{struc_centr}\\ 
$\theta$ & Magnetic field angle & Eq. \eqref{eq:tantheta}\\
$\phi$ & Polarization angle & Eq. \eqref{eq.stokes}\\ \hline
$M_s$ & Sonic Mach number & Eq.\eqref{eq:alf}\\ 
$M_A$ & Alfv\'enic Mach number & Eq.\eqref{eq:BB}\\ 
$f$ & Weighting factor of the DCF/DMA Equation & Eq.\eqref{mean_B}\\
$\bar{f}_0$ & Weighting factor for the zeroth moment DMA equation & Eq.\eqref{mean_B}\\ 
 \hline
$\langle A \rangle_{x}$ & average of the quantity $A$ over variable $x$ & Eq.\eqref{d_theta}\\
$D_{2D/3D}\{A\}$& 2-D/3-D Structure Function of variable $A$  & Eq \eqref{d_theta} \\
\hline
$\gamma$ & Angle between the line of sight and symmetry axis & Eq. \eqref{eq:BBsin}\\
$\mu$ & $=\widehat{k}\cdot \widehat{B}$, $\mu=cos(\gamma)$ & Eq. \eqref{eq:app_espec_prelim}\\
$\mathcal{D}^{\phi}({\bf R})$ & = $D_2\{\phi\}({\bf R})$, of polarization angle structure function & Eq. \eqref{eq:Dphi_def}\\
$\mathcal{D}^v_n(R)$ & Multipole moment of centroid structure function ($SF_2\{C\}({\bf R})$) & Eq. \eqref{eq:csf}\\
$\mathcal{D}^{\phi}_n(R)$ & Multipole moment of polarization angle structure function & Eq. \eqref{eq:phimultipole}\\\hline
 $C_n(m)$ & $
-\frac{i^n\Gamma\left[\frac{1}{2}(|n|-m-1)\right]}{2^{2+m}\Gamma\left[\frac{1}{2} (|n|+m+3)\right]} $ & Eq.\eqref{eq:phimultipole}\\
$G^{(A,F,S)}_n(\gamma)$ & Multipole decomposition of the geometric functions of polarization angles, defined in \cite{LP12} & Eq.\eqref{eq:phimultipole} \\
$\mathcal{W}^{(A,F,S)}_n(\gamma)$ & Multipole decomposition of the geometric functions of velocity centroids, defined in \cite{KLP17a} & Eq.\eqref{eq:csf} \\
$W_I(M_A)$ & Weight of the isotropized spectral part & Eq.\eqref{eq:WIWL}\\
$W_L(M_A)$ & Weight of the local anisotropic spectral part& Eq.\eqref{eq:WIWL}\\
\hline \hline
\enddata
\end{deluxetable}

\section{WAYS OF MEASURING MAGNETIC FIELD USING POLARIZATION}
\label{sec:waystomeasure}

\subsection{ Polarization from Aligned Dust}
Dust polarization arises from emission of non-spherical grains aligned with long axes perpendicular to the ambient magnetic field (see \citealt{Aetal15}). Similarly, polarization of starlight arises from the differential extinction by aligned grains. The processes of dust alignment is generally believed to happen due to radiative torques (RATs) (see \citealt{1976Ap&SS..43..257D,1996AAS...189.1602D} ). The theory of the RAT alignment have is based on the analytical model in \cite{2007MNRAS.378..910L} and further studies e.g. in \cite{2008MNRAS.388..117H,2016ApJ...831..159H}.

The RAT alignment theory at its present form (see \citealt{2019ApJ...883..122L}) can account for the major observational features of grain alignment. In particular, in typical conditions of diffuse ISM the silicate grains are nearly perfectly aligned, while in dense molecular clouds the degree of alignment depends on the grain illumination mostly by embedded stars. In other words, the existing grain alignment theory can evaluate in what conditions one should expect the polarization arising due to the aligned dust to trace magnetic fields.  With more polarization measurements obtained using starlight and with more distances to stars measured there is a possibility to trace magnetic field in 3D.\\

\subsection{Goldreich-Kylafis Effect} 

\citeauthor{1981ApJ...243L..75G} (\citeyear{1981ApJ...243L..75G,1982ApJ...253..606G}, henceforth GK) effect provides a viable way of tracing magnetic fields in molecular clouds. The polarization arises due to the differences of the radiation transfer in the media with anisotropies or shear. The resulting polarization is either parallel or perpendicular to the magnetic field. In spite of this ambiguity, the effect has been successfully employed to trace magnetic field structure of molecular clouds (Girart et al 1999, \citealt{li}) {\torefereeone A more recent study reveals that there are extra complication in understanding the molecular line polarization based on GK effect. It is proposed that the use of circular polarization (Houde et.al 2013, Chamma et.al 2018) can remove the correlation between the polarization angle and the magnetic field (See also \citealt{2013A&A...558A..45H}) In any case,} combining GK with velocity gradients (Yuen \& Lazarian 2017ab, Lazarian \& Yuen 2018a, Hu et al. 2019, \citealt{VDA}) one can remove the 90 degree ambiguity in the magnetic field direction.  {\torefereeone  A recent work in analysing the polarization in the emission of atomic or molecular (sub)millimeter lines has been developed by \cite{2020A&A...636A..14L}}.\\

\subsection{Ground State Alignment} 

A promising development in the RMS  of magnetic field tracing is presented by the atomic/ionic ground state alignment (GSA) effect suggested and quantified for use in astrophysical conditions by \citep{2006ApJ...653.1292Y,2007ApJ...657..618Y,2008ApJ...677.1401Y,2012JQSRT.113.1409Y,2015ApJ...804..142Z,2018MNRAS.475.2415Z,2018MNRAS.479.3923Z}.\footnote{{\torefereeone For further improving of the technique, influence of additional effects, e.g. how the effects of collisions and the stimulated emission \citep{2020A&A...636A..14L} affect the GSA, can be taken into account.}} The GSA employs atoms/ions with fine and hyperfine split levels. The atoms/ions get aligned in the ground or metastable state by external anisotropic radiation. The Larmor precession in the ambient magnetic field re-aligns the atoms/ions imprinting its direction on polarization. The atoms/ions stay in ground or metastable state long and thus they can trace very weak magnetic fields.  The effect has been recently confirmed with observations \citep{2020ApJ...902L...7Z}, opening a wide avenue of applying it for tracing magnetic fields in various environments. The difference in distribution of atoms and conditions for atomic alignment in space provides a way to get the 3D distribution of magnetic field in diffuse medium. The technique is especially interesting for probing magnetic field direction near bright sources.

\section{DESCRIPTION OF COMPRESSIBLE MHD TURBULENCE}
\label{app:mhdturb}

\subsection{General}

Here we briefly summarize the scaling laws {\it in the local frame of reference} for compressible MHD turbulence as we did in \cite{2018ApJ...865...46L}. If the energy is injected {  with the injection velocity $V_L$ that is} less than {  the} Alfv\'en speed {  $V_A$}, the turbulence is {\it sub-Alfv\'enic}. In the opposite case it is {\it super-Alfv\'enic}. { The illustration of turbulence scaling for different regimes can be found in Table \ref{tab:regimes}. We briefly describe the regimes below. A more extensive discussion can be found in the review by \cite{BL13} or Beresnyak \& Lazarian (2019) monograph.  

\begin{table*}[t]
\caption{Regimes and ranges of MHD turbulence. \label{tab:regimes}}
\centering
\begin{tabular}{lllll}
\hline
\hline
Type                        & Injection                                                 &  Range   & Motion & Ways\\
of MHD turbulence  & velocity                                                   & of scales & type         & of study\\
\hline
Weak                       & $V_L<V_A$ & $[L_{inj}, l_{trans}]$          & wave-like & analytical\\
\hline
Strong                      &                      &                                        &                 &                \\
sub-Alfv\'{e}nic            &  $V_L<V_A$ & $[l_{trans}, l_{diss}]$ & eddy-like & numerical \\
\hline
Strong                    &                        &                                          &                 &                   \\
super-Alfv\'{e}nic       & $V_L > V_A$ & $[l_A, l_{diss}]$                    & eddy-like & numerical \\
\hline
& & & \\
\multicolumn{5}{l}{\footnotesize{$L_{inj}$ and $l_{diss}$ are injection and dissipation scales, respectively}}\\
\multicolumn{5}{l}{\footnotesize{$M_A\equiv u_L/V_A$, $l_{trans}=L_{inj}M_A^2$ for $M_A<1$ and $l_A=L_{inj}M_A^{-3}$ for $M_A>1$. }}\\
\end{tabular}
\end{table*}
}

{   \subsection{Sub-Alfv\'enic Turbulence}} 

In the case the Alfv\'enic Mach number $M_A=V_L/V_A < 1$. The turbulence in the range from the injection scale {  $L_{inj}$} to the transition scale 
\begin{equation}
l_{trans}=L_{inj}M_A^2
\label{trans}
\end{equation}
is termed the weak Alfv\'enic turbulence. This type of turbulence keeps the $l_{\|}$ scale stays the same while the velocities change as {  $v_\perp\approx V_L (l_\perp/L_{inj})^{1/2}$} \citep{LV99} The cascading results in the change of the perpendicular scale of eddies {  $l_\perp$} only.  With the decrease of $l_\perp$ the turbulent velocities  {$v_\perp$} decreases. Nevertheless, the strength of non-linear interactions of Alfv\'enic wave packets increases (see \citealt{L16}). Eventually, at the scale $l_{trans}$,  the turbulence turns into the strong regime which obeys the GS95 critical balance. 

The situations when the $l_{trans}$ is less than the turbulence dissipation scale $l_{diss}$ require $M_A$ that is unrealistically small for the typical ISM conditions. Therefore, typically the ISM turbulence transits to the strong regime. If the telescope resolution is enough to resolve scales less than $l_{trans}$ then we should observe the signature of strong turbulence in observation.

The anisotropy of the eddies for sub-Alfv\'enic turbulence is larger than in the case of trans-Alfv\'enic turbulence described by GS95. The following expression was derived in LV99:
\begin{equation}
\label{anis}
l_{\|}\approx L_{inj}\left(\frac{{  l_\perp}}{L_{inj}}\right)^{2/3} M_A^{-4/3}
\end{equation}
where $l_{\|}$ and $l\perp$ are given in the local system of reference. For $M_A=1$ one returns to the GS95 scaling.
The turbulent motions at scales less than $l_{trans}$ obey:
\begin{equation}
{  v_\perp}=V_L \left(\frac{{  l_\perp}}{L_{inj}}\right)^{1/3} M_A^{1/3},
\label{vel_strong}
\end{equation}
i.e. they demonstrate Kolmogorov-type cascade perpendicular to {  local} magnetic field. For the magnetic field it is more natural to rewrite this expression as
\begin{equation}
 b_\perp=B_0 \left(\frac{{  l_\perp}}{L_{inj}}\right)^{1/3} M_A^{4/3},
\label{b_strong}
\end{equation}
 where the value of mean field $B_0$ enters explicitly. 

{  In the range of $[L_{inj}, l_{trans}]$ the direction of magnetic field is weakly perturbed and the local and global system of reference are identical. Therefore the velocity gradients calculated at scales larger than $l_{trans}$ are perpendicular to the large scale magnetic field.} {  While at} scales smaller than $l_{trans}$ the velocity gradients follow the direction of the local magnetic fields, similar to the case of trans-Alfv\'enic turbulence that we discuss in the main text.

Turbulent models are characterized by an inertial range over which power-spectrum is power-law. The inertial scale
is bounded by the injection scale $L_{inj}$ above which the spectrum is assumed to sufficiently quickly flatten out,
or even decrease, and short dissipation scale $L_{dis}$ below which there is a fast cutoff of the power. For sub-Alfv\'enic turbulence, characterized by Alfv\'enic Mach number $M_A < 1$, magnetic field correlation length is similar to injection scale,
however the cascade has two regimes, of weak turbulence between $L_{inj}$ and $L_{trans} \approx L_{inj} M_A^2$ and that of strong turbulence, for scales between
$L_{trans}$ and $L_{dis}$.
For super-Alfv\'enic case, $M_A > 1$, correlation scale is shorter 
$L_{A} \approx L_{inj} M_A^{-3}$, and the magnetic field spectrum is flattened before reaching $L_{inj}$.

{\subsection{Super-Alfv\'enic Turbulence}}
\label{superAlf}

If $V_L>V_A$, at large scales magnetic back-reaction is not important and up to the scale (see Lazarian 2006):
\begin{equation}
l_A=L_{inj}M_A^{-3},
\label{la}
\end{equation}
the turbulent cascade is essentially hydrodynamic Kolmogorov cascade. At the scale $l_A$, the turbulence transfers to the trans-Alfv\'enic turbulence described by GS95 scalings, i.e. anisotropy of turbulent eddies start to occur at scales smaller than $l_A$.

\section{STATISTICAL VARIATIONS IN THE CASE OF SMALL DATA SETS}
\label{sec:variations}

\begin{figure*}[th]
\centering
\includegraphics[width=0.90\textwidth]{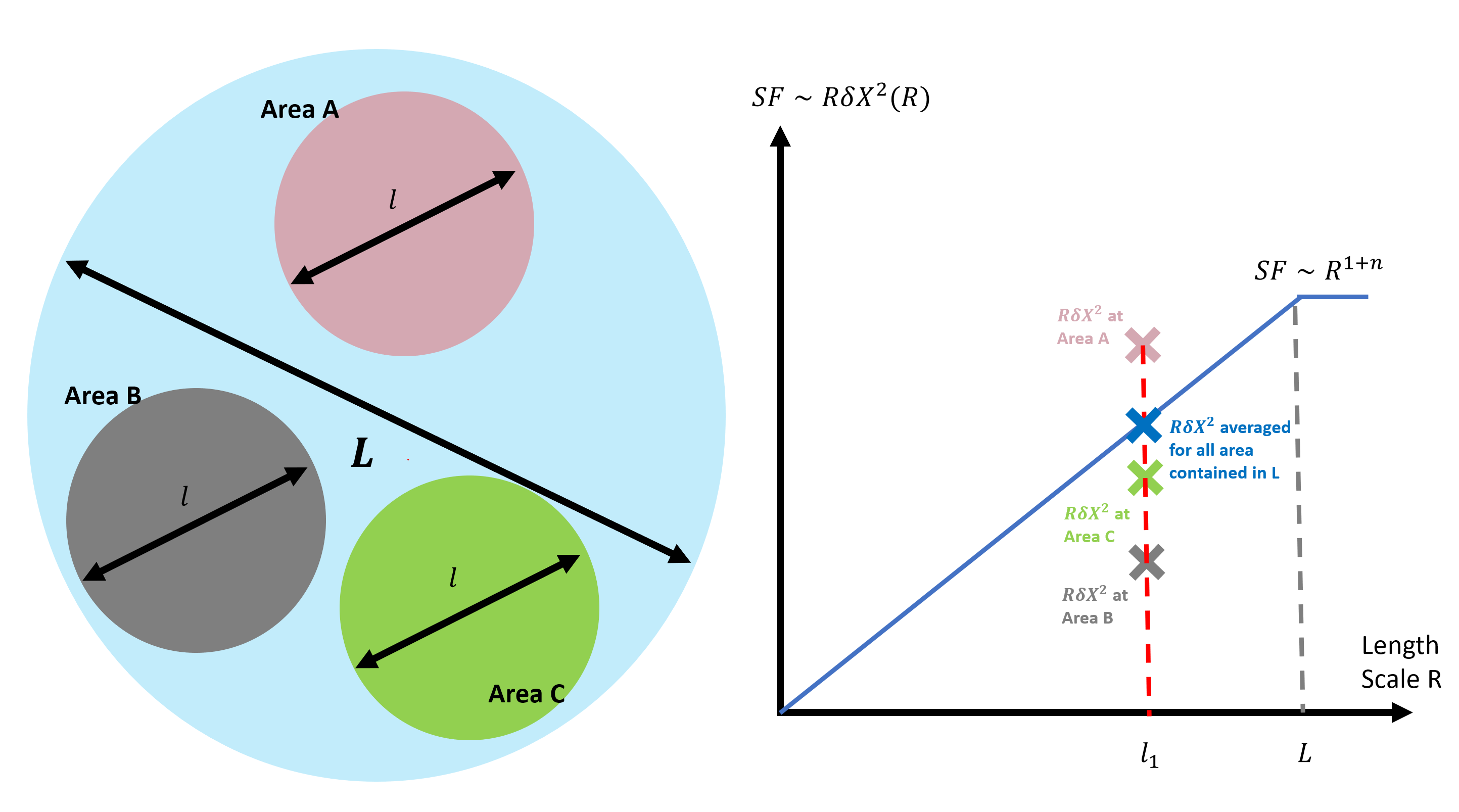}
\caption{\label{fig:statistical_condition} An illustrative figure showing how the dispersion values computed in localized regions with size $l$ is related to the values of the structure function at $R=l$.}
\end{figure*}

The two point global structure function at $R=l$ itself represents the dispersion of the observable with block size of $l$ during computation averaged across the maps with length $L$, which we draw an illustrative figure in Fig.\ref{fig:statistical_condition}. Suppose we have an observable $X$ within the circular area $L$ and we sample the dispersion of $X$ by selecting a smaller area with length scale $l$ ignoring all other interfering factors (e.g. shot noise, dissipation ranges etc, telescope beams smoothing). The individual dispersion values of $X$ within the length scale $l$ will not coincide with the structure function at $R=l$. However, if we take adequate averaging of the areas with $R=l$, the value of such averaging will return to the values as drawn in structure function (blue point).

What do we meany by "adequate" here? Beresnyak \& Lazarian (2005) considered the structure function calculations by selecting a partial number of statistics in simulations, and illustrate that the "adequacy" in recovering the structure functions is just a few $\%$ of the data. \cite{2018ApJ...865...54Y} further showed that even punching $\sim 50\%$ of the data the anisotropy of structure function  would not be altered. This means that {\it potentially} we can divide our map into sub-regions and compute the structure functions in obtaining a distribution of values of $\delta X$ for magnetic field strength estimations.

\section{NUMERICAL TESTING}
\label{sec:method}

\begin{deluxetable}{c c c c c}
\tablecaption{\label{tab:sim} Description of MHD simulation cubes {  which some of them have been used in the series of papers about VGT \citep{YL17a,YL17b,LY18a,LY18b}}.  $M_s$ and $M_A$ are the R.M.S values at each the snapshots are taken. The incompressible simulation is the same as used in Lazarian et.al 2017.   }
\tablehead{Model & $M_S$ & $M_A$ & $\beta=2M_A^2/M_S^2$ & Resolution }
\startdata \hline
b11                     & 0.41  & 0.04 & 0.02 & $480^3$ \\
b12                     & 0.92  & 0.09 & 0.02 & $480^3$ \\
b13                     & 1.95  & 0.18 & 0.02 & $480^3$ \\
b14                     & 3.88  & 0.35 & 0.02 & $480^3$ \\
b15                     & 7.14  & 0.66 & 0.02 & $480^3$ \\ 
b21                     & 0.47  & 0.15 & 0.22 & $480^3$ \\
b22                     & 0.98  & 0.32 & 0.22 & $480^3$ \\
b23                     & 1.92  & 0.59 & 0.22 & $480^3$ \\ 
b31                     & 0.48  & 0.48 & 2.0 & $480^3$ \\
b32                     & 0.93  & 0.94 & 2.0 & $480^3$ \\ 
b41                    & 0.16  & 0.49 & 18 & $480^3$ \\
b42                     & 0.34  & 1.11 & 18 & $480^3$ \\ 
b51                     & 0.05  & 0.52 & 200 & $480^3$ \\
b52                     & 0.10  & 1.08 & 200 & $480^3$ \\ \hline
huge-0                  & 6.17  & 0.22 & 0.0025 & $792^3$ \\
huge-1                  & 5.65  & 0.42 & 0.011 & $792^3$ \\
huge-2                  & 5.81  & 0.61 & 0.022 & $792^3$ \\
huge-3                  & 5.66  & 0.82 & 0.042 & $792^3$ \\
huge-4                  & 5.62  & 1.01 & 0.065 & $792^3$ \\
huge-5                  & 5.63  & 1.19 & 0.089 & $792^3$ \\
huge-6                  & 5.70  & 1.38 & 0.12 & $792^3$ \\
huge-7                  & 5.56  & 1.55 & 0.16 & $792^3$ \\
huge-8                  & 5.50  & 1.67 & 0.18 & $792^3$ \\
huge-9                  & 5.39  & 1.71 & 0.20 & $792^3$ \\ \hline
h0-1200                 & 6.36  & 0.22 & 0.0025 & $1200^3$ \\
h9-1200                 & 10.79 & 2.29 & 0.098 & $1200^3$ \\
e5r2                    & 0.13  & 5.99 & 4363 & $1200^3$ \\
e5r3                    & 0.61  & 0.63 & 2.09 & $1200^3$ \\
e6r3                    & 5.45  & 0.50 & 0.017 & $1200^3$ \\
e7r3                    & 0.53  & 3.24 & 73.64 & $1200^3$ \\\hline
Ms0.2Ma0.2              & 0.2   & 0.2  & 2 & $480^3$ \\
Ms0.4Ma0.2              & 0.57   & 0.28  & 0.48 & $480^3$ \\ 
Ms4.0Ma0.2              & 3.81   & 0.18  & 0.00446 & $480^3$ \\
Ms20.0Ma0.2             & 20.59  & 0.18  & 0.00015 & $480^3$ \\ \hline 
incompressible & 0 & 0.61 & $\infty$ & $512^3$\\
 \hline\hline
\enddata
\end{deluxetable}
Most of the numerical data cubes {\re are} obtained by 3D MHD simulations that is from a single fluid, operator-split, staggered grid MHD Eulerian code ZEUS-MP/3D to set up a three dimensional, uniform, isothermal turbulent medium. To simulate the part of the interstellar cloud, periodic boundary conditions are applied. These simulations use the Fourier-space forced driving solenoidal driving.\footnote{Our choice of force stirring over the other popular choice, i.e. of the decaying turbulence, is preferable because only the former exhibits the full characteristics of turbulence statistics, e.g power law, turbulence anisotropy, extended from $k=2$ to {\re a dissipation scale of $12$ pixels}  in a simulation , and matches with what we see in observations (e.g. \citealt{1995ApJ...443..209A,2010ApJ...710..853C}). } For isothermal MHD simulation without gravity, the simulations are scale-free. If $V_{inj}$ is the injection velocity, while $V_A$ and $V_s$ are the Alfv\'en and sonic velocities respectively, then the two parameters, namely, the Alfv\'en Mach numbers $M_A=V_{inj}/V_A$ and sonic Mach numbers $M_s=V_{inj}/V_s$, determine all properties of the numerical cubes and the resultant simulation is universal in the inertial range. That means one can easily transform to any arbitrary units as long as the dimensionless parameters $M_A,M_s$ are not changed. The chosen $M_A$ and $M_s$ are listed in Table \ref{tab:sim}. For the case of $M_A<M_s$, it corresponds to the simulations of turbulent plasma with thermal pressure smaller than the magnetic pressure, i.e. plasma with $\beta/2=V_s^2/V_A^2<1$. In contrast, the case that is $M_A>M_s$ corresponds to the {\torefereeone thermal} pressure dominated plasma with $\beta/2>1$. To investigate the behavior of the incompressible case, we adopt the incompressible cube our previous work \cite{Letal17}. Further we refer to the simulations in Table \ref{tab:sim} by their model name. The selected ranges of $M_s, M_A, \beta$ are determined by possible scenarios of astrophysical turbulence from subsonic to supersonic cases.  

As our derivations employed the properties of velocities and magnetic field, we first deal with the numerical testing of our predictions without accounting for the effects of density fluctuations. This is possible in observational applications due to the development of \cite{VDA}. \cite{VDA} allows the separation of the density fluctuations from channel and centroid maps, meaning that all constant density approximations in the series of \citeauthor{LP00} papers are now applicable to the maps after the applications of \cite{VDA}.

\section{TURBULENT SPECTRUM AND THE STRUCTURE FUNCTION}
\label{sec:statistics_LP}

\subsection{Turbulent Velocities}
Two point statistics of general random vector field, such as the turbulent velocity field,
can described by the structure function that in Fourier space 
is given by three power-spectra, correspondent to orthogonal motions of the medium that can
be considered statistically independent
\begin{equation}
\begin{aligned}
D^v_{ij}(\mathbf{r}) \equiv \left\langle \left( v_i(\mathbf{x}_1) - v_i(\mathbf{x}_2) \right)
\right. & \left . \left( v_j(\mathbf{x}_1) - v_j(\mathbf{x}_2) \right) \right\rangle = \\
& = \frac{2}{(2\pi)^3} \! \int \!\! d^3 k \left( 1 - e^{i \mathbf{k} \cdot \mathbf{r}}\right)
\left[ \mathcal{A}^v(\mathbf{k}) \widehat{\xi}^{\mathcal{A}}_i \widehat{\xi}^{\mathcal{A}}_j + 
\mathcal{S}^v(\mathbf{k}) \widehat{\xi}^{\mathcal{S}}_i \widehat{\xi}^{\mathcal{S}}_j + \mathcal{F}^v(\mathbf{k}) \widehat{\xi}^{\mathcal{F}}_i \widehat{\xi}^{\mathcal{F}}_j 
\right]
\label{eq:app_Dv}
\end{aligned}
\end{equation}
Which orientation of the orthogonal  triad $\widehat{\xi}^{\mathcal{A}}, \widehat{\xi}^{\mathcal{S}}, \widehat{\xi}^{\mathcal{F}}$  of unit vectors describes independent motions depends on the
physical properties of the medium and velocity excitations.
In the magnetized turbulence three independent modes are identified to be Alfv\'en, Slow, and
Fast (which explains our notation with $\mathcal{A},\mathcal{S}$ and $\mathcal F$ calligraphic labels, see Table.\ref{tab:notations}) which have particular orientations w.r.t. to the direction of the mean magnetic field $\hat{\lambda}$.
MHD motions are statistically anisotropic, with both the mode decomposition dependent
structurally on the direction of the magnetic field, and power spectra, generally,
dependent on the relative angle of the wave vector with the mean field axis.

In the high-$\beta$ plasma,  $\beta \gg 1$,  the Alfv\'en, Slow, Fast decomposition is
\begin{equation}
\widehat{\xi}^{\mathcal{A}}  =  \frac{ \hat{k} \times \hat{\lambda}}{\sqrt{1-\mu^2}} \equiv \widehat{\xi}^{A}, \quad
\widehat{\xi}^{S}  =  \frac{ \hat {\lambda} - \mu \hat{k}}{\sqrt{1-\mu^2}} \equiv \widehat{\xi}^F ~,
\quad
\widehat{\xi}^{\mathcal{F}} =  \hat{k} \equiv \widehat{\xi}^P  ~, \quad
\mathrm{where} ~ \mu \equiv \hat k \cdot \hat\lambda 
\label{eq:highbeta_xi}
\end{equation}
and coincides with $A,F,P$ decomposition into two solenoidal and one potential modes introduced
in the main text.
Namely, in this regime Fast motions are purely potential, while Slow mode together with
Alfv\'en mode are two components of the solenoidal motion. 

For plasma $\beta$ finite, and, in particularly in low-$\beta$ plasma, Alfv\'en modes remain
the same,  but Slow and Fast directions are rotated relative to $\widehat{\xi}^F$ and $\widehat{\xi}^P$
\begin{equation}
\widehat{\xi}^{\mathcal{A}}  =  \frac{ \hat{k} \times \hat{\lambda}}{\sqrt{1-\mu^2}}, \quad
\widehat{\xi}^{\mathcal{S}}  =  \cos(\alpha_\beta) \widehat{\xi}^F + \sin(\alpha_\beta) \widehat{\xi}^P
\quad
\widehat{\xi}^{\mathcal{F}} =  \cos(\alpha_\beta) \widehat{\xi}^P - \sin(\alpha_\beta) \widehat{\xi}^F
\end{equation}
where the rotation angle $\alpha_\beta$ depends on the plasma constant in such a way that
$\alpha_\beta \to 0$ as $\beta \to \infty $ and $\alpha_\beta \to \mathrm{arcsin}(\mu) $ as $  \beta \to 0 $.
The following formula gives a qualitatively correct fit
\begin{equation}
\alpha_\beta = \frac{2}{\pi} \mathrm{arcctg}(\beta) \mathrm{arcsin}(\mu) 
\end{equation}

In the low-$\beta$ plasma,  $\beta \to 0$,  the Alfv\'en, Slow, Fast decomposition is then
\begin{equation}
\widehat{\xi}^{\mathcal{A}}  =  \frac{ \hat{k} \times \hat{\lambda}}{\sqrt{1-\mu^2}}, \quad
\widehat{\xi}^{\mathcal{S}}  =    \sqrt{1-\mu^2} \; \widehat{\xi}^F +  \mu \; \widehat{\xi}^P 
= \hat {\lambda} ~,
\quad
\widehat{\xi}^{\mathcal{F}} =  
 \sqrt{1-\mu^2} \; \widehat{\xi}^P -\mu  \; \widehat{\xi}^F =
\frac{ \hat {k} - \mu \hat{\lambda}}{\sqrt{1-\mu^2}} 
\label{eq:lowbeta_xi}
\end{equation}

\subsection{Magnetic Field Turbulent Fluctuations}
Properties of the magnetic field fluctuations $\delta B$ are linked to medium velocities $\mathbf v$  in MHD turbulence by the field frozen condition
\begin{equation}
\delta \mathbf{B}_\mathbf{k} \propto \mathbf{k} \times ( \mathbf{v}_\mathbf{k} \times \overline{\mathbf{B}} ) / \omega(\mathbf{k})
\label{eq:frozen}
\end{equation}
where the wave frequency $\omega(k) \propto \mathbf{k} \cdot \hat \lambda$ for Alfv\'en and Slow modes. For the fast modes,  $\omega(k) \propto k$ and is angle independent.

Magnetic field is solenoidal in nature and contains only two degrees of freedom which 
are conveniently described via $A$-type $\hat\xi^A$ and $F$-type $\widehat{\xi}^F$ 
contributions, as in the following explicit expression
\footnote{This representation  is superior to $E(=F+A)$ and $F$ decomposition that was used
in LP12, since the isotropic  $E(\mathbf{k}) \left( \delta_ij - \hat{k}_i \hat{k}_j\right)$
spectral part of LP12  describes two degrees of freedom, while $F$ part
also describes one, although there are only two physical degrees of freedom. Here in Eq.~(\ref{eq:app_Dij})
we have a clean separation of two physical degrees of freedom of a solenoidal field in the presence of
a preferred axis.  Each, $A$ and $F$, tensor has only one eigenvector with non-zero eigenvalue, describing
one of the two directions of the perturbed field. Algebraically, though, both representations are equivalent. LP12 approach has advantages for studying effects of symmetry axis wandering, which leads
to isotropization of the correlations.}
\begin{eqnarray}
D_{ij}(\mathbf{r})  && \equiv \left\langle \left( B_i(\mathbf{x}_1) - B_i(\mathbf{x}_2) \right)  \left( B_j(\mathbf{x}_1) - B_j(\mathbf{x}_2) \right) \right\rangle 
 = \\
&& = \frac{2}{(2\pi)^3} \! \int \!\! d^3 \mathbf{k}
 \left( 1 - e^{i \mathbf{k} \cdot \mathbf{r}}\right) 
 \left[ A(k,\mu) \! 
\left( \frac{ (\hat k \times \hat \lambda)_i (\hat k \times \hat \lambda)_j}{1-\mu^2}
\right) 
+ F(k,\mu)
\left(\frac{ (\hat{\lambda}_i - \mu \hat{k}_i) (\hat{\lambda}_j - \mu \hat{k}_j)}{1 - \mu^2}
\right) 
 \right]
 \nonumber
\label{eq:app_Dij}
\end{eqnarray}

The frozen condition Eq.~(\ref{eq:frozen}) projects three, $\mathcal{A},\mathcal{S},\mathcal{F}$
modes of motion into two types of perturbations of the magnetic field, $A$ and $F$.
While {\torefereetwo Alfv\'en} motions $\mathcal{A}$ give rise directly to $A$-type perturbations of $B$-field,
both Slow and Fast motions cause $F$-type magnetic perturbations only.  This is true for
all values of plasma $\beta$, but what relative contributions of Slow and Fast modes
are in $F$-type $\delta \mathrm{\mathbf{B}}$, depends on $\beta$.

Namely, in the high-$\beta$ plasma,  $\beta \gg 1$, we obtain, using  Eqs.~(\ref{eq:frozen})
and (\ref{eq:highbeta_xi}) 
\begin{equation}
A(k,\mu) \propto \mathcal{A}^v(k,\mu)~, \quad F(k,\mu) \propto \mathcal{S}^v(k, \mu) + 
(1-\mu^2) \mathcal{F}^v(k,\mu) (v_A/v_\mathcal{F})^2
\end{equation}
where $v_A$ is Alfv\'en velocity, which in the regime is also the speed of slow modes, while
$v_\mathcal{F}$ is the speed of fast modes.

In the low-$\beta$ regime Slow modes do not perturb the magnetic field 
\begin{equation}
A(k,\mu) \propto \mathcal{A}^v(k,\mu)~, \quad F(k,\mu) \propto   \mathcal{F}^v(k,\mu)
\end{equation}

An important case of high-$\beta$ turbulence is that of strong, nearly incompressible
turbulence, when $\mathcal{A}^v(k,\mu)=\mathcal{S}^v(k,\mu)$ and the magnetic perturbations arising from fast modes are negligible.\footnote{Fast modes in this regime are very similar to sound waves.}
In this case the turbulence is described by a single power spectrum with isotropic 
tensor structure
\begin{equation}
D_{ij}^v (\mathbf{r}) \propto D_{ij}(\mathbf{r}) =  \frac{2}{(2\pi)^3} \! \int \!\! d^3 \mathbf{k}
 \left( 1 - e^{i \mathbf{k} \cdot \mathbf{r}}\right) 
  A(k,\mu) \left( \delta_{ij} - \hat{k}_i \hat{k}_j
 \right)
\label{eq:app_Dij_I}
\end{equation}
where Alfv\'en and Slow modes give two polarizations of a transverse solenoidal wave.

The variance of the magnetic field fluctuations can be obtained as half of the trace of the structure function at large separations $\left\langle \delta B^2 \right\rangle = \frac{1}{2} \sum_i D_{ii}(r \to \infty)$ which gives
\begin{equation}
\left\langle \delta B^2 \right\rangle =
\frac{1}{(2\pi)^3} \int \!\! d^3 k \left[ A(k,\mathbf{\hat k} \cdot \hat \lambda) +  F(k,\mathbf{\hat k} \cdot \hat \lambda) \right]
= \frac{1}{2 \pi^2} \int \!\! k^2 dk \left[ A_0(k) +  F_0(k) \right]
\end{equation}
where $A_0(k)$ is the monopole of the power spectrum 
$A$, $A_0=\frac{1}{4\pi} \int d\Omega_k A(k,\mathbf{\hat k} \cdot \hat \lambda)$, and, correspondingly, $F_0(k)$ is the monopole of the spectrum $F$. 

\subsection{2D Projected Structure Functions}
\label{sec:Appendix_Dyy} 

The projected 2D  structure function is obtained by integration over LOS $z$ direction,
\begin{equation}
\widetilde{D}_{ij}(\mathbf{R}) = \int dz \left( D_{ij}(\mathbf{R},z) - D_{ij}(0,z) \right)=
\frac{2}{(2\pi)^3} \int d^3 k \left( 1 - e^{i \mathbf{K} \cdot \mathbf{R}} \right) \int dz e^{i k_z z} \left[ \cdots \right] ~,
\end{equation}
where 2D vectors orthogonal to LOS are introduced in capitalized notation as $\mathbf{k}=(\mathbf{K},k_z)$ and $\mathbf{r}=(\mathbf{R},z)$.
The integral over $z$ translates into
$\delta^D(k_z)$-like behaviour if the LOS integration range exceeds the POS scale $R$
under study. In this
approximation, subsequent integration over $k_z$ amounts to setting $k_z=0$, and the result
of LOS projection can be obtained from
Eq.~(\ref{eq:app_Dij}) by substituting  $\mathbf{k} \to (\mathbf{K},0)$, including $\mathbf{\hat{k}}
\to (\mathbf{\hat{K},0)}$, and $\hat{\lambda}=(\sin\gamma \hat\Lambda, \cos\gamma)$ where $\hat{\Lambda}$ 
is the 2D direction of the  mean field on the sky.  Under this transformation $\mu = \sin\gamma \cos\phi_K$ where $\cos\phi_K = \mathbf{\hat K} \cdot \hat \Lambda$ is the cosine of the angle between
POS projections of the wavevector and the symmetry axis. 

We will give the result for the magnetic field explicitly, in $A,F$ decompositions. For POS components ($i,j=x,y)$
\begin{align}
\widetilde{D}_{ij}(\mathbf{R}) = & \frac{1}{2\pi^2} \! \int \!\! d^2 K \left( 1 - e^{i \mathbf{K} \cdot \mathbf{R}}\right) 
\left[ \vphantom{\frac{(\mathbf{\hat K} \cdot \hat \Lambda)^2 \hat{K}_i  \hat{K}_j + \hat{\Lambda}_i \hat{\Lambda}_j - (\mathbf{\hat K} \cdot \hat \Lambda)(\hat{K}_i \hat{\Lambda}_j +   \hat{K}_j \hat{\Lambda}_i)}
{1 - \sin^2\gamma (\mathbf{\hat K} \cdot \hat \Lambda)^2} }
A(K, \sin\gamma \cos\phi_K) \!
\left( \delta_{ij} - 
\frac{\hat{K}_i \hat{K}_j + \sin^2\gamma \hat{\Lambda}_i \hat{\Lambda}_j - \sin^2\gamma \cos\phi_K (\hat{K}_i\hat{\Lambda}_j + \hat{K}_j\hat{\Lambda}_i)}
{1 - \sin^2\gamma\cos^2\phi_K}
\right)  + \right.
\nonumber \\
& \left. +  F(K,\sin\gamma\cos\phi_K)  \left( -\hat{K}_i \hat{K}_j +
\frac{\hat{K}_i \hat{K}_j + \sin^2\gamma \hat{\Lambda}_i \hat{\Lambda}_j - \sin^2\gamma \cos\phi_K (\hat{K}_i\hat{\Lambda}_j + \hat{K}_j\hat{\Lambda}_i)}
{1 - \sin^2\gamma \cos^2\phi_K}
\right) \right]
\label{eq:app_tildeDij}
\end{align}

For this paper we are interested in the structure function of the perturbations of the magnetic field component
perpendicular to the mean field.  Considering projected mean field to be along the $x$-direction, $\hat\Lambda=(1,0)$, this is the y-component for which
\begin{align}
\widetilde{D}_{yy}(\mathbf{R}) = & \frac{1}{2\pi^2} \! \int \!\! d^2 K \left( 1 - e^{i \mathbf{K} \cdot \mathbf{R}}\right) 
\left[ 
A(K,\sin\gamma \cos\phi_K) \frac{ \cos^2\gamma \cos^2\phi_K}{1 - \sin^2\gamma \cos^2\phi_K}  +  \; F(K,\sin\gamma \cos\phi_K)
\frac{\sin^2\gamma \cos^2\phi_K \sin^2\phi_K }{1 - \sin^2\gamma \cos^2\phi_K} \right]
\label{eq:app_tildeDyy}
\end{align}
For completeness we shall also quote
\begin{align}
\widetilde{D}_{xx}(\mathbf{R}) = & \frac{1}{2\pi^2} \! \int \!\! d^2 K \left( 1 - e^{i \mathbf{K} \cdot \mathbf{R}}\right) 
\left[ 
A(K,\sin\gamma \cos\phi_K) \frac{ \cos^2\gamma \sin^2\phi_K}{1 - \sin^2\gamma \cos^2\phi_K}  +  \; F(K,\sin\gamma \cos\phi_K)
\frac{\sin^2\gamma \sin^4\phi_K }{1 - \sin^2\gamma \cos^2\phi_K} \right]
\label{eq:app_tildeDxx} \\
\widetilde{D}_{zz}(\mathbf{R}) = &  \frac{1}{2\pi^2} \! \int \!\! d^2 K \left( 1 - e^{i \mathbf{K} \cdot \mathbf{R}}\right) 
\left[ 
A(K,\sin\gamma \; \cos\phi_K) +  F(K,\sin\gamma\; \cos\phi_K)
\frac{\cos^2\gamma}{1 - \sin^2\gamma \cos^2 \phi_K} \right]
\label{eq:app_tildeDzz}
\end{align}

In what follows we shall consider two specific models of the turbulence.
One of them is model (I) of strong incompressible turbulence where Alfv\'en and slow modes are equally distributed.
This corresponds to $A(\mathbf{k})=F(\mathbf{k})$. The other, model (A), is a pure Alfv\'enic turbulence $F(\mathbf{k})=0$. In addition, we also give the expressions for $A(\mathbf{k})=0$ case (F) which
represents magnetic field perturbations in both slow and fast modes in high and low $\beta$ regimes.
To simplify notation, let us use $E(\mathbf{k})$ to designate the power spectrum generally,
without specific reference to one of the modes. Then we can write
\begin{align}
\widetilde{D}_{yy}(\mathbf{R}) = & \frac{1}{2 \pi^2} \! \int \!\! d^2 K \left( 1 - e^{i \mathbf{K} \cdot \mathbf{R}}\right) E(K,\sin\gamma \cos\phi_K) G^{(I,A,F)}(\gamma, \phi_K) ~,~ \mathrm{where} ~
\left\{
\begin{array}{lll}
G^I &=& \cos^2\phi_K \\
G^A &=& \frac{\cos^2\gamma \; \cos^2\phi_K }{1-\sin^2\gamma\cos^2\phi_K}\\
G^F &=& \frac{\sin^2\gamma \; \cos^2\phi_K \sin^2 \phi_K}{1-\sin^2\gamma\cos^2\phi_K}
\end{array}
\right.
\label{eq:app_tildeDijIA}
\end{align}
One can point out that the model of strong turbulence utilizes all two degrees of freedom that 
general representation allows for, while the purely Alfv\'enic model is degenerate, using only one degree of freedom. This leads to some artifacts in mathematical approximations, as for instance in our idealized axis-symmetric approximation  the vanishing of the projected structure function if $\cos\gamma=0$, i.e when the mean magnetic field is exactly perpendicular to LOS. In reality this means that in such particular configuration Alfv\'enic mode will give subdominant contribution.

We shall focus on the multipole coefficients $\widetilde{D}_{yy}^n (R) = 
\frac{1}{2\pi} \int_0^{2\pi} d\phi_R  \widetilde{D}_{yy}(R,\phi_R) e^{-i n \phi_R}$ for which we obtain, after performing integration
over $\phi_K$
\begin{align}
\widetilde{D}_{yy}^n (R) = & \frac{1}{\pi} \! \int \!\! K d K \left( \delta_{n0} - i^n J_n(K R) \right) 
\sum_{p=-\infty}^\infty E_p^{2D} (K,\sin\gamma)\; G_{n-p}^{(I,A,F)}(\gamma) ~,
\label{eq:app_tildeDij0IA}
\end{align}
where $E_n^{2D} (K,\sin\gamma)$ are coefficients of 2D multipole expansion on the sky of the projected $k_z=0$
power spectrum (see \citealt{LP12,KLP17b}) with respect to angle $\phi_K$, while $G_n^{(I,A,F)}(\gamma)$ are similar multipoles for geometrical functions in Eq.~(\ref{eq:app_tildeDijIA}), with both sets of coefficients potentially dependent on $\sin\gamma$. Decomposition of the geometrical factor is particularly simple in the strong incompressible case, 
$G_n^{I} =\frac{1}{2} \delta_{n0} + \frac{1}{4} \left(\delta_{2n}+\delta_{-2n}\right)$.

In the following Section~\ref{sec:E_anistropy} we discuss that 
the level of anisotropy of the power spectrum can be often considered
scale-independent, i.e  the dependence of the power on the direction of the wave-vector can be separated as
\begin{equation}
E(\mathbf{k}) = E_0(k) \widehat{E}(\widehat{\mathbf{k}} \cdot \widehat{\lambda})
, \quad \frac{1}{4\pi} \int \widehat{E}(\widehat{\mathbf{k}} \cdot \widehat{\lambda}) 
d \Omega_{\widehat{\mathbf{k}}} = 1
\end{equation}

With this factorization, 
Eq.~(\ref{eq:app_tildeDijIA}) can then be presented in a general form
\begin{equation}
\widetilde{D}_{yy}^n (R) \approx  \left\langle \delta B^2 \right\rangle
\mathcal{I}_n(R)
\sum_{p=-\infty}^\infty \widehat{E}_p^{2D} (\sin\gamma)\; G_{n-p}^{(I,A,F)}(\gamma)
\label{eq:app_tildeDRgen}
\end{equation}
where the formal expression for the scaling functions $\mathcal{I}_n(R)$ is
\begin{equation}
\mathcal{I}_n(R) = \frac{\pi \int \! K d K \left( \delta_{n0} - i^n J_n(K R) \right)E_0(K) }
{ \int \! k^2 dk E_0(k)}
\label{eq:scalingfcn}
\end{equation}

\subsection{2D Structure Function of LOS Turbulent Velocities}
\label{sec:app_turbv}

In velocity centroid studies (see \citealt{KLP17a} for details) one looks at the projection of the line-of-sight component of the velocity, $ \int dz v_z(z) $. In this projection the potential mode vanishes, so the relevant 2D structure function has the form
\begin{equation}
\begin{aligned}
\widetilde{D}^v_{zz}(\mathbf{R}) & =   \frac{1}{2\pi^2} \! \int \!\! d^2 K \left( 1 - e^{i \mathbf{K} \cdot \mathbf{R}}\right) \times \\
& \times \left[ 
A^v(K,\sin\gamma\cos\phi_K) \left(1 - \frac{\cos^2\gamma}{1 - \sin^2\gamma \cos^2\phi_K}\right)
+  F^v(K,\sin\gamma\cos\phi_K)
\frac{\cos^2\gamma}{1 - \sin^2\gamma \cos^2\phi_K} \right]
\label{eq:app_tildeDzzv}
\end{aligned}
\end{equation}

In high-$\beta$ case,  with velocity spectra $A^v=F^v$ as usual for 
strong nearly incompressible regime (I) or, individually, for $A$ and $F$ modes, we get
\begin{align}
\widetilde{D}^v_{zz}(\mathbf{R}) = & \frac{1}{2\pi^2} \! \int \!\! d^2 K \left( 1 - e^{i \mathbf{K} \cdot \mathbf{R}}\right) E^v(K,\sin\gamma \cos\phi_K) \mathcal{W}^{(I,A,F)}(\gamma, \phi_K) ~,~ \mathrm{where} ~
\left\{
\begin{array}{lll}
\mathcal{W}^I &=& 1 \\
\mathcal{W}^A &=& \frac{\sin^2\gamma \sin^2\phi_K}{1-\sin^2\gamma\cos^2\phi_K}\\
\mathcal{W}^F &=& \frac{\cos^2\gamma}{1-\sin^2\gamma\cos^2\phi_K}
\end{array}
\right.
\label{eq:app_tildeDzzIAv}
\end{align}
that differs from Eq.~(\ref{eq:app_tildeDijIA}) only in geometrical functions $\mathcal{W}$.
We note that in LOS projection, the $F$ mode in the high-$\beta$ regime is provided completely 
by the slow mode motions, while the fast mode motions, that are potential, are projected out.

In low-$\beta$ case, while the form Eq.~(\ref{eq:app_tildeDzzIAv}) for the $\widetilde{D}^v_{zz}(\mathbf{R}) $ remains the same, the relevant geometrical functions are the Alfv\'en $\mathcal{W}^A$ that is unchanged from Eq.~(\ref{eq:app_tildeDzzIAv}) and, now separately,
slow and fast mode contributions to the $F$ part 
\begin{equation}
\mathcal{W}^{\mathcal{S}} = (1-\mu^2) \mathcal{W}^F = \cos^2\gamma ~. \quad
\mathcal{W}^{\mathcal{F}} = \mu^2 \mathcal{W}^F = 
\frac{ \cos^2\gamma \sin^2\gamma\cos^2\phi_K}{1-\sin^2\gamma\cos^2\phi_K}
\label{eq:W_lowbeta}
\end{equation}
that could be easily found from Eq.~(\ref{eq:lowbeta_xi}).

Velocities have the same scaling as the magnetic field fluctuations, 
$E^v(\mathbf{k}) \propto E(\mathbf{k})$.
Following the steps of the previous section, the multipole representation of $\widetilde{D}^v_{zz}$ 
is then
\begin{equation}
\widetilde{D}^v_n (R) \approx  \left\langle \delta v^2 \right\rangle
\mathcal{I}_n(R)
\sum_{p=-\infty}^\infty \widehat{E}_p^{2D} (\sin\gamma)\; \mathcal{W}_{n-p}^{(I,A,F,\mathcal{S}, \mathcal{F})}(\gamma)
\label{eq:app_tildeDRvgen}
\end{equation}
with the same $\mathcal{I}_n(R)$ scaling function as in Eq,~(\ref{eq:app_tildeDRgen}).
Here we have dropped the $zz$ label as we are always dealing only with $z$ component of the velocity.  In the strong incompressible case $\mathcal{W}^I_n = \delta_{n0}$.

\subsection{Angular Anisotropy of the Turbulent Power Spectrum}
\label{sec:E_anistropy}

Let us focus on sub-Alfv\'enic case in strong turbulence regime. 
We can model $E(\mathbf{k})$ power spectrum as 
\begin{equation}
E(k,\mu = \widehat{\mathbf{k}}\cdot \widehat{\lambda}) = A_{trans} (k L_{trans})^{-3-m} \widehat{E}(k, \mu), \quad k L_{trans} > 1
\label{eq:app_espec_prelim}
\end{equation}
with $m$ being index of power scaling, and angular dependence factored out in 
$\widehat{E}(k, \mu)$ that is defined normalized to have 3D monopole equal to unity, $\frac{1}{2}\int_{-1}^1 d\mu \widehat{E}(k,\mu)=1$. 
The spectral amplitude $A_{trans}$ at the transition scale can be linked to
the amplitude at injections scale via weak turbulence scaling with index $m_{w} \approx 1$.
\begin{equation}
A_{trans} = A_{inj} (L_{trans}/L_{inj})^{m_w}
\label{eq:AtranstoAinj}
\end{equation}
$A_{inj}$, in turn, determines the variance of the magnetic field
\begin{equation}
\left\langle \delta B^2 \right\rangle
\approx \frac{A_{inj}}{\pi^2 L_{inj}^3 m_w} ~, 
\label{eq:var_B}
\end{equation}
thus
\begin{equation}
E(k,\mu = \widehat{\mathbf{k}}\cdot \widehat{\lambda}) = 
\left\langle \delta B^2 \right\rangle \pi^2 m_w L_{inj}^3  (L_{trans}/L_{inj})^{m_w}
(k L_{trans})^{-3-m} \widehat{E}(k, \mu), \quad k L_{trans} > 1
\label{eq:app_espec}
\end{equation}

The level of anisotropy, 
in particular the quadrupole of $ \widehat{E}(k,\mu)$ can be scale dependent.
The strong turbulence relation Eq.~(\ref{anis}) for $m=2/3$
gives ratio $l_\parallel/l_\perp$ of individual eddies in configuration space to be scale dependent, increasing for smaller ones $\propto l^{-1/3}$. However, CLV02 has
argued that if one considers the ensemble averaged power spectrum in the volume 
$L_c^3 < L_{trans}^3$, the averaging of orientation of individual small eddies results
in the scale independent anisotropy of the spectrum, determined by the shape of the largest eddies at scale $L_c$, which satisfies Eq.~(\ref{anis}). Switching to the angle coordinates of the wavevectors relative to the  magnetic field, so that $k = L_c^{-1}$,  $l_\parallel = k_\parallel^{-1} = L_c/\mu $ and
$l_\perp = k_\perp^{-1} = L_c/\sqrt{1-\mu^2}$, we have
\begin{equation}
\frac{L_c}{\mu}  \approx L_{inj} \left(\frac{L_c}{L_{inj} \sqrt{1-\mu^2} } \right) ^{2/3} M_A^{-4/3} \quad
\Longrightarrow \quad (L_{inj}/L_c)^{1/3} M_A^{-4/3} \frac{\mu}{(1-\mu^2)^{1/3}} \approx 1
\label{eq:strong_ani}
\end{equation}

If $L_c$ is raised to $L_{trans} = L_{inj} M_A^2 $, Eq.~(\ref{eq:strong_ani}) suggests the following form
for the angular dependence of the power spectrum 
\begin{equation}
    \widehat{E}_{L_{trans}}\left(\mu\right) = 
    \frac{ \exp \left[-M_A^{-2} \frac{|\mu|}{(1-\mu^2)^{1/3}}\right]}
    {\int_0^1  \exp \left[-M_A^{-2} \frac{|\mu|}{(1-\mu^2)^{1/3}}\right] d\mu}, \quad k > L_{trans}^{-1}
    \label{eq:hatE_mu}
\end{equation}
This form is valid for each separate $L_{trans}^3$ volume, however each different volume
will have different direction of the volume-averaged magnetic field, i.e., the axis from
which $\mu$ angle is measured.  Variations of this mean axis are determined by
long fluctuating modes $ k < L_{trans}$ which are in the weak turbulence regime.
Global definition of the power spectrum will average over such long-wave 
``wandering'' of the preferred axis and will describe the residual anisotropy of the spectrum with respect to the globally mean magnetic field.

The effect of long wave field wandering can be estimated by averaging the power spectrum
over the distribution of $\widehat{\lambda}$ directions. We will model the distribution of $\lambda$ direction by assuming that the perturbations in the magnetic field 
$\delta \mathbf{b}$ are Gaussian and over large scales 
distributed in 3D isotropically around $\widehat{\lambda}_0$ global mean direction 
with the variance 
$ \langle \delta b^2 \rangle /B_0^2 = M_A^2$,
which gives\footnote{This expression retains qualitative features  but is an improvement on LP12}

\begin{eqnarray}
\overline{\widehat{E}(\widehat{\mathbf{k}} \cdot \widehat{\lambda}_0)} & = & 
 \frac{1}{2 \pi} \int_0^{2\pi} \!\!\!\!\! d \phi_{\widehat{\lambda}} 
 \int_0^{\frac{\pi}{2}} \!\!\!\! \sin\theta_{\widehat{\lambda}} d \theta_{\widehat{\lambda}} 
 \widehat{E}_{L_{trans}}(\widehat{\mathbf{k}} \cdot \widehat{\lambda}) \mathcal{P}(\theta_{\widehat{\lambda}})
\label{eq:E_averaging} \\
\mathcal{P}(\theta_{\widehat{\lambda}}) & \equiv & \left(1 + \frac{2 \cos^2\theta_{\widehat{\lambda}}}{M_A^2} \right)
\exp\left[-\frac{ \sin^2\theta_{\widehat{\lambda}}}{ M_A^2}  \right]
\label{eq:P_lambda}
\end{eqnarray}
where $\theta_{\widehat{\lambda}} \in (0,\pi/2) $ and 
$\phi_{\widehat{\lambda}} \in (0,2\pi)$ are spherical angles
of the headless vector $\widehat{\lambda}$ with respect to $\widehat{\lambda}_0$ axis.
Let us note that in the isotropic limit $M_A \to \infty$, $\langle \cos^2 \theta_{\widehat{\lambda}} \rangle \to 1/3$ with respect to an \textit{arbitrary} axis. In the opposite limit,
$M_A \to 0$, $\mathcal{P}(\theta_{\widehat{\lambda}}) \to \delta_D(\theta_{\widehat{\lambda}})$
and $\widehat{\lambda} \to \widehat{\lambda}_0 $ with $\langle \sin^2(\theta_{\widehat{\lambda}}) \rangle \sim M_A^2$.  For $M_A < 1$, the result
of the averaging can be approximated as
\begin{equation}
\overline{\widehat{E}(\widehat{\mathbf{k}} \cdot \widehat{\lambda}_0)} 
\approx \frac{2}{\sqrt{\pi} M_A \mathrm{erf}(1/M_A)} \exp \left( - \frac{(\widehat{\mathbf{k}} \cdot \widehat{\lambda}_0)^2}{M_A^2} \right) ~.
\label{eq:hatEmu_averaged}
\end{equation}
Let us also quote a multipole expansion of the 2D projection of this function
\begin{equation}
\overline{\widehat{E}_p^{2D}} = (-1)^{p/2} \; \frac{2 \exp\left(-\onehalf M_{A\perp}^{-2}\right) I_{p/2}\left(\onehalf M_{A\perp}^{-2}\right)}{\sqrt{\pi} M_A \mathrm{erf}(1/M_A)}  ~.
\label{eq:A2D_p}
\end{equation}

\section{RELATIVE ROLE OF $\parallel$ AND $\bot$ MAGNETIC FIELD FLUCTUATIONS FOR LOW $\beta$ CASE }
\label{sec:numerical_modes}
While the main body of the paper deals with the scaling of the projected fluctuations of magnetic field, here we will discuss the properties of 3D fluctuations. These fluctuations are known to be very anisotropic in MHD turbulence and it is only perpendicular to the mean magnetic field component of magnetic fluctuations that contributes to the observed fluctuations in angle.\footnote{This point was initially missed in ST21 and corrected in SX21. There it was assumed that turbulent fluctuations are statistically isotropic, i.e. $\delta B_\|=\delta B_\bot$ for all $M_A<1$. This is a very strong assumption that contradicts to what we know about MHD turbulence. For instance, Alfv\'enic fluctuations are perpendicular to the local direction of magnetic field. The contribution of slow modes is mostly to $\delta B_{\|}$ and it is marginal for turbulence in low-$\beta$ plasmas, while it is only fast modes that are isotropic.}   

For our MHD simulations with solenoidal driving, we show our results for the relative amplitude of magnetic fluctuations in Fig. \ref{fig:fraction}. We see that for low $\beta$ supersonic $M_s=6$ and $M_A=0.1$ turbulence, the fraction of energy in fast modes is around $6 \%$ of Alfv\'en mode energy. At the same time, while the total fraction of energy in slow modes can be comparable or even exceed the energy in Alfv\'en modes, the contributions of the slow modes in the 3D fluctuations of magnetic field is also marginal.   Thus, in our simulations for turbulence in low $\beta$ medium the dispersion of $\delta B_\bot$ scales as $\propto M_A$ in agreement with the DCF approach.

\begin{figure}[ht]
    \centering
    \includegraphics[width=0.66\textwidth]{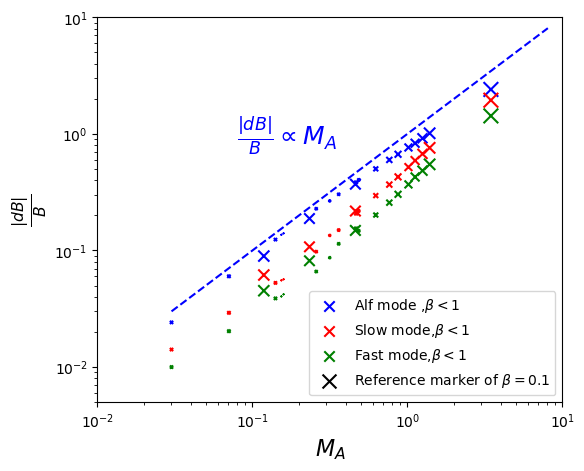}
    \caption{\label{fig:db2b_MA} A figure showing how the 3D magnetic field fluctuations for each fundamental MHD mode (in the global frame of reference) varies as a function of $M_A$ for low $\beta$ simulations. Blue points: Alfv\'en mode, green points: slow mode, red points: fast modes. The trend lines are added for readers' references. The size of the marker is $\propto \beta$ , where we provide a reference marker of size $\beta=0.1$ at the legend.  }

\end{figure}

 Fig.\ref{fig:db2b_MA} shows how the 3D magnetic field fluctuations relative to the guide field vary as a function of the global $M_A$ in the numerical cubes. For low $\beta$ case we can observe the fluctuations arising from slow modes being almost parallel to the ambient field. The contributions of these fluctuations to the variations of magnetic field direction is negligible and for simulations at hand is on the order of accuracy of the decomposition technique. At the same time, we observe that the fluctuations from Alfv\'en modes that are perpendicular to the mean field, i.e. responsible for the variations of $\phi$, scale $\propto M_A$. 

\section{SIMILARITIES AND DIFFERENCES OF SYNCHROTRON AND DUST POLARIZATION STATISTICS}
\label{app:lp12_vs_ch5}

In the current paper we discuss about mainly the statistics of dust polarization, which has been covered in \S 4 in extended detail. However readers might recognize that the fundamental theory of \cite{LP12} relies on the "synchrotron" formalism. Readers might wonder how similar and different are these formalism in our development of DMA?

Let us recall how the dust and synchrotron polarization depends is theoretically written in the form of $P=Q+iU$:
\begin{equation}
\begin{aligned}
    P_{dust} \propto \int dz n_{dust} e^{2i\theta_B}\sin^2\gamma\\
    P_{synch} \propto \int dz n_{re} B_{tot}^2 e^{2i\theta_B}sin^2\gamma
\end{aligned}
\label{eq:dust_vs_synch}
\end{equation}
where we have to remind our readers that $\tan\theta_B = B_y/B_x$, $\gamma$ is the line of sight angle, and $n_{re}$ is the density of relativistic electrons. Notice that the relativistic electron density is well known to be less correlated to the ISM density, while the dust density is believed to the highly correlated to the ISM density (See Draine 2011) and a justification is required to disregard the dust fluctuations (either via \cite{VDA}, or via the treatment in \S 4). The fundamental differences between the dust and synchrotron polarization formulae (Eq.\ref{eq:dust_vs_synch}) is the emergence of the $\sin^2\gamma$ term in the dust polarization formula, while the strong $\propto B_{tot}^2$  dependence as discussed in \cite{LP12}.

\begin{figure}[th]
    \centering
    \includegraphics[width=0.49\textwidth]{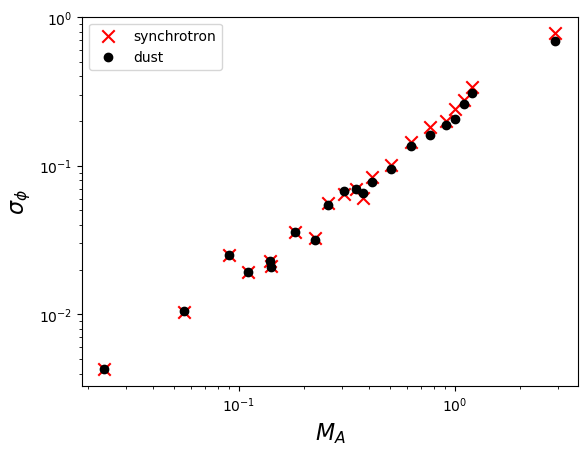}
    \caption{\label{fig:dust_vs_synch} A figure showing how the dispersion of dust (black dot) and synchrotron polarization angle (red cross) varies as a function of $M_A$.}
\end{figure}

In our current paper, we are not concerned with the actual statistics of Q and U in general, but the polarization angle $\phi = \tan_2^{-1}(U/Q)$. The easiest way to compare their performance is to compute the dispersion of polarization angle using the two formula (Eq.\ref{eq:dust_vs_synch}). Fig.\ref{fig:dust_vs_synch} shows how the dispersion of polarization angles vary as a function of $M_A$. We can see that when $M_A\ll 1$ the difference between the dust and synchrotron formalism is infinitesimal, which is expected by theory (See \citealt{LP12}). However, when $M_A>1$, they start have have slight divergence.  When $M_A \ll 1$, the synchrotron polarization formula could be approximated as $P_{synch} \approx B_{tot}^2 \int dz ...$ which the remaining factor within the integral carries the same factor as the dust formula, and the $B_{tot}^2$ term cancels when computing $\phi$. However, when $M_A > 1$ the aforementioned relation does not hold, and therefore we expect deviations between the two expressions. However as one can see from Fig.\ref{fig:dust_vs_synch} that the difference between them is rather minor, we expect the scaling in the main text would change only slightly.

\section{REMOVING EFFECTS OF DENSITY FROM THE POLARIZATION DATA}
\label{polar_VDA}

In \cite{VDA}  provided techniques of removing the density contributions from both the centroid and Stokes parameters. Here we discuss how the corresponding procedure can it be implemented for polarization data and what is the constrains the procedure accuracy.  The form of the decomposition equation that is employed for the polarization is  the same its velocity version. In \cite{VDA} it was shown that the employed linear-algebra construction works best for subsonic media. We illustrate this effect using our numerical simulations.

From \cite{VDA}, we can write the velocity caustics $p_v$ as 
\begin{equation}
\begin{aligned}
    p_v &= p - \left( \langle pI\rangle-\langle p\rangle\langle I \rangle\right)\frac{I-\langle I\rangle}{\sigma_I^2}
\end{aligned}
\label{eq:VDA_ld2}
\end{equation}
where $p$ is the channel map, and $I$ is the total column density map. Very similarly, we can also compute the Stokes variant version:

\begin{equation}
\begin{aligned}
    Q_v &= Q - \left( \langle QI\rangle-\langle Q\rangle\langle I \rangle\right)\frac{I-\langle I\rangle}{\sigma_I^2}\\
    U_v &= U - \left( \langle UI\rangle-\langle U\rangle\langle I \rangle\right)\frac{I-\langle I\rangle}{\sigma_I^2}
\end{aligned}
\label{eq:VDA_stokes}
\end{equation} 
where $I$ here is the unpolarized Stokes intensity. Notice that the formulae (Eq.\ref{eq:VDA_stokes}) has to be tested to see whether they work.

\begin{figure}[th]
    \centering
    \includegraphics[width=0.49\textwidth]{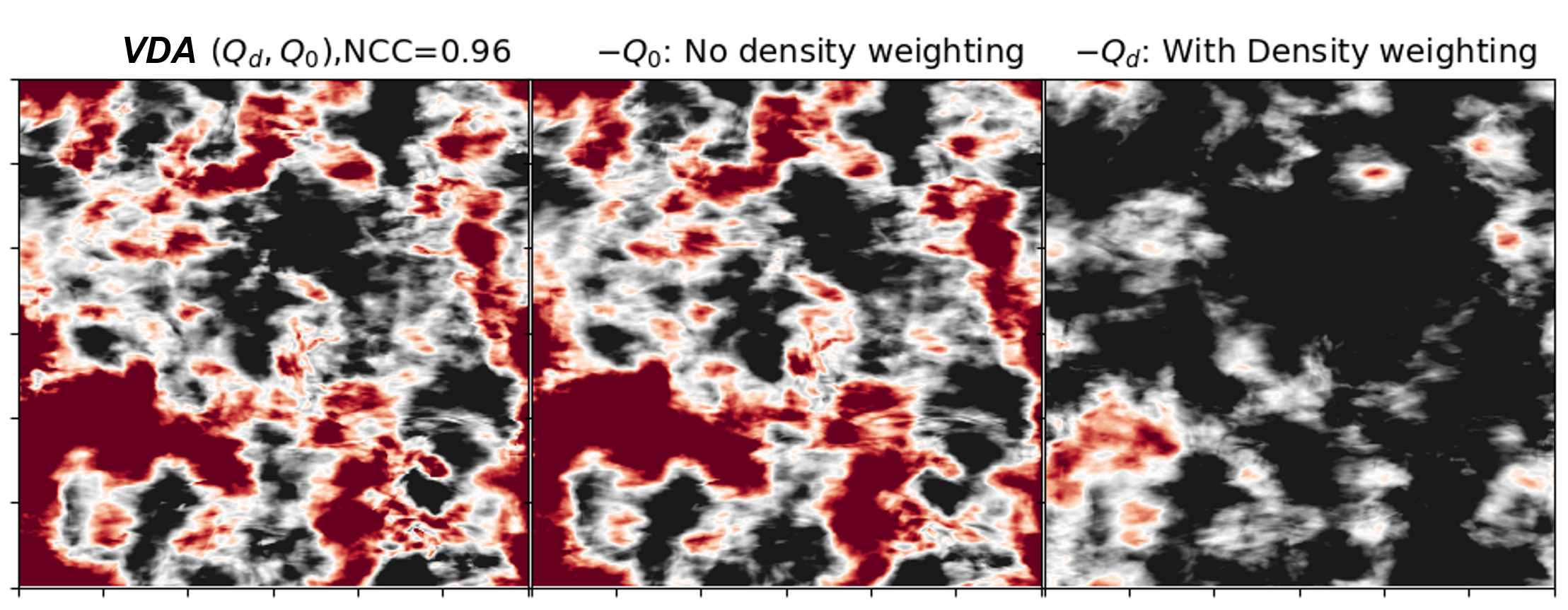}
    \includegraphics[width=0.49\textwidth]{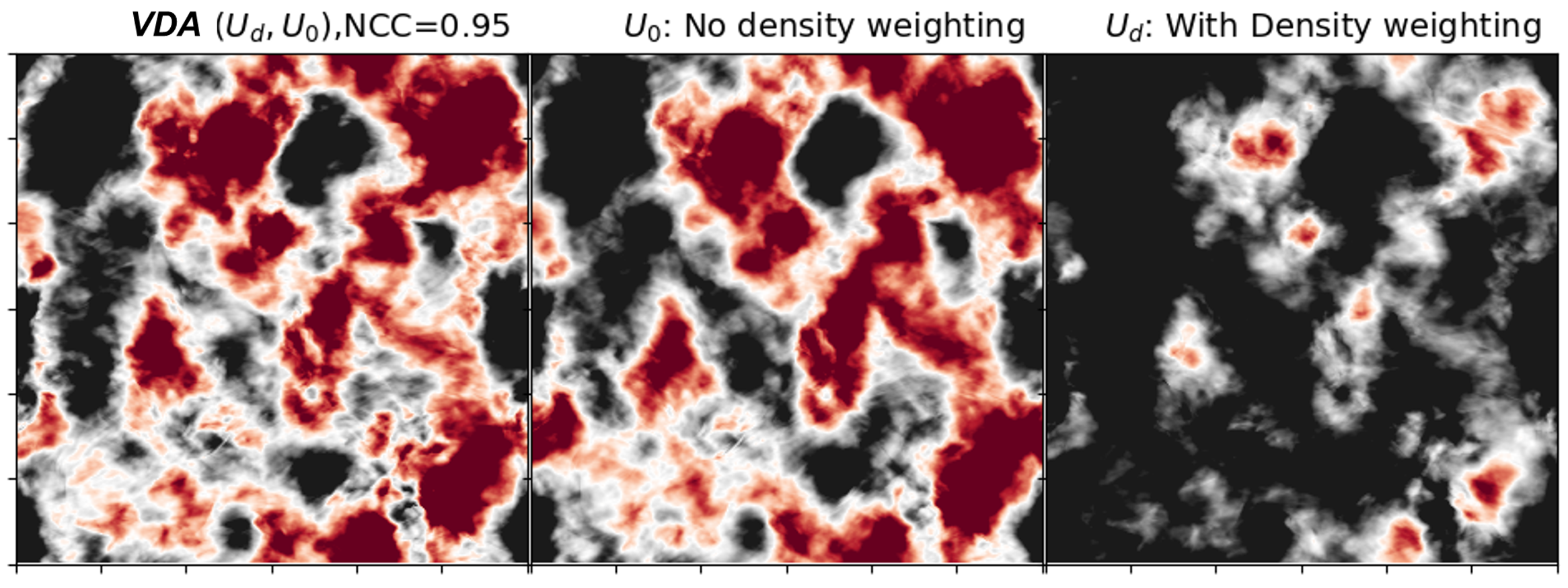}
    \caption{\label{fig:StokesDDA_2D_e5r3} A simple decomposition using \cite{VDA} for dust polarization (Q,U) from subsonic turbulence. From the left: The decomposed velocity-only Q (upper row) and U (lower row); middle: the constant density Stokes parameter; right: the density-weighted Stokes parameter from dust emission. }
\end{figure}

\begin{figure}[th]
    \centering
    \includegraphics[width=0.49\textwidth]{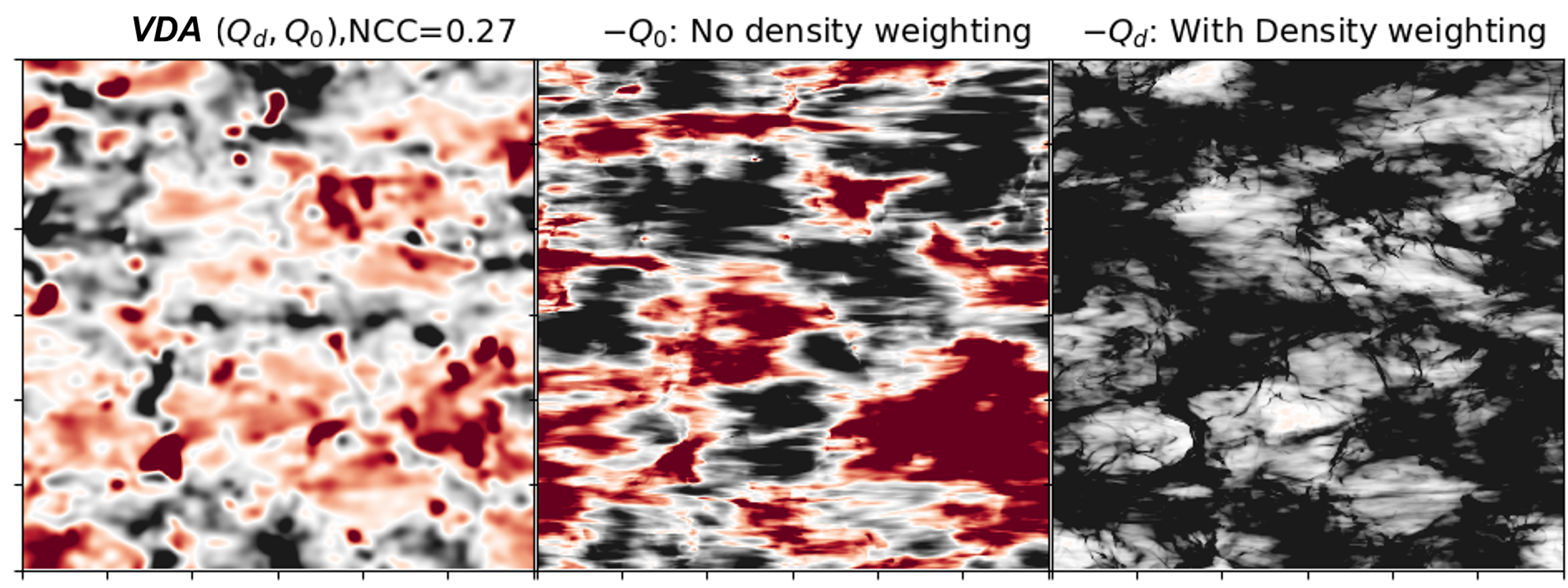}
    \includegraphics[width=0.49\textwidth]{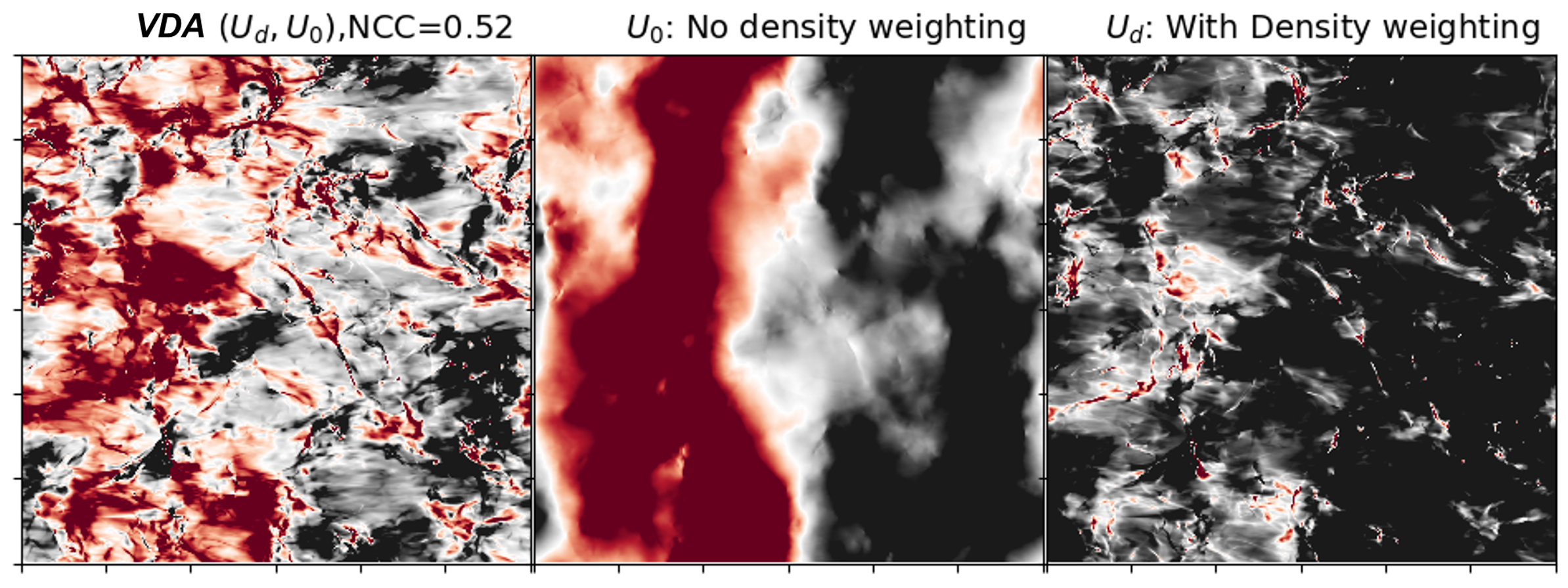}
    \caption{\label{fig:StokesDDA_2D_h01200} A simple decomposition using \cite{VDA} for dust polarization (Q,U) from supersonic turbulence. From the left: The decomposed velocity-only Q (upper row) and U (lower row); middle: the constant density Stokes parameter; right: the density-weighted Stokes parameter from dust emission.}
\end{figure}

Fig.\ref{fig:StokesDDA_2D_e5r3} shows the decomposition of Q and U in subsonic turbulence while that of Fig.\ref{fig:StokesDDA_2D_h01200} shows the same decomposition in Stokes parameters synthesized from supersonic turbulence. We can see that the result that we obtained here is consistent with the spectroscopic equivalent in \cite{VDA}: The subsonic decomposition can recover the constant density Stokes parameter with the Normalized Correlation Coefficient (NCC, see \citealt{VDA}) to be very close to 1, while that of the supersonic case we can somehow recover the structure but the NCC drops below 0.6. This simple illustration suggests that we can remove the density contribution easily and with high accuracy in the case of subsonic turbulence. 

Notice that our technique also applies to any single-frequency synchrotron emissions {\it without significant Faraday rotation}, because the density dependence in both dust and synchrotron emissions are the same. However, without multi-frequency emissions it is impossible for us to obtain the density-free information in supersonic turbulence.

\begin{thebibliography}{}
\providecommand\natexlab[1]{#1}
\providecommand\JournalTitle[1]{#1}
\bibitem[Alfv{\'e}n(1942)]{1942Natur.150..405A} Alfv{\'e}n, H.\ 1942, \nat, 150, 405
\bibitem[Andersson et al.(2015)]{Aetal15} Andersson, B.-G., Lazarian, A., \& Vaillancourt, J.~E.\ 2015, \araa, 53, 501 
\bibitem[Armstrong et al.(1995)]{1995ApJ...443..209A} Armstrong, J.~W., Rickett, B.~J., \& Spangler, S.~R.\ 1995, \apj, 443, 209
\bibitem[Brandenburg \& Lazarian(2013)]{BL13} Brandenburg, A., \& Lazarian, A.\ 2013, \ssr, 178, 163
\bibitem[Beattie et al.(2022)]{2022arXiv220213020B} Beattie, J.~R., Krumholz, M.~R., Skalidis, R., et al.\ 2022, arXiv:2202.13020
\bibitem[Beresnyak \& Lazarian(2019)]{2019tuma.book.....B} Beresnyak, A., \& Lazarian, A.\ 2019, Turbulence in Magnetohydrodynamics
\bibitem[Biskamp(2003)]{2003matu.book.....B} Biskamp, D.\ 2003, Magnetohydrodynamic Turbulence
\bibitem[Brunetti \& Lazarian(2007)]{BL07} Brunetti, G., \& Lazarian, A.\ 2007, \mnras, 378, 245 
\bibitem[Burkhart et al.(2010)]{2010ApJ...708.1204B} Burkhart, B., Stanimirovi{\'c}, S., Lazarian, A., et al.\ 2010, \apj, 708, 1204
\bibitem[Burkhart \& Lazarian(2012)]{BL12} Burkhart, B., \& Lazarian, A.\ 2012, \apjl, 755, L19 
\bibitem[Burkhart et al.(2014)]{Betal14} Burkhart, B., Lazarian, A., Le{\~a}o, I.~C., de Medeiros, J.~R., \& Esquivel, A.\ 2014, \apj, 790, 130
\bibitem[Caldwell et al.(2017)]{2017ApJ...839...91C} Caldwell, R.~R., Hirata, C., \& Kamionkowski, M.\ 2017, \apj, 839, 91
\bibitem[Chandrasekhar \& Fermi(1953)]{CF53} Chandrasekhar, S., \& Fermi, E.\ 1953, \apj, 118, 113 
\bibitem[Chepurnov \& Lazarian(2009)]{CL09} Chepurnov, A., \& Lazarian, A.\ 2009, \apj, 693, 1074 
\bibitem[Chepurnov \& Lazarian(2010)]{2010ApJ...710..853C} Chepurnov, A., \& Lazarian, A.\ 2010, \apj, 710, 853
\bibitem[Cort{\'e}s et al.(2021)]{2021ApJ...923..204C} Cort{\'e}s, P.~C., Sanhueza, P., Houde, M., et al.\ 2021, \apj, 923, 204. doi:10.3847/1538-4357/ac28a1
\bibitem[Chitsazzadeh et al.(2012)]{2012ApJ...749...45C} Chitsazzadeh, S., Houde, M., Hildebrand, R.~H., \& Vaillancourt, J.\ 2012, \apj, 749, 45
\bibitem[Cho \& Lazarian(2002)]{CL02} Cho, J., \& Lazarian, A.\ 2002, Physical Review Letters, 88, 245001 
\bibitem[{Cho \& Lazarian(2003)}]{CL03} \mnras, \ 2003, 345, 325
\bibitem[Cho \& Vishniac(2000)]{CV00} Cho, J., \& Vishniac, E.~T.\ 2000, \apj, 539, 273 
\bibitem[Cho \& Yoo(2016)]{CY16} Cho, J., \& Yoo, H.\ 2016, \apj, 821, 21 
\bibitem[Cho (2017)]{Cho17} Cho, J.\ 2017, Journal of Physics Conference Series, 012002
\bibitem[Cho (2019)]{Cho19} Cho, J.\ 2019, \apj, 874, 75
\bibitem[Cho et al.(2002)]{CLV02} Cho, J., Lazarian, A., \& Vishniac, E.~T.\ 2002, \apjl, 566, L49 
\bibitem[Clark et al.(2015)]{2015PhRvL.115x1302C} Clark, S.~E., Hill, J.~C., Peek, J.~E.~G., Putman, M.~E., \& Babler, B.~L.\ 2015, Physical Review Letters, 115, 241302 
\bibitem[Crutcher(2004)]{2004Ap&SS.292..225C} Crutcher, R.~M.\ 2004, \apss, 292, 225
\bibitem[Crutcher(2010)]{C10} Crutcher, R.\ 2010, From Stars to Galaxies: Connecting our Understanding of Star and Galaxy Formation, 3 
\bibitem[Davis(1951)]{1951PhRv...81..890D} Davis, L.\ 1951, Physical Review, 81, 890
\bibitem[Dolginov \& Mytrophanov(1976)]{1976Ap&SS..43..257D} Dolginov, A.~Z., \& Mytrophanov, I.~G.\ 1976, \apss, 43, 257
\bibitem[Draine \& Weingartner(1996)]{1996AAS...189.1602D} Draine, B.~T., \& Weingartner, J.~C.\ 1996, American Astronomical Society Meeting Abstracts 189, 16.02
\bibitem[Draine(2006)]{2006ApJ...636.1114D} Draine, B.~T.\ 2006, \apj, 636, 1114
\bibitem[Elmegreen \& Scalo(2004)]{2004ARA&A..42..211E} Elmegreen, B.~G., \& Scalo, J.\ 2004, \araa, 42, 211
\bibitem[Esquivel \& Lazarian(2005)]{EL05} Esquivel, A., \& Lazarian, A.\ 2005, \apj, 631, 320
\bibitem[Esquivel \& Lazarian(2010)]{EL10} Esquivel, A., \& Lazarian, A.\ 2010, \apj, 710, 125
\bibitem[Esquivel \& Lazarian(2011)]{EL11} Esquivel, A., \& Lazarian, A.\ 2011, \apj, 740, 117 
\bibitem[Esquivel et al.(2015)]{Eetal15} Esquivel, A., Lazarian, A., \& Pogosyan, D.\ 2015, \apj, 814, 77 
\bibitem[Eswaraiah et al.(2021)]{2021arXiv210302219E} Eswaraiah, C., Li, D., Furuya, R.~S., et al.\ 2021, arXiv:2103.02219
\bibitem[Falceta-Gon{\c c}alves et al.(2008)]{Fal08} Falceta-Gon{\c c}alves, D., Lazarian, A., \& Kowal, G.\ 2008, \apj, 679, 537-551
\bibitem[Falcon et al.(2007)]{2007PhRvL..98o4501F} Falcon, E., Fauve, S., \& Laroche, C.\ 2007, \prl, 98, 154501
\bibitem[Gaensler et al.(2011)]{2011Natur.478..214G} Gaensler, B.~M., Haverkorn, M., Burkhart, B., et al.\ 2011, \nat, 478, 214
\bibitem[Galli et al.(2006)]{2006ApJ...647..374G} Galli, D., Lizano, S., Shu, F.~H., et al.\ 2006, \apj, 647, 374
\bibitem[Girart et al.(1999)]{1999ApJ...525L.109G} Girart, J.~M., Crutcher, R.~M., \& Rao, R.\ 1999, \apjl, 525, L109. doi:10.1086/312345

\bibitem[Girart et al.(2006)]{2006Sci...313..812G} Girart, J.~M., Rao, R., \& Marrone, D.~P.\ 2006, Science, 313, 812

\bibitem[Goldreich, \& Kylafis(1981)]{1981ApJ...243L..75G} Goldreich, P., \& Kylafis, N.~D.\ 1981, \apjl, 243, L75
\bibitem[Goldreich, \& Kylafis(1982)]{1982ApJ...253..606G} Goldreich, P., \& Kylafis, N.~D.\ 1982, \apj, 253, 606
\bibitem[Goldreich \& Sridhar(1995)]{GS95} Goldreich, P., \& Sridhar, S.\ 1995, \apj, 438, 763 
\bibitem[Gonz{\'a}lez-Casanova \& Lazarian(2017)]{GCL17} Gonz{\'a}lez-Casanova, D.~F., \& Lazarian, A.\ 2017, \apj, 835, 41 
\bibitem[Ha et al.(2021)]{2021ApJ...907L..40H} Ha, T., Li, Y., Xu, S., et al.\ 2021, \apjl, 907, L40. %doi:10.3847/2041-8213/abd8c9

\bibitem[Haugen \& Brandenburg(2004)]{2004PhRvE..70c6408H} Haugen, N.~E.~L. \& Brandenburg, A.\ 2004, \pre, 70, 036408.% doi:10.1103/PhysRevE.70.036408

\bibitem[Heitsch et al.(2001)]{2001ApJ...561..800H} Heitsch, F., Zweibel, E.~G., Mac Low, M.-M., Li, P., \& Norman, M.~L.\ 2001, \apj, 561, 800

\bibitem[Heyer \& Brunt(2004)]{2004ApJ...615L..45H} Heyer, M.~H., \& Brunt, C.~M.\ 2004, \apjl, 615, L45

{\torefereetwo
\bibitem[Hezareh et al.(2013)]{2013A&A...558A..45H} Hezareh, T., Wiesemeyer, H., Houde, M., et al.\ 2013, \aap, 558, A45. %doi:10.1051/0004-6361/201321900
}
\bibitem[Hildebrand et al.(2009)]{2009ApJ...696..567H} Hildebrand, R.~H., Kirby, L., Dotson, J.~L., Houde, M., \& Vaillancourt, J.~E.\ 2009, \apj, 696, 567
\bibitem[Hoang \& Lazarian(2008)]{2008MNRAS.388..117H} Hoang, T., \& Lazarian, A.\ 2008, \mnras, 388, 117
\bibitem[Hoang \& Lazarian(2016)]{2016ApJ...831..159H} Hoang, T., \& Lazarian, A.\ 2016, \apj, 831, 159
\bibitem[Houde(2004)]{2004ApJ...616L.111H} Houde, M.\ 2004, \apjl, 616, L111
\bibitem[Houde et al.(2009)]{2009ApJ...706.1504H} Houde, M., Vaillancourt, J.~E., Hildebrand, R.~H., Chitsazzadeh, S., \& Kirby, L.\ 2009, \apj, 706, 1504
\bibitem[Houde et al.(2011)]{2011ApJ...733..109H} de, M., Rao, R., Vaillancourt, J.~E., \& Hildebrand, R.~H.\ 2011, \apj, 733, 109 

\bibitem[Houde et al.(2013)]{2013ApJ...766...49H} Houde, M., Fletcher, A., Beck, R., et al.\ 2013, \apj, 766, 49


\bibitem[Houde et al.(2016)]{2016ApJ...820...38H} Houde, M., Hull, C.~L.~H., Plambeck, R.~L., Vaillancourt, J.~E., \& Hildebrand, R.~H.\ 2016, \apj, 820, 38

\bibitem[Ho et al. (2021)]{HO21} Ho,K. W. , Yuen,K. H. , Lazarian, A., 2021, submitted to \mnras

\bibitem[Ho et al. (2022)]{HO22} Ho,K. W. , Yuen,K. H. , Lazarian, A., 2022,in perp

\bibitem[Hu et al.(2019a)]{survey} Hu, Y., Yuen, K. H.,  Lazarian V., et al. \ 2019, Nature Astronomy

\bibitem[Hu et al.(2019b)]{velac} Hu, Y., Yuen, K.~H., Lazarian, A., et al.\ 2019, \apj , arXiv:1904.04391.
\bibitem[Hu et al.(2019c)]{IGvHRO} Hu, Y., Yuen, K.~H., \& Lazarian, A.\ 2019, \apj, 886, 17

\bibitem[Johns-Krull(2007)]{2007IAUS..243...31J} Johns-Krull, C.~M.\ 2007, Star-disk Interaction in Young Stars, 31

\bibitem[Kandel et al.(2016)]{KLP16} Kandel, D., Lazarian, A., \& Pogosyan, D.\ 2016, \mnras, 461, 1227 
\bibitem[Kandel et al.(2017a)]{KLP17a} Kandel, D., Lazarian, A., \& Pogosyan, D.\ 2017, \mnras, 464, 3617 
\bibitem[Kandel et al.(2017b)]{KLP17b} Kandel, D., Lazarian, A., \& Pogosyan, D.\ 2017, \mnras, 470, 3103 
\bibitem[Kandel et al.(2018)]{KLP18} Kandel, D., Lazarian, A., \& Pogosyan, D.\ 2018, \mnras, 478, 530
\bibitem[Kobulnicky et al.(1994)]{1994AJ....107.1433K} Kobulnicky, H.~A., Molnar, L.~A., \& Jones, T.~J.\ 1994, \aj, 107, 1433
\bibitem[Kowal \& Lazarian(2007)]{2007ApJ...666L..69K} Kowal, G., \& Lazarian, A.\ 2007, \apjl, 666, L69
\bibitem[Kowal \& Lazarian(2010)]{2010ApJ...720..742K} Kowal, G., \& Lazarian, A.\ 2010, \apj, 720, 742
\bibitem[Lankhaar \& Vlemmings(2020)]{2020A&A...636A..14L} Lankhaar, B., \& Vlemmings, W.\ 2020, \aap, 636, A14
\bibitem[Larson(1981)]{1981MNRAS.194..809L} Larson, R.~B.\ 1981, \mnras, 194, 809
\bibitem[Lazarian \& Esquivel(2003)]{LE03} Lazarian, A., \& Esquivel, A.\ 2003, \apjl, 592, L37 
\bibitem[Lazarian \& Hoang(2007)]{2007MNRAS.378..910L} Lazarian, A., \& Hoang, T.\ 2007, \mnras, 378, 910
\bibitem[Lazarian \& Hoang(2019)]{2019ApJ...883..122L} Lazarian, A., \& Hoang, T.\ 2019, \apj, 883, 122

\bibitem[Lazarian \& Pogosyan(2000)]{LP00} Lazarian, A., \& Pogosyan, D.\ 2000, \apj, 537, 720 
\bibitem[Lazarian \& Pogosyan(2004)]{LP04} Lazarian, A., \& Pogosyan, D.\ 2004, \apj, 616, 943 
\bibitem[Lazarian \& Pogosyan(2006)]{LP06} Lazarian, A., \& Pogosyan, D.\ 2006, \apj, 652, 1348 
\bibitem[Lazarian \& Pogosyan(2008)]{LP08} Lazarian, A., \& Pogosyan, D.\ 2008, \apj, 686, 350 
\bibitem[Lazarian \& Pogosyan(2012)]{LP12} Lazarian, A., \& Pogosyan, D.\ 2012, \apj, 747, 5 (LP12)
\bibitem[Lazarian \& Pogosyan(2016)]{LP16} Lazarian, A., \& Pogosyan, D.\ 2016, \apj, 818, 178 (LP16)
\bibitem[Lazarian \& Vishniac(1999)]{LV99} Lazarian, A., \& Vishniac, E.~T.\ 1999, \apj, 517, 700 
\bibitem[Lazarian \& Yuen(2018a)]{LY18a} Lazarian, A., \& Yuen, K.~H.\ 2018, \apj, 853, 96 
\bibitem[Lazarian \& Yuen(2018b)]{LY18b} Lazarian, A., \& Yuen, K.~H.\ 2018, arXiv:1802.00028 
\bibitem[Lazarian et al.(2002)]{Letal02} Lazarian, A., Pogosyan, D., \& Esquivel, A.\ 2002, Seeing Through the Dust: The Detection of HI and the Exploration of the ISM in Galaxies, 276, 182 
\bibitem[Lazarian et al.(2017)]{Letal17} Lazarian, A., Yuen, K.~H., Lee, H., \& Cho, J.\ 2017, \apj, 842, 30
\bibitem[Lazarian et al.(2018)]{2018ApJ...865...46L} Lazarian, A., Yuen, K.~H., Ho, K.~W., et al.\ 2018, \apj, 865, 46

\bibitem[Lazarian(2007)]{L07} Lazarian, A.\ 2007, \jqsrt, 106, 225 
\bibitem[Lazarian(2016)]{L16} Lazarian, A.\ 2016, \apj, 833, 131 
\bibitem[Li et.al (2011)]{li} Li, Hua-Bai \& Henning, T. 2011, Nature, 479, 499
\bibitem[Li et al.(2021)]{2021MNRAS.tmp.3119L} Li, P.~S., Lopez-Rodriguez, E., Ajeddig, H., et al.\ 2021, \mnras. doi:10.1093/mnras/stab3448

\bibitem[Lithwick \& Goldreich(2001)]{LG01} Lithwick, Y., \& Goldreich, P.\ 2001, \apj, 562, 279 
\bibitem[Lu et al.(2019)]{Lu19} Lu, Z., Lazarian, A., \& Pogosyan, D.\ 2019, arXiv e-prints, arXiv:1910.02226
\bibitem[Makwana \& Yan (2020)]{2020PhRvX..10c1021M} Makwana, K., \& Yan, H.\ 2020, Physics Review X, 10,
\bibitem[Maron \& Goldreich(2001)]{MG01} Maron, J., \& Goldreich, P.\ 2001, \apj, 554, 1175 

\bibitem[Mestel \& Spitzer(1956)]{1956MNRAS.116..503M} Mestel, L., \& Spitzer, L.\ 1956, \mnras, 116, 503
\bibitem[McKee \& Ostriker(2007)]{MO07} McKee, C.~F., \& Ostriker, E.~C.\ 2007, \araa, 45, 565 
\bibitem[Monin \& Yaglom (1975)]{MY75} Monin, A. \& Yaglom, A. 1975, Statistical Fluid Mechanics: Mechanics of Turbulence, MIT
\bibitem[Mouschovias et al.(2006)]{2006ApJ...646.1043M} Mouschovias, T.~C., Tassis, K., \& Kunz, M.~W.\ 2006, \apj, 646, 1043
\bibitem[Padoan et al.(2009)]{2009ApJ...707L.153P} Padoan, P., Juvela, M., Kritsuk, A., et al.\ 2009, \apjl, 707, L153
\bibitem[Peek et al.(2018)]{susantail} Peek, J.~E.~G., Babler, B.~L., Zheng, Y., et al.\ 2018, \apjs, 234, 2 
\bibitem[Skalidis \& Tassis(2021)]{2021A&A...647A.186S} Skalidis, R. \& Tassis, K.\ 2021, \aap, 647, A186. doi:10.1051/0004-6361/202039779
\bibitem[Soler et al.(2013)]{Setal13} Soler, J.~D., Hennebelle, P., Martin, P.~G., et al.\ 2013, \apj, 774, 128 
\bibitem[Truelove et al.(1997)]{1997ApJ...489L.179T} Truelove, J.~K., Klein, R.~I., McKee, C.~F., et al.\ 1997, \apjl, 489, L179
\bibitem[Tofflemire et al.(2011)]{2011ApJ...736...60T} Tofflemire, B.~M., Burkhart, B., \& Lazarian, A.\ 2011, \apj, 736, 60
\bibitem[Xu \& Zhang(2016a)]{2016ApJ...824..113X} Xu, S., \& Zhang, B.\ 2016, \apj, 824, 113
\bibitem[Xu \& Zhang(2016b)]{2016ApJ...832..199X} Xu, S., \& Zhang, B.\ 2016, \apj, 832, 199

\bibitem[Yan \& Lazarian(2002)]{YanL02} Yan, H., \& Lazarian, A.\ 2002, Physical Review Letters, 89, 281102 
\bibitem[Yan \& Lazarian(2003)]{YanL03} Yan, H., \& Lazarian, A.\ 2003, arXiv:astro-ph/0311369
\bibitem[Yan \& Lazarian(2004)]{YanL04} Yan, H., \& Lazarian, A.\ 2004, \apj, 614, 757
\bibitem[Yan \& Lazarian(2006)]{2006ApJ...653.1292Y} Yan, H., \& Lazarian, A.\ 2006, \apj, 653, 1292
\bibitem[Yan \& Lazarian(2007)]{2007ApJ...657..618Y} Yan, H., \& Lazarian, A.\ 2007, \apj, 657, 618
\bibitem[Yan \& Lazarian(2008)]{2008ApJ...677.1401Y} Yan, H., \& Lazarian, A.\ 2008, \apj, 677, 1401
\bibitem[Yan \& Lazarian(2012)]{2012JQSRT.113.1409Y} Yan, H., \& Lazarian, A.\ 2012, \jqsrt, 113, 1409
\bibitem[Yoo \& Cho(2014)]{YC14} Yoo, H., \& Cho, J.\ 2014, \apj, 780, 99 
\bibitem[Yoon \& Cho(2019)]{YC19} Yoon, H., \& Cho, J.\ 2019, \apj, 880, 137
\bibitem[Yuen \& Lazarian(2017a)]{YL17a} Yuen, K.~H., \& Lazarian, A.\ 2017, \apjl, 837, L24 
\bibitem[Yuen \& Lazarian(2017b)]{YL17b} Yuen, K.~H., \& Lazarian, A.\ 2017, arXiv:1703.03026 
\bibitem[Yuen et al.(2018)]{2018ApJ...865...54Y} Yuen, K.~H., Chen, J., Hu, Y., et al.\ 2018, \apj, 865, 54.

\bibitem[Yuen et al.(2019)]{reply19} Yuen, K.~H., Hu, Y., Lazarian, A., \& Pogosyan, D.\ 2019, arXiv:1904.03173 

\bibitem[Yuen \& Lazarian (2020a)]{GA} Yuen, K.~H. \&  Lazarian, A., 2020, \apj, 898, 65


\bibitem[Yuen \& Lazarian (2020b)]{curvature} Yuen, K.~H. \& Lazarian, A., 2020, \apj, 898,66
\bibitem[Yuen et.al (2021)]{VDA} Yuen, K.~H. Ho, K.~W. \& Lazarian, A., 2021, \apj, 910,2 (VDA)

\bibitem[Yuen et al. (2022a)]{spectrum} 
Yuen, K.~H.,Ho, K.~W., Law, C.~Y, Chen,A. \& Lazarian, A., 2022, submitted

\bibitem[Yuen et al. (2022b)]{leakage} 
Yuen, K.~H.,Yan, H. \& Lazarian, A., 2022, submitted to \mnras

\bibitem[Yuen et al. (2022c)]{cattail} 
Yuen, K.~H.,Chen,A., Ho, K.~W., \& Lazarian, A., 2022, submitted to \mnras

\bibitem[Zhang et al.(2015)]{2015ApJ...804..142Z} Zhang, H., Yan, H., \& Dong, L.\ 2015, \apj, 804, 142. doi:10.1088/0004-637X/804/2/142

\bibitem[Zhang \& Yan(2018)]{2018MNRAS.475.2415Z} Zhang, H. \& Yan, H.\ 2018, \mnras, 475, 2415. doi:10.1093/mnras/stx3164


\bibitem[Zhang et al.(2018)]{2018MNRAS.479.3923Z} Zhang, H., Yan, H., \& Richter, P.\ 2018, \mnras, 479, 3923. doi:10.1093/mnras/sty1594



\bibitem[Zhang et al.(2020a)]{2020ApJ...902L...7Z} Zhang, H., Gangi, M., Leone, F., et al.\ 2020, \apjl, 902, L7. doi:10.3847/2041-8213/abb8e1


\bibitem[Zhang et al.(2020b)]{2020NatAs.tmp..174Z} Zhang, H., Chepurnov, A., Yan, H., et al.\ 2020, Nature Astronomy. doi:10.1038/s41550-020-1093-4

\end{thebibliography}
\end{document}